\documentclass[usenatbib]{mn2e}
\usepackage{graphicx}
\usepackage{times}
\usepackage{color}
\usepackage{amsmath}
\usepackage{epstopdf}

\DeclareGraphicsExtensions{.pdf,.png,.jpg}

\title[High-z galaxies with mechanical supernova feedback]
{Towards simulating star formation in turbulent high-$z$ galaxies with mechanical supernova feedback }
\author[T. Kimm et al.]{  
\parbox[t]{\textwidth}{
Taysun Kimm$^{1}$\thanks{e-mail: kimm@astro.princeton.edu}, 
Renyue Cen$^{1}$,
Julien Devriendt$^{2,3}$, 
Yohan Dubois$^{4,5}$,
and Adrianne Slyz$^2$
} 
\vspace*{6pt} \\
$^1$ Princeton University Observatory, Peyton Hall, 4 Ivy Lane, Princeton, NJ 08544-1001, USA\\
$^2$ Astrophysics, University of Oxford, Denys Wilkinson Building, Keble Road, Oxford OX1 3RH, UK\\
$^3$ Observatoire de Lyon, UMR 5574, 9 avenue Charles Andr\'{e}, Saint Genis Laval 69561, France \\
$^4$ Sorbonne Universit\'{e}s, UPMC Univ. Paris 06, UMR 7095, Institut d'Astrophysique de Paris, F-75005 Paris, France \\
$^5$ CNRS, UMR 7095, Institut d'Astrophysique de Paris, 98 bis Boulevard Arago, F-75014 Paris, France
}

\begin{document}
\maketitle

\def\lesssim{\lower.5ex\hbox{$\; \buildrel < \over \sim \;$}}
\def\gtrsim{\lower.5ex\hbox{$\; \buildrel > \over \sim \;$}}
\newcommand{\fbar}{\mbox{$f_{\rm bar}$}}
\newcommand{\mbar}{\mbox{$m_{\rm bar}$}}
\newcommand{\nH}{\mbox{$n_{\rm H}$}}
\newcommand{\kms}{\mbox{${\rm km\,s^{-1}}$}}
\newcommand{\msun}{\mbox{$\rm M_\odot$}}
\newcommand{\msunyr}{\mbox{$\rm M_\odot\,{\rm yr^{-1}}$}}
\newcommand{\mvir}{\mbox{$M_{\rm vir}$}}
\newcommand{\mgas}{\mbox{$M_{\rm gas}$}}
\newcommand{\mstar}{\mbox{$M_{\rm star}$}}
\newcommand{\sfr}{\mbox{$\dot{M}_{\rm star}$}}
\newcommand{\mhalo}{\mbox{$M_{\rm halo}$}}
\newcommand{\rvir}{\mbox{$R_{\rm vir}$}}
\newcommand{\mn}{\mbox{{\sc \small Horizon}-MareNostrum}}
\newcommand{\nut}{\mbox{{\sc \small Nut}}}
\newcommand{\ramses}{\mbox{{\sc \small Ramses}}}
\newcommand{\nth}{\mbox{$n_{\rm SF}$}}
\newcommand{\cmq}{\mbox{${\rm cm^{-3}}$}}

\newcommand{\araa}{\mbox{ARA\&A}}
\newcommand{\aap}{\mbox{A\&A}}
\newcommand{\apj}{\mbox{ApJ}}
\newcommand{\aj}{\mbox{AJ}}
\newcommand{\apjl}{\mbox{ApJL}}
\newcommand{\apjs}{\mbox{ApJS}}
\newcommand{\mnras}{\mbox{MNRAS}}
\newcommand{\nat}{\mbox{Nature}}
\newcommand{\pasp}{\mbox{PASP}}

\newcommand{\TK}{\bf \color{blue} }

\begin{abstract}
To better understand the impact of supernova (SN) explosions on the evolution of galaxies, 
we perform a suite of high-resolution (12 pc), zoom-in cosmological simulations of a Milky 
Way-like galaxy at $z=3$ with adaptive mesh refinement. We find that SN explosions can 
efficiently regulate star formation, leading to the stellar mass and metallicity consistent with
 the observed mass-metallicity relation and stellar mass-halo mass relation at $z\sim3$. 
 This is achieved by making three important changes to the classical feedback scheme: i) the 
 different phases of SN blast waves are modelled directly by injecting radial momentum expected 
 at each stage, ii) the realistic time delay of SNe is required to disperse very dense gas 
 before a runaway collapse sets in, and iii) a non-uniform density distribution of the interstellar 
 medium (ISM) is taken into account below the computational grid scale for the cell in 
 which a SN explodes. The simulated galaxy with the SN feedback model shows strong outflows, 
 which carry approximately ten times larger mass than star formation rate, as well as smoothly 
 rising circular velocity. Although the metallicity of the outflow depends sensitively on the feedback 
 model used, we find that the accretion rate and metallicity of the cold flow around the virial radius 
 is impervious to SN feedback. Our results suggest that understanding the structure of the turbulent
 ISM may be crucial to assess the role of SN and other feedback processes in galaxy formation theory.
\end{abstract}


\voffset=-0.6in
\hoffset=0.2in

\section{Introduction}

Outflows are common in actively star-forming galaxies.
Multi-wavelength images of a local starburst galaxy, M82, 
exhibit a clear bipolar outflow, with X-ray emitting gas at the centre 
\citep[e.g.][]{mccarthy87,strickland09}. 
Lyman-break galaxies (LBGs) at high redshifts ($z\sim2-3$) ubiquitously reveal 
significant blueshifted absorption of low-ionization metal lines, such 
as {\sc cii} $\lambda$1334 \citep{shapley03,steidel10}.
These outflows often carry a large amount of gas that is comparable to or even several 
times larger than the mass of newly formed stars \citep{martin99,martin05,steidel10,newman12,chisholm14}, 
indicating that stellar feedback is the main process shaping galaxy properties. 
However, little is yet known about which types of stellar feedback mechanisms are 
most important in regulating star formation in the baryon-driven
picture of galaxy formation

Studies show that a variety of galactic phenomena can naturally be explained by 
the presence of feedback from stars. \citet{shetty12,kim13} demonstrate that 
energetic supernova (SN) explosions can drive turbulence in the interstellar 
medium (ISM), resulting in a low star formation efficiency of $\sim1\%$ per dynamical 
timescale of a galaxy, 
consistent with galactic scale observations \citep{kennicutt98,bigiel08,evans09}.
The evolution of the mass-metallicity relation \citep{erb06,maiolino08,zahid13,steidel14}  
also seems to require that metals are not instantaneously recycled to form the next 
generation stars, but  instead are first blown away from the galaxy \citep{dalcanton07,finlator08}.
Finally, semi-analytic calculations of galaxy formation suggest that, in order to match the 
galaxy luminosity functions below $L_*$  \citep[e.g.][]{bell03} or the stellar mass to halo 
mass relation \citep[e.g.][]{moster10,guo10}, the regulation of star formation by SN
needs to be very effective so that only a small fraction ($\lesssim 20\%$) of baryons  
turns into stars \citep[e.g.][references therein]{cole00,khochfar07,somerville08}.

Attempts have been made to understand the formation of galaxies with SN feedback 
in a fully cosmological context using numerical methods \citep{katz92,cen92,navarro93}.
However, \citet{navarro00} find that simulated galaxies with kinetic feedback
rotate too fast, inconsistent with the local Tully-Fisher relation \citep{tully77}. 
The angular momentum catastrophe 
is attributed in part to the fact that dynamical friction of infalling gas is 
overestimated as standard smooth particle hydrodynamics (SPH) cannot capture 
hydrodynamic instabilities at contact discontinuities accurately and gas comes in 
cold and clumpy \citep{navarro00,sijacki12}. However, simulations with 
grid-based codes and weak SN feedback also find that galaxies are too compact \citep{joung09,kimm11a,hummels12}, 
as gas cools excessively and forms stars at the galaxy centre. 
How to overcome the overcooling problem \citep{katz92} and robustly include the effects of SNe 
are important tasks for all numerical galaxy formation studies.

In order to ensure that energy from SNe is not radiated away artificially, 
several authors introduced the cooling suppression model in which 
gas near young stars is assume to be adiabatic for several to tens of 
Myr \citep{governato10,guedes11}. With this feedback model, 
 \citet{guedes11} find that a realistic disc galaxy with rising rotation 
curves can be produced within a $\Lambda$CDM paradigm if stars are permitted to 
form only in dense environments ($\nH=5\,\cmq$), so that star formation and outflows 
are strongly clustered. They claim that the collective winds are efficient at removing 
low-angular momentum gas that is accreted at high redshifts \citep[see also][]{brook11}.
\citet{oh11} analyse the simulation of a dwarf galaxy by \citet{governato10}, and 
find that the simulated inner slope of mass density as well as the shape of rotation curves 
are consistent with those of THINGS galaxies \citep{walter08}, demonstrating that 
strong feedback is crucial to reproduce the observations. 
Yet the timescale imposed for the adiabatic phase is controversial, 
given that it relies on the density of the ISM, which is largely 
under-resolved in the simulations.

Recently, other forms of stellar feedback have been suggested as  
important mechanisms to control the growth of galaxies. \citet{walch12,dale14} show 
that photoionisation is capable of dispersing a small cloud of mass $10^4\,\msun$ in 
$\sim$ 3 Myr by generating an over-pressurised HII region 
\citep[see also][]{vazquez-semadeni10,sales14}. \citet{dale14} point out that winds 
from massive stars can create a cavity of $\sim$ 10 pc at the centre of 
$10^{4}$--$10^{5}\,\msun$ clouds, but they alone cannot  unbind the gas from the 
system \citep[see][for a uniform case]{geen15}.  One also expects that radiation 
pressure from ionizing photons has a similar effect as stellar winds (for metal-rich gas), 
given that their momentum injection rates are comparable \citep{leitherer99}. 
However, \citet{sales14} assert that the inclusion of radiation pressure by ionising photons 
on top of photoionisation has a subdominant effect, because gas near a massive 
star is transparent to the Lyman-limit photons once ionised and cannot be accelerated 
continuously \citep[c.f.][]{wise12a}. 
The presence of dust can change the conclusion though, as it can also absorb the ionising 
radiation. More importantly, when gas is optically thick to infrared (IR) photons, 
the latter can in principle be multiply scattered supplying 10--100 times the original 
momentum of the ionizing radiation \citep{murray10}. Hopkins and collaborators conduct
systematic studies to gauge the relative importance of photo-heating, radiation pressure, 
and SN explosions in isolated discs \citep{hopkins11,hopkins12a,hopkins12b}, 
and find that radiation pressure with a momentum boost of 10--100 is required 
to drive violent galactic outflows in actively star-forming galaxies \citep[c.f.][]{rosdahl15}. A similar conclusion is 
reached by \citet{agertz13,aumer13}.

It should be noted, however, that several inaccurate assumptions  
are made concerning SN explosions in previous cosmological simulations.
First, even though the initial explosion energy of $10^{51}\,{\rm erg}$ is used, 
as guided by observations \citep[e.g.][]{nomoto93}, 
the momentum transfer from SN to the ISM is likely 
to be underestimated due to limited resolution.
Because a large amount of energy is deposited in a very concentrated form, 
SN blast waves must undergo an adiabatic phase during which radial momentum should
increase by up to an order of magnitude ($\approx3\times10^5\,\kms\msun$), compared 
to the initial ejecta momentum \citep{sedov59,taylor50,chevalier74,cioffi88,blondin98}.  
Without modelling this phase properly, hydrodynamic simulations underestimate 
the momentum transfer from SN. \citet{thornton98} performed 
a parameter study of the explosion in uniform media using one dimensional 
hydrodynamic simulations, and showed that the final radial momentum is a weak 
function of the background density ($\propto \rho^{-0.12}$) and metallicity ($\propto Z^{-0.14}$)
\citep[see also][]{kim15,geen15}.  The final momentum transferred to 
an inhomogeneous medium turns out to be surprisingly similar 
\citep{kim15,iffrig15,martizzi15}, as it has a weak dependence on density. 
Although this has been appreciated in studies of the generation of turbulence in the ISM 
\citep[e.g.][]{shetty08}, only recently a few cosmological simulations begun to 
take into account the momentum at different phases of SN expansions 
\citep{kimm14,hopkins14}.
Second, each star particle is often assumed to release SN energy in a single event 
at a fixed time delay of $\sim10$ Myr \citep{dubois08}. This may overestimate the 
impact of SN, as clustered explosions can work together to create galactic fountains. 
But at the same time, it is likely that the regulation of star formation 
is mistimed when the delay for SN explosions is fixed to a single time such as 10 Myr.
\citet{aumer13} reported that in the absence of feedback mechanisms active
before SNe, to regulate star formation in their simulations, they needed to 
use an early time delay of 3 Myr. This time delay appeared to leave galaxy star formation 
histories unchanged when radiation pressure was included in the simulations.
Lastly, even though up-to-date cosmological simulations can now employ a spatial 
resolution as high as 10 parsecs, this is not enough to resolve the turbulent 
structure of the ISM where the volume-filling density is smaller than the mean 
\citep[e.g.][]{federrath12}.  \citet{iffrig15} demonstrate that the expansion of a SN out to 
some fixed radius depends on the local environments of the explosion. This implies that 
determining the mass swept up from the host cell of a SN may not be trivial. 
In this respect, questions remains regarding how momentum should be injected on 
resolved scales.

The aim of this paper is to understand the impact of SN by realistically modelling 
the momentum injection from SN. Can the momentum input from SN alone possibly
explain the observed relations, such as the mass-metallicity relation or stellar mass to halo mass
relation? Why do the classical feedback schemes fail to reproduce these? 
In order to address these questions, we perform a suite of zoom-in, cosmological
hydrodynamic simulations with  adaptive mesh refinement (AMR).
We describe our simulation set up and a variety of feedback models tested in this work 
in Section 2. Our results on the evolution of stellar mass, metallicity, and kinematic properties 
are presented in Section 3. We discuss the effects of each prescription and why some 
feedback models are more efficient at suppressing star formation in Section 4. 
Finally, we summarise our findings in Section 5.

\section{Simulations}

\begin{figure*}
   \centering 
\includegraphics[width=5.8cm]{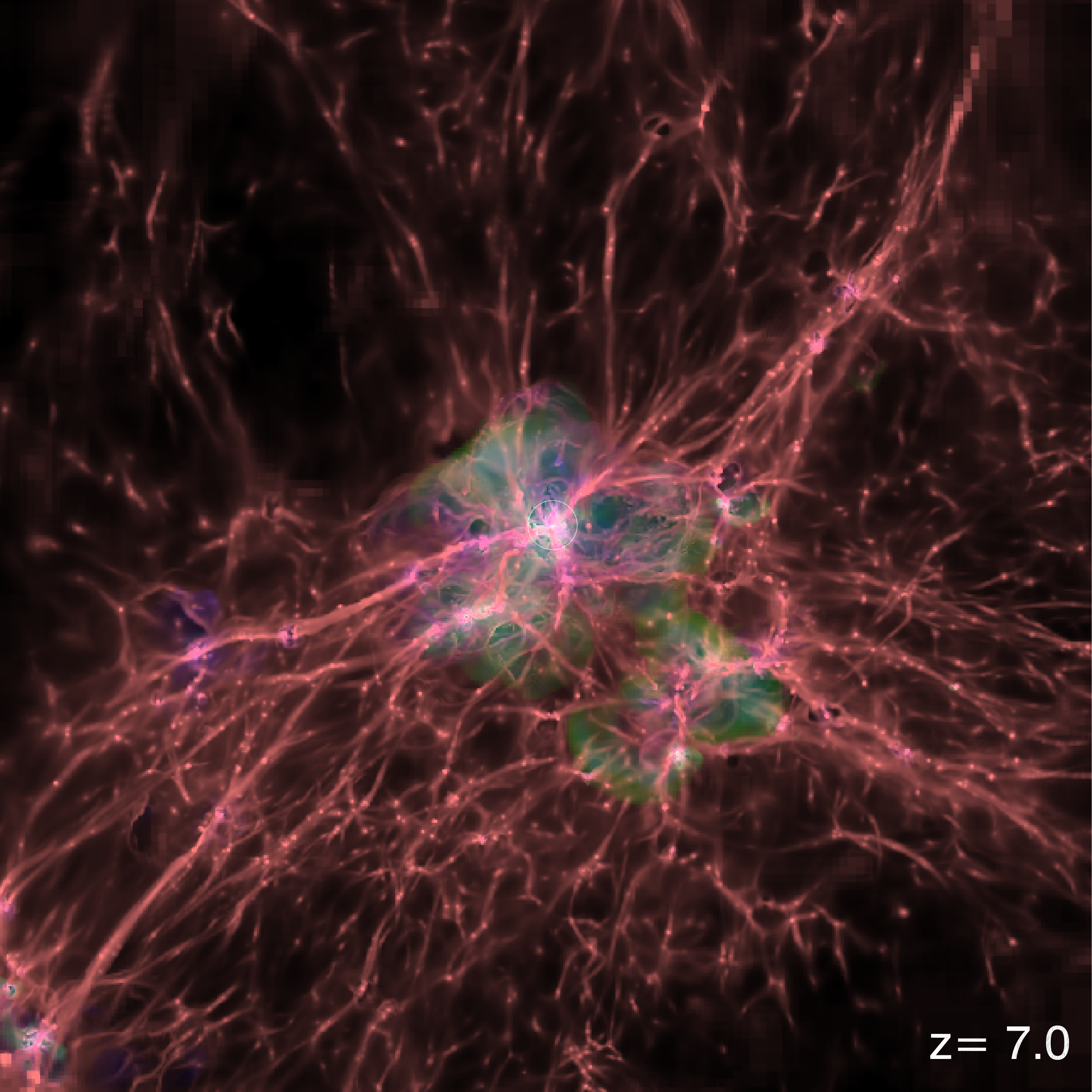} 
\includegraphics[width=5.8cm]{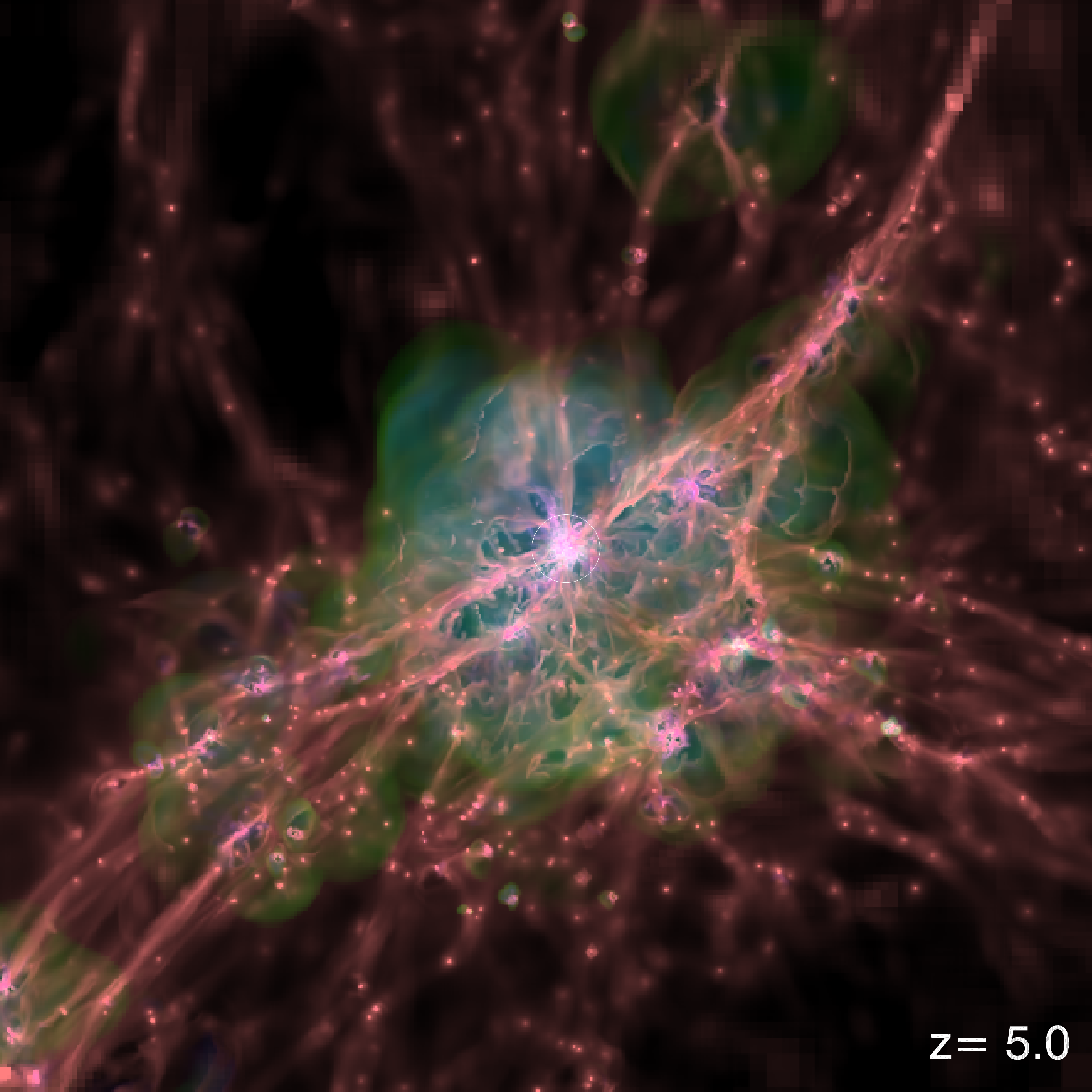} 
\includegraphics[width=5.8cm]{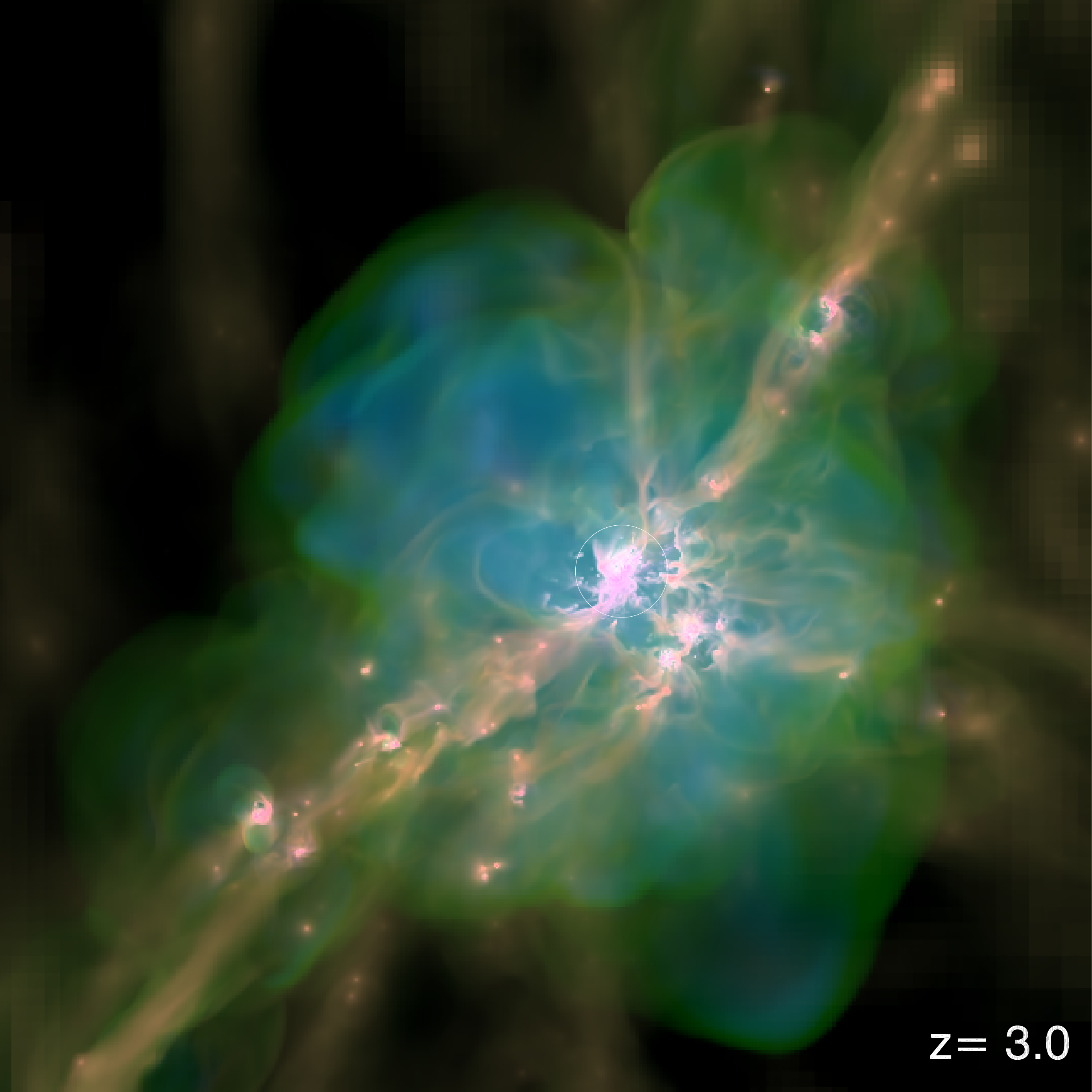} 
   \caption{Composite images of the Nut galaxy at $z=7$, $5$, and $3$ from the NutMFBmp run. 
   The pink, blue, and green colours represent the projected distributions of density, 
   temperature, and metallicity, respectively. The image measures 2.715 Mpc (comoving) on 
   a side. The white circle denotes the virial radius of the host halo of the Nut galaxy 
   (41 kpc at $z=3$, physical), which is fed by roughly three large-scale filaments. 
   It can be seen that the metal-enriched SN-driven winds extend out to several virial 
   radii of the halo at $z=3$.
   }
   \label{fig:nut}   
\end{figure*}

We use the tree-based Eulerian AMR hydrodynamics code, \ramses\ \citep{teyssier02}, 
to study the impact of SN feedback on the evolution of a Milky Way-type galaxy. 
The initial conditions for the simulations are identical to those used for the 
\nut\ suite presented in \citet{powell11,geen13}, which were generated 
using the {\sc mpgrafic} software \citep{prunet08,bertschinger01}
with WMAP5 cosmological parameters ($\Omega_m=0.258,\Omega_\Lambda=0.742, 
\Omega_{\rm b}=0.045, \sigma_8=0.8,H_0=72\,{\rm km\, s^{-1}\, Mpc^{-1}}$)  \citep{dunkley09}.  
The computational domain represents a periodic box of 9 $h^{-1}$ Mpc on a side,
covered with $128^3$ cells. Within this volume, three nested grids are placed in a 
spherical region of radius 2 Mpc (comoving) encompassing a dark matter halo of 
$\mvir=5.5\times10^{11}\,\msun$ at $z=0$. 
The corresponding mass resolution of a dark matter particle is $5.5\times10^4\,\msun$. 
Only the spherical region is allowed to be further refined.
Figure~\ref{fig:nut} shows the composite images of density, temperature, and 
metallicity around the Nut halo at three different redshifts ($z=$3, 5, and 7) from 
a run with SN feedback.

The Poisson equation is solved using a multi-grid method \citep{guillet11} 
for grids with level 7 $\le lv \le$ 9, while a conjugate gradient method is 
used for more refined grids ($lv \ge 10$, \citealt{teyssier02}). The Euler equations 
are integrated in super-comoving coordinates \citep{martel98} with a {\sc hllc} 
Riemann solver \citep{toro94}. We adopt the typical courant factor of 0.7.

Metal-dependent radiative cooling is modeled with cooling curves from 
\citet{sutherland93}, down to $\sim10^4\,{\rm K}$. Gas cooler than 
$10^{4}\,{\rm K}$ can lose energy further via metal fine structure transitions, 
following \citet{rosen95}. A uniform ultraviolet background is turned 
on at $z=8.5$ using \citet{haardt96}, and modeled as a heating term 
in the energy equation. However, gas denser than $\nH=0.01\,{\rm cm^{-3}}$ is assumed to be 
self-shielded from the radiation \citep{rosdahl12,faucher-giguere10}. 
The simulations start  with gas of metallicity $Z=2\times10^{-5}$ at $z=499$, 
as the primordial SN explosions in mini-haloes are not resolved in our simulations 
\citep[e.g.][]{whalen08,wise12a}. 

In order to better resolve the ISM in galaxies, 
we employ a maximum physical spatial resolution of 12 pc\footnote{
We perform a convergence test by comparing the MFBmp run and its twin version with a higher resolution (6 pc) 
down to $z=5.5$, while keeping other parameters fixed, except for the mass of a star particle (203\,\msun). 
We find that the stellar mass of the main galaxy is converged within 10\% at $z=5.5$.
} ($=12.5\,{\rm Mpc}/2^{20}$) by 
triggering 10 more levels of refinement when baryonic plus dark matter mass in each cell
 becomes greater than $4.4\times10^5\,\msun$. In addition, we also enforce that gas denser 
 than $\nH=1$, $42.5$, $340\,{\rm cm^{-3}}$ is always 
resolved at least to resolutions of 48 pc, 24 pc, and 12 pc cell, respectively. 
At the final redshift ($z=3$), the main halo is resolved with $3.0$--$3.7\times10^6$ computational cells, depending on the feedback model. We find that 59\%, 14\%, and 5\% of the gas mass 
inside $0.1\,\rvir$ ($\approx3.6\,{\rm kpc} \sim 2 R_{\rm e}$) is resolved on 48, 24, and 12 pc resolution 
elements in the case of the NutMFBmp run. The main galaxy in the NutTFB run uses more high-resolution 
cells (40\%, 32\%, and 24\% of the gas mass is resolved on at least 48, 24, and 12 pc cells, respectively), 
as the ISM becomes denser than that of MFBmp due to the overcooling.

Dark matter haloes are identified with the HaloMaker \citep{tweed09}.
The virial radius of each halo is defined as the radius within which 
the mean density is equal to the critical density of the Universe 
($\rho_{\rm crit} = 3 H(z)^2 / 8 \pi G$) times the virial over-density 
\citep[$\Delta_{\rm crit} \equiv 18\pi^2+ 82 x -39 x^2$,][]{bryan98},
where $x\equiv \Omega_m/(\Omega_m + a^3\Omega_\Lambda)-1$ and $a$ is the scale factor.

\subsection{Star formation}

We model star formation as a stochastic process \citep[e.g.][]{katz92,rasera06}, 
based on a Schmidt law \citep{schmidt59},
\begin{equation}
\dot{\rho}_{\rm star} = \epsilon_{\rm ff} \rho_{\rm gas} / t_{\rm ff}, 
\end{equation}
where $\rho_{\rm gas}$ is the gas density, $t_{\rm ff}=\sqrt{ 3\pi/32G\rho_{\rm gas}}$ is 
the free-fall time of the gas, and  $\epsilon_{\rm ff}$ is the efficiency of star formation per 
free-fall time.
Specifically, we examine the probability of forming a star particle in a cell denser than 
a certain threshold density (\nth) in each fine time step. 
We adopt $\nth=42.5\,{\rm cm^{-3}}$, motivated by the Larson-Penston density 
($\rho_{\rm LP}\simeq 8.86 \,c_s^2 /\, \pi \,G \,\delta x^2$) above which 
isothermal gas undegoes a runaway collapse \citep{larson69,penston69,gong11,gong13},
where $c_s$ is the sound speed and $\delta x$ is the scale of the gas cloud. For a typical temperature of the cold ISM ($\sim$ 30 K) and the finest size of our computational grid (12 pc), this yields $\nH=42.5\,{\rm cm^{-3}}$.
Although the threshold density is slightly lower than the typical mean
density of a giant molecular cloud ($\nH\sim100\,\cmq$),  we note that the majority 
($\gtrsim70\%$) of stars form in such environments,
as star formation is generally a slow process \citep[][see below]{kennicutt98,krumholz07,evans09}. 

The number of stars formed ($N_p$) is determined by drawing a random number 
from a Poisson distribution with a mean of 
\begin{equation}
\lambda \equiv \epsilon_{\rm ff} (\rho \Delta x^3/m_{\star,{\rm min}}) \left(\Delta t_{\rm sim} / t_{\rm ff} \right),
\end{equation}
where $\Delta x$ is the size of a computational cell, and 
$m_{\star, {\rm min}}=\alpha \,N_p \,\nth \,m_{\rm H} \,\Delta x_{\rm min}^3/X_{\rm H}$ is 
the minimum mass of a star particle. Here $m_{\rm H}$ is the hydrogen mass, $X_{\rm H}$ 
(=0.76) is the mass fraction of hydrogen for the primordial composition, and
$\alpha$ is a parameter that controls the minimum mass, which we 
use $\alpha=0.264$ to achieve $m_{\star,{\rm min}}=610\,\msun$.
We assume that 2 per cent ($\epsilon_{\rm ff}=0.02$) of the dense gas ($\nH\ge42.5\,\cmq$) 
turns into stars per free-fall time \citep{krumholz07}.

\subsection{Feedback models}

We examine six different feedback models to investigate the impact of core-collapse 
SN on the evolution of galaxies.  We use the Chabrier initial mass function 
\citep[IMF,][]{chabrier05} throughout this study. We assume that 31\% of the total mass is 
lost via Type II SN explosion. For a progenitor mass of 19.1 \,\msun, this means
that a star particle of 610 \msun\ would host 10 SNe. The corresponding specific 
energy per $\msun$ at the time of explosion is $1.6\times10^{49}\,{\rm erg}$.
The metallicity of the ejecta is taken to be 0.05,
which is equivalent to a yield of 0.016 for a simple stellar population with 1 \msun. 

Feedback models may be distinguished by how the energy or momentum 
is put into the ISM. Note that we do not turn off the radiative cooling \citep{mori97,stinson06} 
or hydrodynamically decouple winds from the ISM \citep{scannapieco06,oppenheimer06}. 
We describe the details of each model below.

\begin{itemize}
  \setlength{\itemsep}{8pt}
\item[(i)] {\bf Reference run} (NutCO)\\
The reference run (NutCO) does not include any energy or 
momentum feedback, but gas and newly processed metals from SNe 
are added to their host cell, once a star particle becomes older 
than 10 Myr (so called ``metal feedback''). Since there is no direct energy input in 
this case, star formation is essentially controlled by the cosmic inflow and 
uniform background UV heating. 
Note that we use a higher minimum mass for star particles ($m_{\star,{\rm min}}=2310\,\msun$) 
than other runs to reduce the computational cost.

\item[(i)] {\bf Thermal feedback} (NutTFB)\\
The total energy of $10^{51}\,{\rm erg}$ per SN is 
simply added to the host cell of the star particle, along with ejected gas and metals. Each 
star particle is assume to generate a single SN event at 10 Myr after the birth. 
This can increase the temperature of a cell up to $\sim3\times 10^8\,{\rm K}$ in a 
low-density medium (i.e. $M_{\rm gas} \ll M_{\rm ej}$). In principle, the over-pressurised 
gas can drive the expanding motion depositing $\sim30\%$ of the total into the kinetic 
energy \citep{chevalier74}. However, as is well known in the literature, 
the atomic and metal cooling processes can rapidly radiate the internal energy away 
before the SN-driven blast wave sweeps up the surrounding medium in dense environments or in 
simulations where the cooling radius is under-resolved \citep{katz92,navarro93,abadi03,slyz05,hummels12,kimm14}. 
To circumvent this issue, some studies assume that gas is adiabatic for several 
tens of Myr \citep{mori97,thacker01,stinson06,governato07,teyssier13}. 
Our thermal feedback model does not include the prescription to 
properly address excessive radiative losses.

\item [(ii)] {\bf Kinetic feedback with a small mass-loading} (NutKFB)\\
We also test the kinetic feedback scheme of \citet{dubois08}, in which 
the initial explosion energy of $E_{\rm SN}=10^{51}\,{\rm erg}$ is modelled by 
increasing the kinetic energy\footnote{When the momentum added is cancelled out, 
the kinetic energy is used to increase the temperature to conserve the total energy. 
This is also the case for the mechanical feedback runs.} of the surrounding gas.
An important parameter in this model is a mass-loading factor 
($\eta_w \equiv M_{\rm load} / M_{\rm ej}$), 
which accounts for how much gas is entrained with the SN ejecta. Notice that a 
mass-loading factor ($\eta$) is often defined as the mass ratio between the outflow 
rate ($\dot{M}_{\rm out}$) and the star formation rate ($\dot{M}_{\rm star}$), 
which is different from what we use in this study. Empirical determinations of the 
mass-loading factor suggest that it varies depending on the mass of a galaxy 
\citep{martin05,rupke05,weiner09,martin12,newman12,chisholm14,arribas14}, 
but is roughly on the order of $\eta_w\sim10$. Motivated by this, we adopt $\eta_{w,{\rm max}}=10$ 
(i.e. $\eta\sim3$). Notice, however, that $\eta_w$ can be as small as zero 
if there is little gas in the host cell of SN.
The velocity of the neighboring gas is then determined by 
$v = \sqrt{2\,E_{\rm SN}/(1+\eta_w)\,M_{\rm ej}}$, hence the input momentum 
can be as 3.3 times as high as the initial momentum if there is enough gas to entrain from
the host cell. Again, SNe are modeled as a single event at 10 Myr per individual star 
particle, but this is evaluated at every coarse time step ($\sim 0.16\, {\rm Myr}$) to 
reduce the message passing interface communications between computing cores.

\item [(iii)] {\bf Mechanical feedback} (NutMFB)\\
\citet{kimm14} introduced a physically based feedback model, 
which can properly explain the transfer of momentum at all stages of a Sedov-Taylor blast 
wave (from the adiabatic to snowplough phase) \citep[see also][]{hopkins14}.  
The main difference from the kinetic feedback model is that the input momentum is 
calculated according to the stage of the Sedov-Taylor blast wave. Although 
the momentum available during the free expansion phase is only 
$\approx 4.5\times10^4\,{\rm km\,s^{-1}}\,\msun$, it increases as more gas is 
swept up by the shock ($p \propto \sqrt{m}$). 
When the energy lost at the shell becomes significant ($\Delta E/E \gtrsim 0.3$) and the pressure inside 
the bubble becomes comparable to that of the pre-shock region, the expansion of the shell becomes 
momentum-conserving. At this point, the shell mass gain is much higher than the 
decrease in velocity resulting in the net increase of the total momentum 
by a factor of $\sim$7-10 at the end of the adiabatic phase. This factor can be smaller 
if the number of SNe is larger or the background density of gas is higher 
or the metallicity is higher, as the radiative cooling becomes more significant. 

Specifically, the input momentum during the 
snowplough phase is calculated as \citep[][see also \citealt{kim15,geen15}]{blondin98,thornton98}
\begin{equation}
p_{\rm SN,snow} \approx 3\times 10^5\,  {\rm km\, s^{-1}} \,\msun\, E_{51}^{16/17} n_{\rm H}^{-2/17} Z'^{-0.14}, 
\label{psn_one}
\end{equation}
where $E_{51}$ is the number of SNe, $n_{\rm H}$ is the hydrogen number density, and $Z'$ is the metallicity 
in solar units, with a minimum of 0.01 ($Z'\equiv {\rm max} \left[Z/Z_\odot,0.01\right]$). 
Whether the expansion is in the adiabatic or momentum-conserving phase is determined 
by the mass swept up from the host and neighboring cell. This is done by calculating 
the mass ratio ($\chi$) for each 48 neighbours ($=N_{\rm nbor}$) per explosion as  
\begin{equation}
\chi\equiv dM_{\rm swept}/dM_{\rm ej},
\end{equation}
where
\begin{align}
&dM_{\rm ej} = (1-\beta_{\rm sn}) M_{\rm ej} / N_{\rm nbor}, \\
&dM_{\rm swept} = \rho_{\rm nbor} \left(\frac{\Delta x}{2}\right)^3 + \frac{(1-\beta_{\rm sn})\rho_{\rm host} \Delta x^3}{N_{\rm nbor}} + dM_{\rm ej},
\label{eq:swept}
\end{align}
$\Delta x$ and $\rho_{\rm host}$ is the size and gas density of the host cell, respectively, 
and $N_{\rm nbor}=48$ is the number of neighbouring 
cells (see Figure 15 in \citealt{kimm14}). Here $\beta_{\rm sn}$ determines the 
mass fraction of ($M_{\rm ej}+\rho_{\rm host}\Delta x^3$) to be left in the host cell.
We take $\beta_{\rm sn}=4/52$ to distribute the mass as evenly as possible 
to the host and neighboring cells when the cells are on the same level.
If $\chi$ is greater than the transition mass ratio, 
\begin{equation}
\chi_{\rm tr}\equiv 69.58 \, E_{\rm 51}^{-2/17} n_{\rm H}^{-4/17} \, Z'^{-0.28},
\end{equation}
we input $p_{\rm SN,snow}$ (Equation~\ref{psn_one}), whereas momentum appropriate 
to the adiabatic phase is injected otherwise,
\begin{align}
p_{\rm SN} = \left\{ \begin{array}{ll}
     p_{\rm SN, ad}=\sqrt{2 \chi\, M_{\rm ej}\,f_e\,E_{\rm SN}}
 & (\chi < \chi_{\rm tr}) \\
    p_{\rm SN, snow}    & (\chi \ge \chi_{\rm tr}) \\ 
   \end{array}
   \right. ,
   \label{eq:psn}
\end{align}
where $f_e = 1-\frac{\chi-1}{3(\chi_{\rm tr}-1)}$ is used to smoothly connect the two regimes.
Note that the local properties of gas at which a SN explodes can be significantly different 
from those of gas at which the star particle is formed, 
as we allow for the ISM to dynamically evolve for 10 Myr before the SN explosion.

\begin{figure}
   \centering
   \includegraphics[width=4.cm]{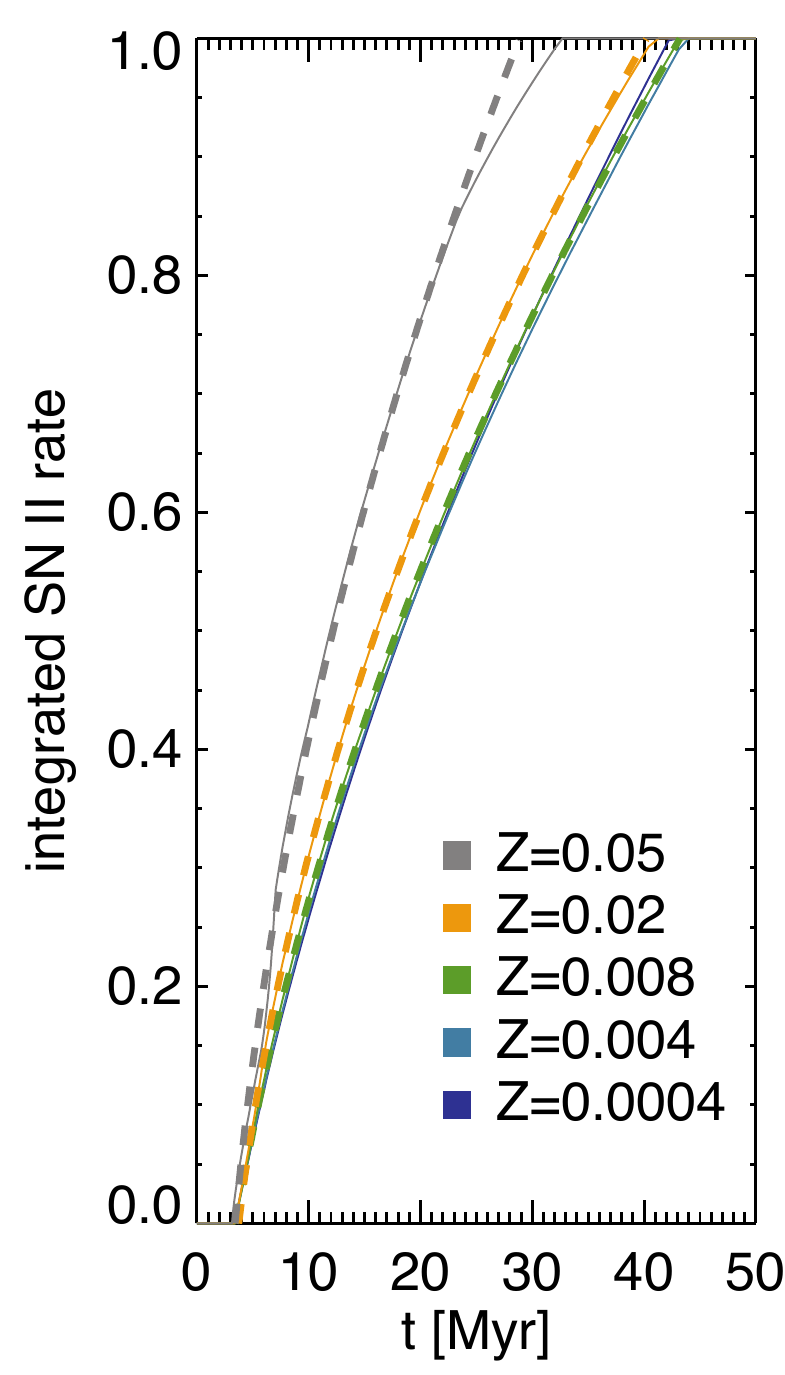}
      \includegraphics[width=4.cm]{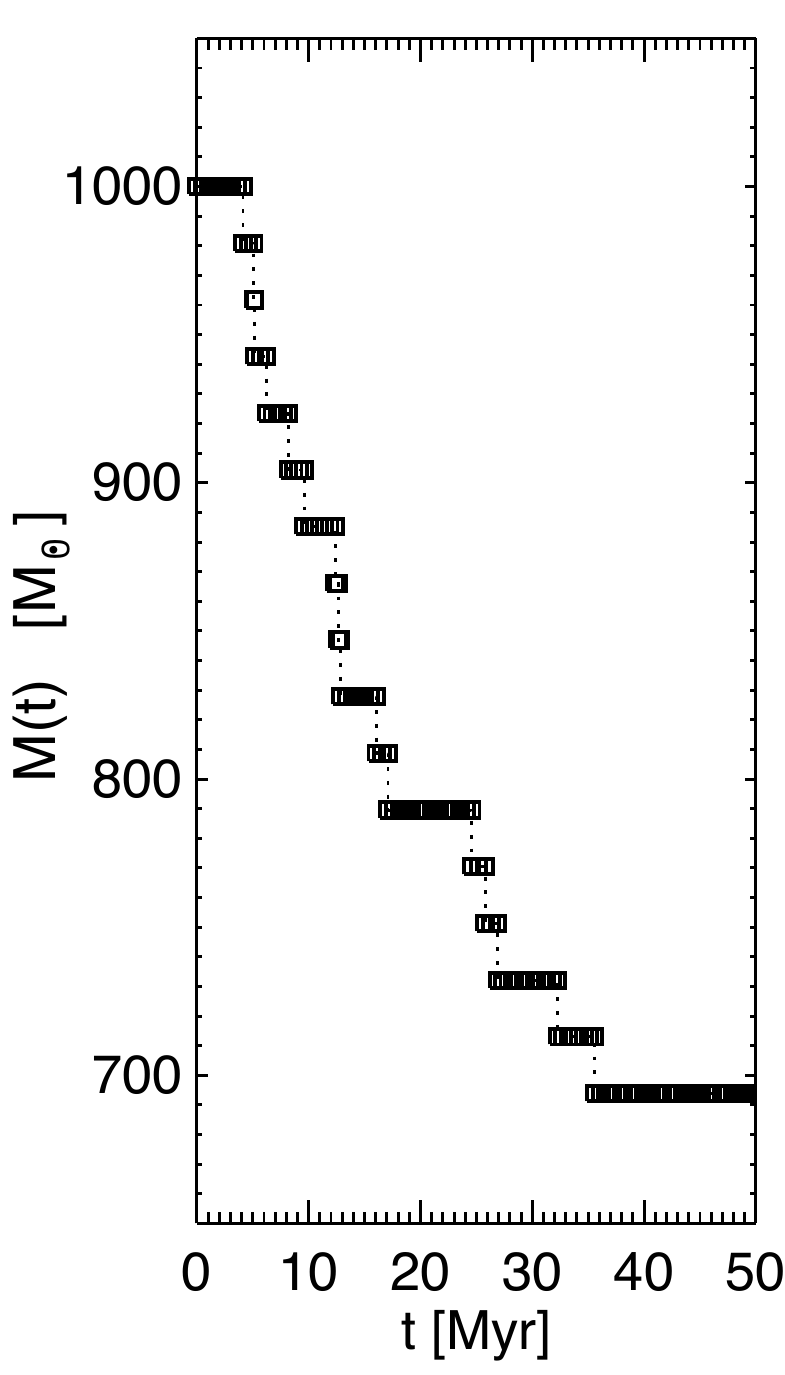}  
   \caption{ {\em Left:} Integrated SN rates based on the population synthesis code 
   Starburst99 (solid) and the corresponding polynomial fits (dashed lines, Equation~\ref{eq:fit}). 
   Different colours indicate different metallicities, as shown in the legend. 
   {\em Right:} An example of the random sampling of discrete SN explosions 
   for a $1000\,\msun$ star particle. 
  }
   \label{fig:SNrate}
\end{figure}

\begin{table*} 
\caption{Summary of {\sc nut42} simulation parameters and physical ingredients. From left to right, columns 
are as follows: simulation name, type of SN feedback,  minimum grid size (physical units), 
mass of a dark matter particle, minimum mass of a star particle, threshold density for star 
formation, star formation efficiency per free-fall time, final redshift of each simulation, 
time delay of a core-collapse SN, and remarks. $\eta_{w,{\rm max}}$ is the maximum 
mass-loading factor with respect to the ejecta mass, used in \citet[][DT08]{dubois08}. 
$f_{\rm w,host}$ indicates the fraction of the mass entrained from the host cell of a SN.}
\label{SimSummary} 
\begin{tabular}{@{}lccccccccl}
\hline 
Simulations & SN II feedback &$\Delta x_{\rm min}$&  $m_{\rm DM}$ & $m_{\rm star,min}$ & $n_{\rm th}$       &  $\epsilon_{\rm ff}$       &$z_{\rm end}$   & $t_{\rm delay}$ & Remarks\\ 
            & & [pc, physical]               & [\msun] & [\msun]  & [${\rm H/cm^{3}}$] &         &   &  [Myr]  & \\ 
\hline 
NutCO     & metals only     & 12 & $5.5\times10^4$ & 2310  & 42  & 0.02 & 3 & 10 & \\
NutTFB  &  thermal  & 12 & $5.5\times10^4$ & 610  & 42  & 0.02 & 3 & 10 & \\
NutKFB  &  kinetic (DT08)  & 12 & $5.5\times10^4$ & 610  & 42  & 0.02 & 3 & 10 & $\eta_{w,{\rm max}}=10$\\
NutMFB  & mechanical    & 12 & $5.5\times10^4$ & 610  & 42 & 0.02 & 3 & 10 & \\
NutMFBm  & mechanical    & 12 & $5.5\times10^4$ & 610  & 42 & 0.02 & 3 & realistic & \\
NutMFBmp  & mechanical    & 12 & $5.5\times10^4$ & 610  & 42 & 0.02 & 3 & realistic & porous ISM ($f_{\rm w,host}=0.1$) \\
\hline 
\end{tabular} 
\end{table*}

\item [(iv)] {\bf Mechanical feedback with multiple explosions} (NutMFBm) \\
This model is based on the same mechanical feedback scheme as NutMFB, 
but differs in that SNe from a star particle are distributed in time, as in the 
real Universe. In cosmological simulations with finite resolution, an individual star 
particle represents a coeval star cluster, which is usually assumed to have a single 
explosion at some fixed age. Our models (NutCO, TFB, KFB, MFB) also assume 
that SNe explode at 10 Myr, meaning that tens of SNe explode simultaneously.
In reality, some SN explosions commence at as early as 3 Myr, 
whereas an 8 \msun\ star evolves off the main sequence only at $\sim$ 40 Myr.  
The multiple SNe model is designed to mimic the discrete nature of the multiple explosions 
by randomly sampling the lifetime of massive stars for each particle. To do so, we use 
the inverse sampling method for a polynomial fit to the integrated rate of SN Type II (Figure~\ref{fig:SNrate}, left panel), 
\begin{equation}
n(<t) = \sqrt{a(Z)+b(Z)\,t_6} + c(Z)
\label{eq:fit}
\end{equation}
where $t_6\equiv  t / 10^6\, {\rm yr}$, $a(Z)  =-0.268 + 0.139\,Z' - 0.575 \, Z'^2$, 
$b(Z) = 0.042 + 0.022\,Z' + 0.0790\,Z'^2$, 
$c(Z) = -0.086 - 0.170\,Z' + 0.187\,Z'^2$,
and $Z' = {\rm max} (Z,0.008)/0.02$.
We use the population synthesis code Starburst99 \citep{leitherer99} to compute 
the SN rates as a function of time and metallicity.
Figure~\ref{fig:SNrate} (right panel) shows an example of the random sampling and 
corresponding mass evolution of a 1000 \msun\ star particle.
Note that, unlike NutKFB and MFB, we evaluate the explosion at every fine time step 
(several hundreds to thousand years) for a better accuracy.

\item [(v)] {\bf Mechanical feedback with multiple explosions in a porous ISM} (NutMFBmp) \\
Particular attention is paid to how momentum is distributed to the neighboring cells 
in this model. In NutMFB and NutMFBm, all of the gas in the host cell of SN is 
re-distributed uniformly to its surroundings. This may be a reasonable assumption for 
an ISM with a smooth structure or for an ISM with a low Mach number where 
the volume filling density is not very different from the average density 
\citep{vazquez-semadeni94,federrath12}. However, high-z galaxies are normally 
turbulent \citep[e.g.][]{forster-schreiber09}, and our resolution elements may not fully 
resolve the porous structure of the turbulent ISM. If a SN explodes in a porous medium, 
energy will preferentially propagate through low-density channels 
\citep{iffrig15,kim15,martizzi15}, unlike SN bubble expansion in a uniform medium.
In this regard, it is likely that {\em too much gas is entrained from the host cell of the SN 
in simulations with limited resolution (including ours), 
leading to the underestimation of the initial gas outflow velocity in the neighboring cells}.
For example, if a SN explodes in a cell with ${\nH=10\, {\rm cm^{-3}}}$ resolved to 24 pc 
and a neighbouring cell has a density of 1\,\cmq, the initial wind velocity would be 
about 30\,\kms\ for a uniform ISM.
However, if the ISM were highly turbulent with a Mach number of 10,
as is often observed in the Galactic star-forming clouds \citep[e.g.][]{roman-duval10},
the volume filling density would be roughly ten times smaller for a collapsing cloud \citep{molina12,federrath12} 
and the wind velocity would therefore be roughly two times faster than the 
estimate with the uniform case. 
Note that the increase in the wind velocity is partially attributed to the fact that 
the input momentum would be 30\% larger in the porous case, as the background density 
is lower. In other words, under-resolving the host cell of SN can artificially 
smooth out the winds, reducing the initial velocity of the neighboring gas, especially
along low-density channels.  

In an attempt to examine the effect of the porous ISM and estimate its possible 
consequence on the impact of SNe, 
we allow for a faster wind by reducing the amount of gas mass entrained from the SN cell. 
We simply assume that only 10\% of the gas from the host cell of SN ($f_{\rm w,host}=0.1$) is 
directly involved in the expansion. This means that Equation~\ref{eq:swept} reads
\begin{equation}
dM_{\rm swept} = \rho_{\rm nbor} \left(\frac{\Delta x}{2}\right)^3 + \frac{f_{\rm w,host} \rho_{\rm host} \Delta x^3}{N_{\rm nbor}} + dM_{\rm ej}.
\label{eq:newloading}
\end{equation}
We emphasise that this number is rather arbitrarily chosen, 
hence the comparison with observations should be taken with caution. Nevertheless, we note 
that it can clearly demonstrate the importance of detailed modelling of momentum injection 
and address the question as to whether the momentum budget from SN suffices to 
regulate star formation.
\end{itemize}

The parameters used in the six runs are summarised in Table~\ref{SimSummary}.

\section{Results}

In this section, we compare the baryonic properties of simulated galaxies to 
evaluate the effect of different SN feedback models.  Our zoom-in region contains 99 
dark matter haloes\footnote{Subhaloes are not included in this sample.} 
with $\mvir\ge10^8\,\msun$ at z=3, including the progenitor of a Milky Way-like galaxy. 
Our analysis mostly focuses on the most massive galaxy, but we also present 
results for other smaller galaxies to see a general trend of the physical properties as a 
function of galaxy mass.

The centre of a galaxy is often taken as the densest region of a dark matter halo in 
large-scale simulations \citep[e.g.][]{tweed09}. But it is not necessarily the case 
in high-resolution simulations where mergers are more frequent 
and a galaxy is often very clumpy. Starting from the stellar plus gas distributions 
within 0.5 \rvir, we redefine the galaxy centre by iteratively calculating the 
centre of baryonic mass within its half-mass radius ($r_{\rm eff,m}$) 
until converged. This ensures that the centre of a galaxy is normally 
taken as the centre of the most massive clump.
The galaxy stellar mass is then computed by summing up the mass of the star 
particles within 0.2 \rvir\ from the centre. 

\begin{figure}
   \centering
   \includegraphics[width=7cm]{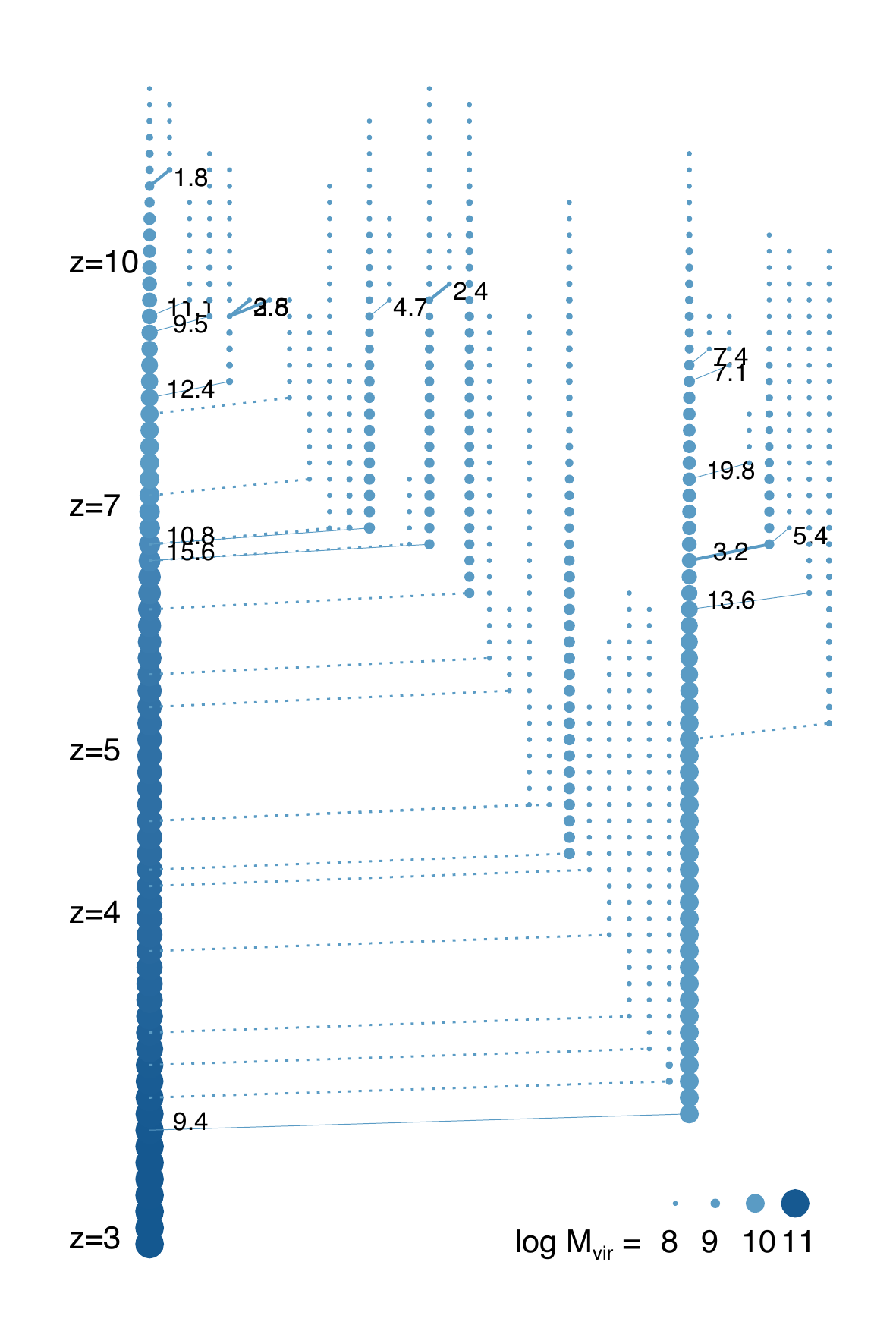}
   \caption{ Halo merger tree of the Nut galaxy. The leftmost branch corresponds to the 
   merger history of the main galaxy. The digits shown in the right side of a merger event 
   indicate a merger ratio ($\gamma=M_{\rm cen}/M_{\rm sat}$) below 20. 
   The ratio for the very minor merger ($20\le \gamma \le 100$) is omitted. 
   Major ($\gamma\le4$) and minor mergers are displayed as thick and thin solid 
   lines, respectively. A bigger symbol indicates a more massive halo. Note that actual
   galaxy mergers take place later than the halo merger.
  }
   \label{fig:tree}
\end{figure}

\begin{figure*}
   \centering
   \includegraphics[width=16cm]{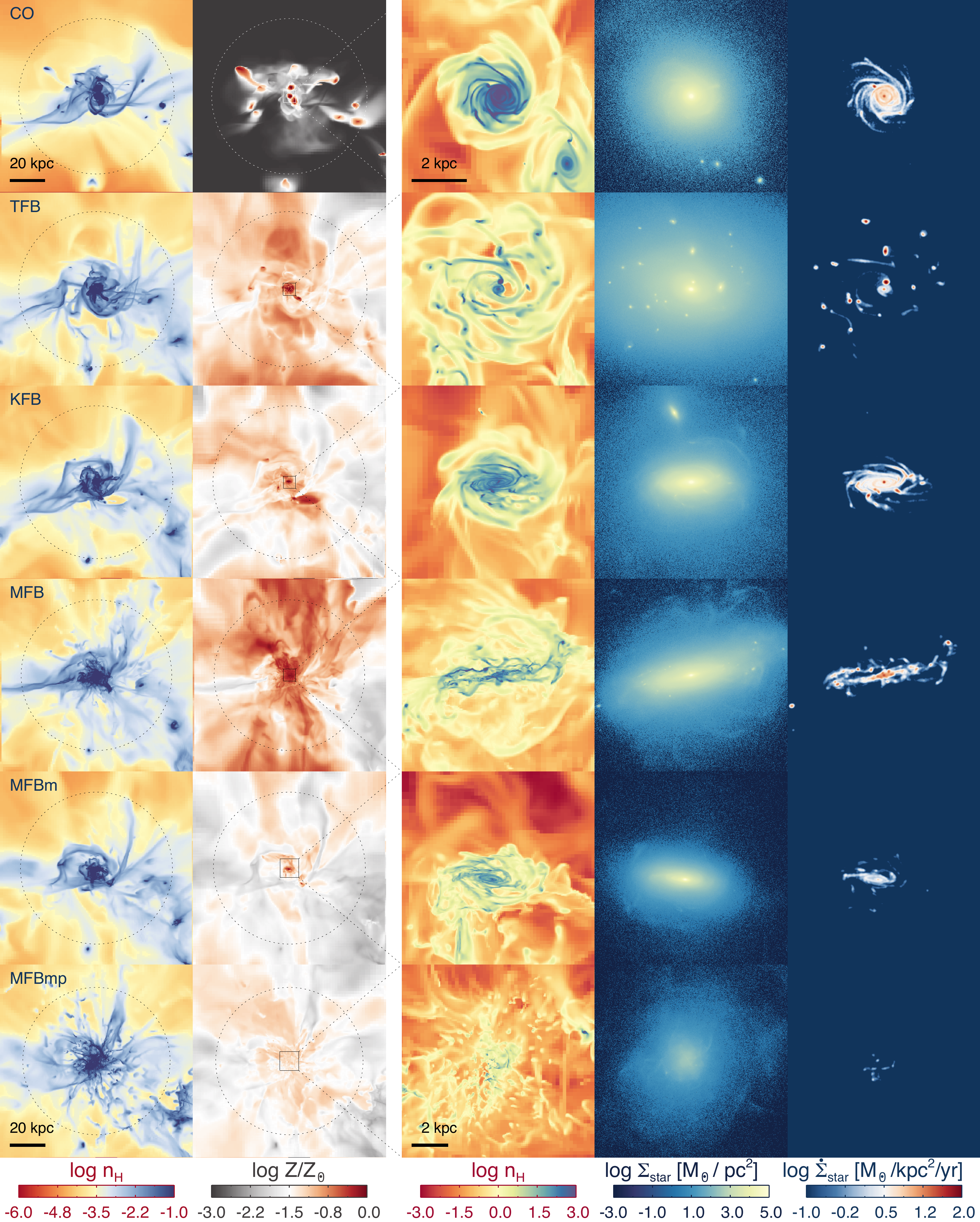} 
   \caption{Projected distributions of the density (first and third columns), 
   metallicity (second column),  stellar density (fourth column), and star formation rate density 
   (fifth column) of the Nut galaxy in runs with different feedback models, as indicated in the top 
   left corner of each panel.  The dotted circles in the first two columns denote the virial radius 
   of the dark matter halo. The last three columns show the central region
   of the halo. Note that all images are produced along the same direction (y-axis)
   for direct comparisons. The stellar component is markedly different 
   from each other depending on the feedback model adopted. The spin direction of the  
   galaxy is particularly sensitive to the choice of the feedback model. The strong outflow in 
   the NutMFBmp run prevents the formation of a well-ordered gaseous disc at $z=3$.    }
   \label{fig:mmp_img}
\end{figure*}

\begin{figure*}
   \centering 
    \includegraphics[width=8.0cm]{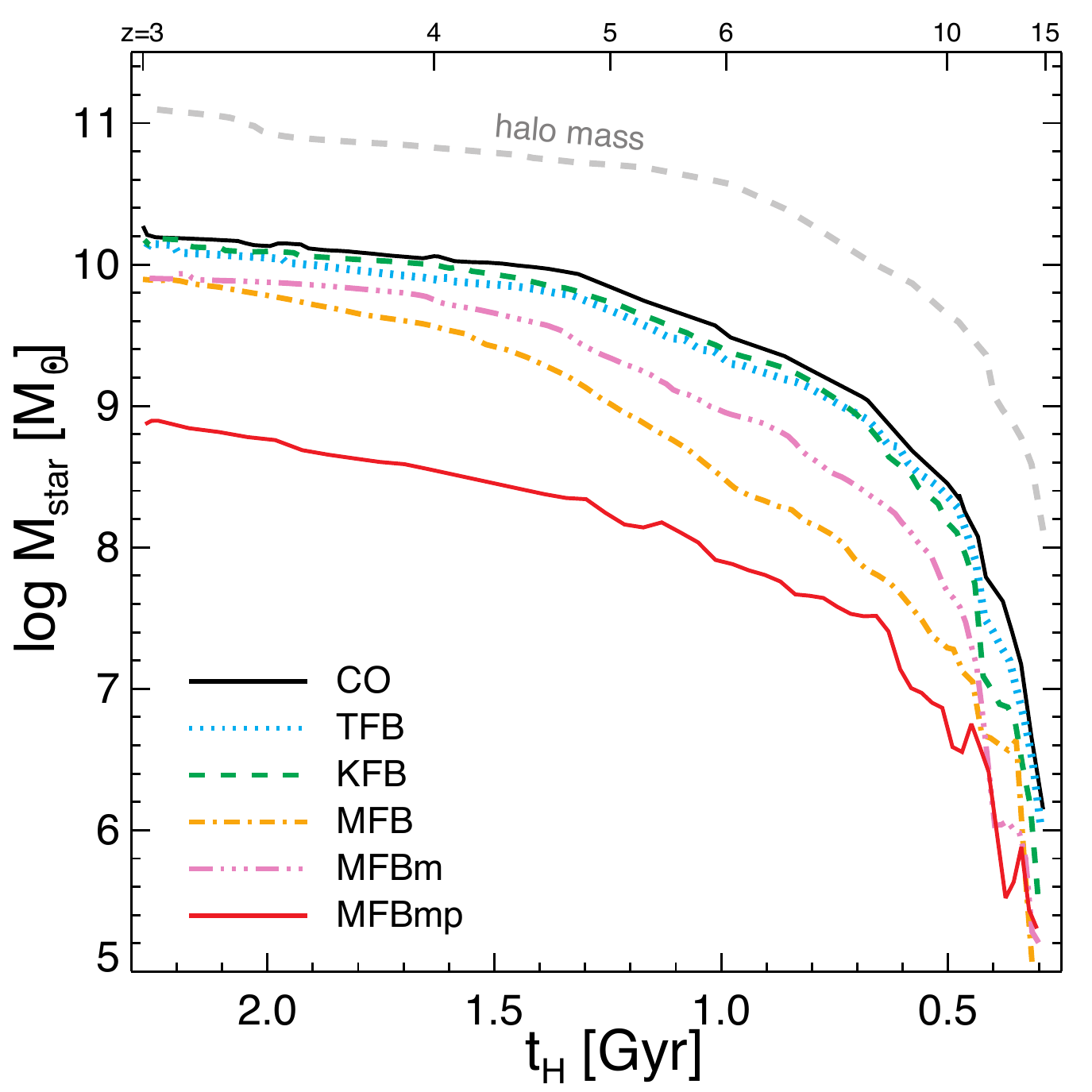} 
    \includegraphics[width=8.0cm]{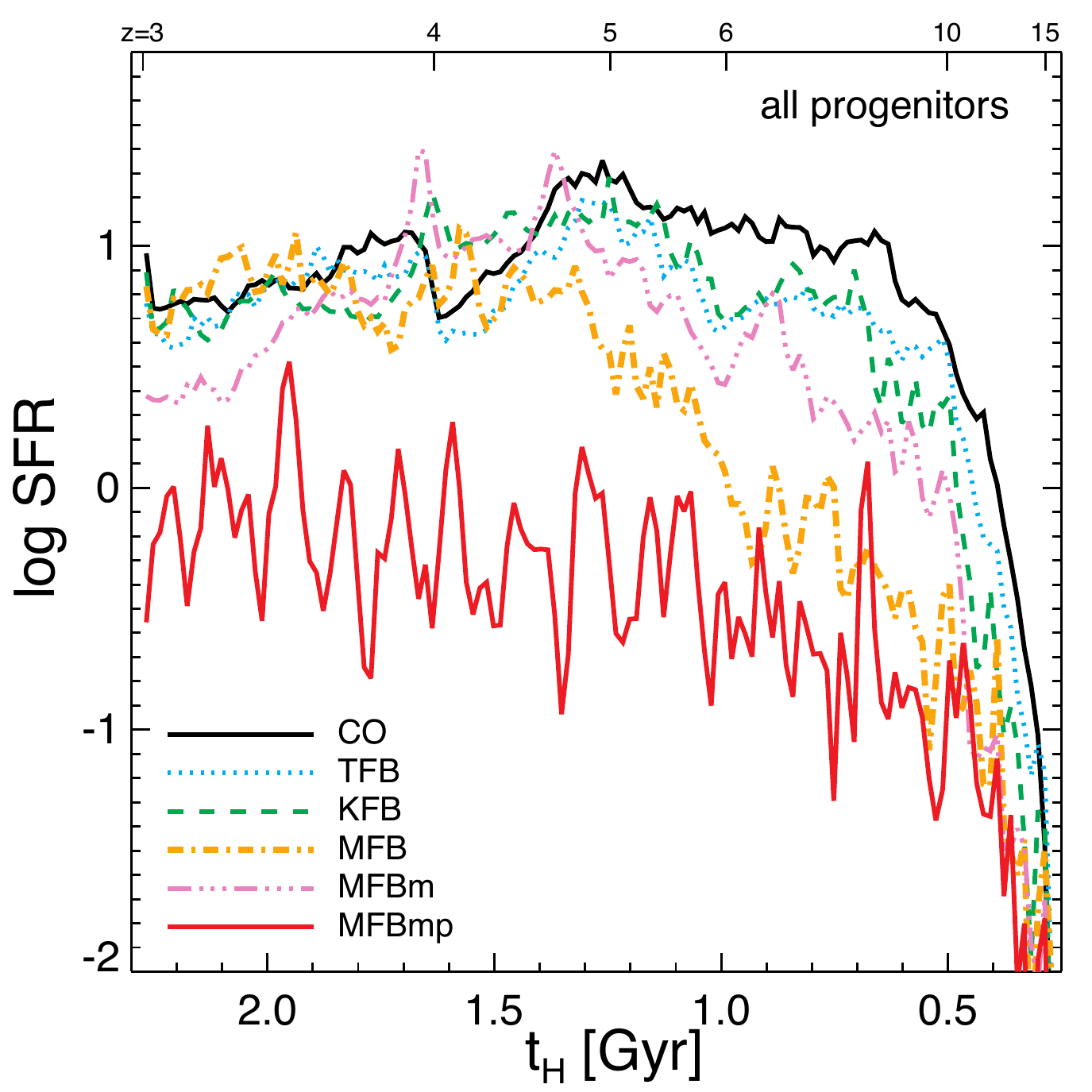} 
    \caption{Comparison of the galaxy stellar mass (left) and star formation history (right) of the Nut 
         galaxy. The abscissa indicates the age of the universe ($t_{\rm H}$). 
         Different colour-codings correspond to different feedback models, as indicated in the legend.
         The right panel shows star formation histories calculated using star particles located within 
         0.2 virial radius of a dark matter halo at $z=3$. Note that this includes stars formed in 
         all progenitors. It can be seen that mechanical SN feedback is more effective at 
         suppressing star formation than either thermal or kinetic feedback. Star formation 
         is very episodic in the MFBmp run, as SNe efficiently destroy star-forming clouds.
         }
   \label{fig:sf}
\end{figure*}

The main galaxy in our simulation is hosted by a dark matter halo with 
$5.5\times10^{11}\,\msun$ at $z=0$ \citep{kimm11a}. At $z=3$, its virial mass is  
$\approx10^{11}\,\msun$, which is the typical host halo mass of  
Lyman alpha emitters \citep[LAE,][]{gawiser07}.
As a comparison, LBGs are thought to reside in more massive haloes
with $\mhalo\gtrsim10^{11.5}$ \citep{giavalisco01,arnouts02,adelberger05,lee09,bielby13}.

Using the two Millennium simulations \citep{springel05,boylan-kolchin09}, 
\citet{fakhouri10} measured that a halo of mass $10^{11}-10^{12}\,\msun$ would typically have 
$\sim$2, 10, and 20 mergers of the ratio 1:3.3, 1:33, and, 1:100 at $3<z<9$, respectively.
The Nut halo undergoes 6  minor mergers (mass ratio of 4--20), and 15 very minor 
(mass ratio of 20--100) mergers at $3<z<10$, hence it has a relatively quiet merger 
history (Figure~\ref{fig:tree}). Note that the only major merger occurs at 
$z=2.2$ \citep{kimm11a}. In Figure~\ref{fig:mmp_img}, we show the projected distributions 
of gas density,  metallicity,  stellar density, and star formation rate density of the Nut halo 
and galaxy at $z=3$ in each run. 

\subsection{Star formation histories}

We begin our investigation of the evolution of galaxy properties with the cooling run (NutCO) 
in which energy from SNe is neglected. Because there is no pressure support from stars and 
the UV background radiation is self-shielded above 0.01 H/cc \citep{faucher-giguere10,rosdahl12}, 
star-forming gas clumps experience a runaway collapse converting a large fraction of the 
total baryons into stars at all redshifts (Figure~\ref{fig:sf}, the left panel). The growth of the 
stellar component closely follows the assembly history of its host dark matter halo (shown as a 
dashed line), indicating that star formation is limited by gas accretion.  
The main progenitor of the Nut galaxy forms stars at a rate of $\sim 10\,\msunyr$ in the last 
$\sim1.5$ Gyr (Figure~\ref{fig:sf}, right panel), resulting in a final stellar mass of 
$\mstar=1.9\times10^{10}\,\msun$ at $z=3$. As a consequence of the efficient star formation, 
the galaxy ends up being gas-poor at $z=3$ 
($f_{\rm gas}=\mgas/\left[\mgas+\mstar\right]\approx0.15$), 
inconsistent with the observations at $z\sim2-3$ that galaxies in a similar mass range  
are typically gas rich ($f_{\rm gas}\gtrsim0.5$) \citep[e.g.][]{mannucci09,tacconi10}.

\begin{figure}
   \centering
   \includegraphics[width=8.5cm]{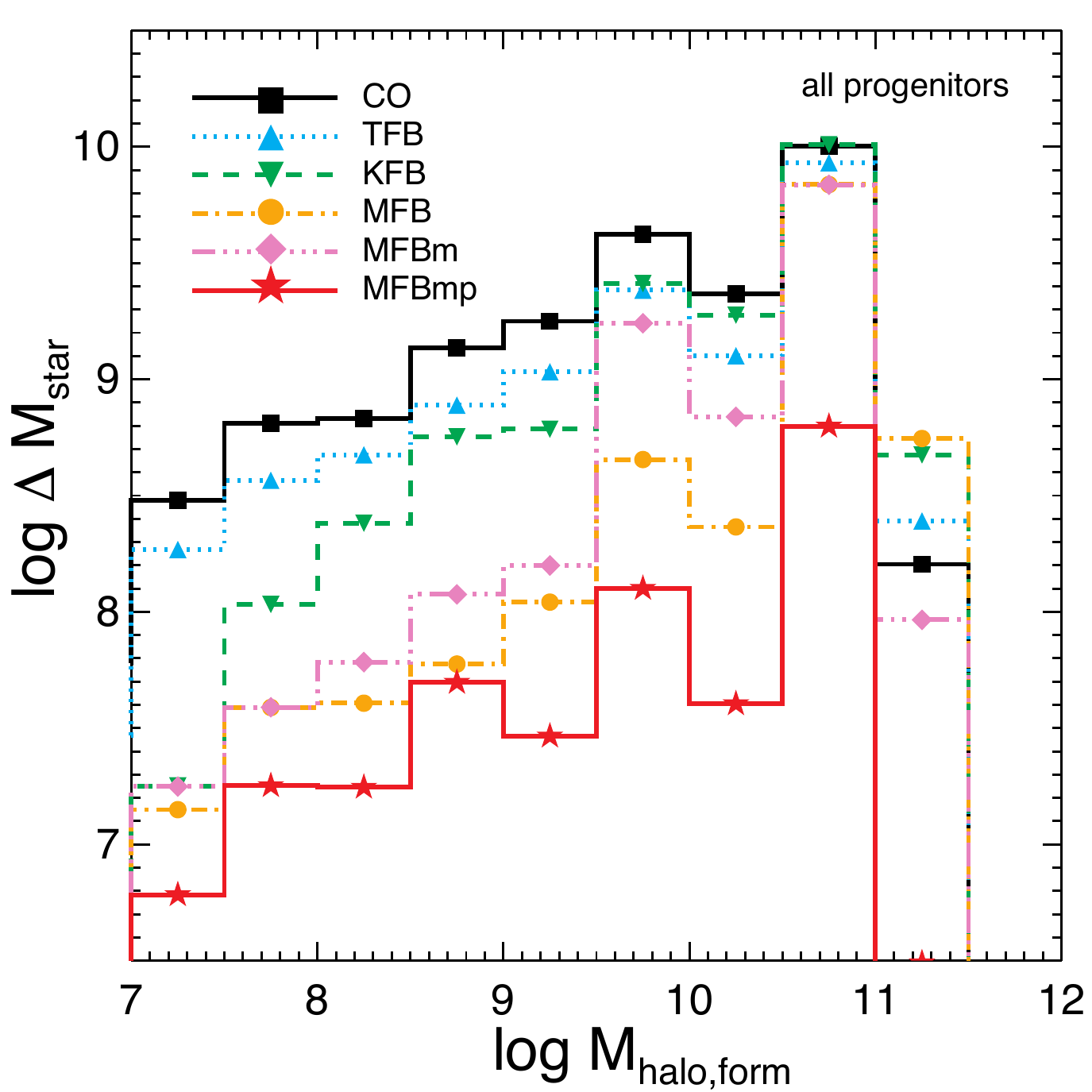} 
   \caption{The mass of dark matter haloes in which each star particle is born.
   The y-axis indicates the amount of stars formed in haloes of different masses. 
   A larger amount of stars form in more massive haloes, consistent with the 
   observed main star formation sequence. Star formation should be suppressed 
   not only in small haloes but also in massive haloes to be consistent with 
   the empirical results that only a small fraction of baryons turns into stars. }
   \label{fig:mform}
\end{figure}

Figure~\ref{fig:mform} shows the mass of the progenitor dark matter halos in which each 
star particle is formed. The plot demonstrates that most of the stars populating the Nut galaxy 
$z=3$ in the NutCO run are formed in massive haloes. The amount of stars formed between 
$3\leq z \leq 6$ (when the halo mass is $10.5 \leq \log \mhalo/\msun \leq 11$) accounts for half 
of the total stellar mass, which reflects that more massive galaxies show a higher level of 
star formation activity \citep{salim07,elbaz07,noeske07,daddi07,gonzalez10,whitaker12,salmon15}.
Thus, feedback must regulate star formation at least in haloes within this mass range, 
if not larger, in order to be consistent with the observational findings that only a small 
fraction ($\lesssim 10\%$) of baryons is turned into stars at  $z\sim3$ \citep{gawiser07,moster10,guo10,behroozi13,moster13}.

The inclusion of thermal SN explosions (NutTFB) has a weak impact on the 
star formation history of our intermediate-size galaxy. This is not unexpected, given that 
thermal energy is susceptible to severe artificial radiative losses. 
Nevertheless, the galaxy stellar mass from the NutTFB run is reduced by 31\% 
($\mstar=1.3\times10^{10}\,\msun$), compared with that from the cooling run.
Figure~\ref{fig:sf} (the right panel) shows that the suppression of star formation occurs 
mostly at an early epoch ($z\gtrsim4$), suggesting that thermal feedback is 
more efficient in dark matter haloes with a shallower potential. 

There is a clear trend that star formation is more suppressed
in smaller haloes with kinetic feedback (Figures~\ref{fig:sf} and ~\ref{fig:mform}). 
This is essentially because the ejecta momentum from SN explosions with a slight boost 
($p\lesssim 3\,p_{\rm ej}$) is significant enough 
to disperse small star-forming clouds present in dwarf-sized galaxies. 
However, as the halo becomes larger and more gas is accreted, 
the galaxy forms a denser and more massive gas cloud at its centre, 
leading to efficient star formation. The resulting stellar mass of the Nut galaxy 
at $z=3$ in the NutKFB run becomes greater 
($1.5\times10^{10}\,\msun$) than that in the NutTFB run. 

\begin{figure}
   \centering
   \vspace{0.1in}
   \includegraphics[width=8.cm]{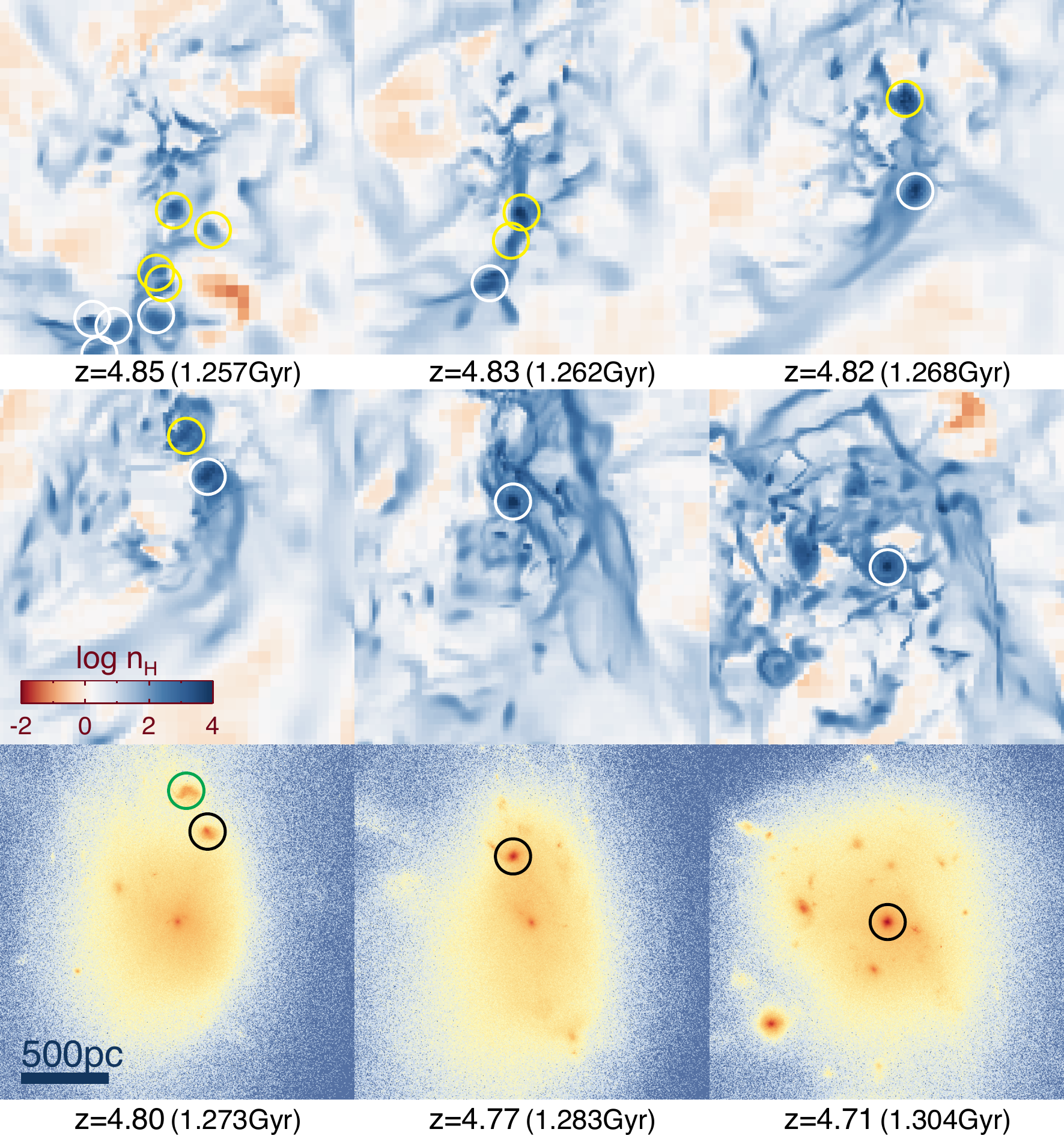} 
   \caption{Formation of a massive stellar core due to the overcooling 
   triggered by mergers of gas clumps in the NutMFB run at $z\approx4.8$. 
   The top and middle panels display the projected gas density at different redshifts. 
   The numbers in the parentheses are the age of the universe at each redshifts. 
   The circles indicate the clumps that lead to a formation of a giant clump 
   of $1.4\times10^8\,\msun$ at $z=4.71$ 
   ($\mstar=8\times10^7\,\msun$, $\mgas=6\times10^7\,\msun$) for which  
   cooling is accelerated.  The clumps with the same colours merge first 
   into a clump before the white and yellow circles merge at $z=4.77$.
   Also included in the bottom panels are the stellar density 
   distributions at $4.71\le z \le 4.80$. The escape velocity of the massive clump 
   is $\sim 110\,\kms$. Note that winds with $v>110\kms$ are 
   difficult to generate inside a dense, gas-rich core via momentum input from SN. 
   A possible solution to prevent the onset of the overcooling is by having early, efficient (SN) feedback. }
   \label{fig:overcool}
\end{figure}

On the other hand, the use of the appropriate momentum from the adiabatic to snowplow 
phase ensures that star formation is notably suppressed in the NutMFB run. 
The galaxy stellar mass in this run is 2.4 times smaller than that of the NutCO run at $z=3$. 
It is worth noting that the mass difference 
is even more significant at high redshifts ($z\gtrsim5$), indicating again that the regulation of
star formation is more effective in small-mass haloes. The amount of stars formed in haloes 
with $\mhalo \lesssim 10^{10.5}\,\msun$ is reduced by an order of magnitude (Figure~\ref{fig:mform}). However, once a giant gas clump forms ($M_{\rm core}\sim 10^8\,\msun$) 
as a result of mergers of massive gas clumps ($\mgas\sim10^{5-6}\,\msun$) 
at the centre of the halo at $z\sim4.8$, even momentum from SNe 
fail to disrupt the cloud (Figure~\ref{fig:overcool}).
We will discuss the cause of the inefficiency in more detail in Section~\ref{sec:discussion}. 

In previous cosmological simulations, owing to finite resolution, assuming 
that each star particle of mass $10^4-10^6\,\msun$ experiences a single explosion 
was unavoidable.  It is now feasible to model discrete, multiple SNe, at least in
zoom-in simulations, and to investigate their effect (NutMFBm). There are two important 
differences in MFBm compared to the aforementioned runs. First, since SNe start occurring 
as early as 3 Myr after the formation of a star particle in the NutMFMm run, the 
self-regulation process starts dispersing dense gas clouds earlier than in the runs with a single event (NutTFB, KFB, and MFB). On the other hand, 
as SN explosions are distributed over time, they become less disruptive in dense environments. This is precisely 
the motivation of using a ``superbubble'' feedback in some studies \citep[e.g.][]{sharma14}.
Figures~\ref{fig:sf} and ~\ref{fig:mform} indeed show that the stellar mass grows faster 
than the NutMFB run, indicating that the feedback is not as efficient as the single explosion 
case (NutMFB).

The last model (NutMFBmp) explores the possibility that SN explosions channel 
preferentially through a lower-density medium which may not be fully resolved in our 
simulations. Note that the momentum input from SNe is not boosted artificially in this run, 
but calculated according to Equation~\ref{eq:psn} with an updated average density,
as a smaller amount of mass from the cell hosting the SN is now entrained 
($f_{\rm host}=0.1$, Equation~\ref{eq:newloading}).
We find that the subtle change in modeling momentum deposition can significantly alter
star formation histories of the main galaxy from  NutMFBmp (Figure~\ref{fig:sf}, red lines). 
As the outflow becomes faster along lower-density channels and momentum 
injection is slightly increased ($\sim$30\%\footnote{Note that 
only a small fraction of cells (10\%) would have 30\% increase in the SN momentum 
by assuming that the volume-filling density is ten times smaller than the mean density 
of the host cell of SN (see Section~\ref{sec:discussion}). The total momentum input 
from SNe in the galaxy is increased only by $\sim$4 \%.},
if the density is reduced by a factor of ten), 
SN explosions ensure to remove gas from  dense star-forming regions continuously
and suppress gas accretion on to the galaxy centre (see Section~\ref{sec:discussion}). 
As a result, the galaxy stellar mass is reduced 
by more than an order of magnitude at $z=3$ ($\mstar=7.4\times10^8\,\msun$),
compared to that of the cooling run. 
Notice that we are not claiming that the outflow along lower-density channels 
would easily disrupt the massive gas clouds with $m_{\rm gas} \gg 10^7\,\msun$ which 
already suffer from the overcooling problem. Rather, we find that the feedback scheme is able 
to prevent the formation of overcooled structures in the first place 
through the continuous outflows and keep the star formation rate low.

We also note that star formation in the NutMFBmp run becomes episodic,
as a small number of gas clumps govern the total star formation rate 
(Figure~\ref{fig:mmp_img}). When these clumps are disrupted by the effective feedback, 
star formation drops rapidly. Such bursty star formation is also seen in 
\citet[][Figure 9]{hopkins14}. At lower redshifts ($z<1$) when the galaxy becomes 
more massive and its specific star formation rate drops, many star-forming clouds are 
likely to form in the galactic disc \citep[see Figure 1 in ][for example]{hopkins14} and 
star formation histories become smoother.

It is worth mentioning that the role of the outflow along lower-density 
channels in regulating star formation may vary under certain circumstances. For example, 
using a set of idealised simulations with an initially turbulent medium, \citet{hennebelle14} 
show that SNe exploding in lower density environments enhance star formation (their model D) 
than the case where SNe always explode in a dense region (their model C1, Figure 13). 
They argue that this is likely because most of the SN energy is carried away with diffuse gas, 
compressing the neighboring dense clouds (model D). This may sound contradictory to 
our findings, but we note that the different conclusions may be reached due to  
different initial conditions and timescales involved. First, while dense clouds develop 
simultaneously over the entire computational domain in \citet{hennebelle14}, gas is accreted 
smoothly through the cosmic web over several Gyr in our simulations. Thus, the diffuse gas 
accelerated by SN explosions is less likely to encounter the collapsing clouds and trigger 
star formation in our case. Second, although the diffuse hot gas may not destroy 
nearby star-forming clouds, it can reduce the gas cooling and accretion onto the galaxy centre 
by interacting with halo gas, leading to a lower star formation activity on cosmological time scales.

 \begin{figure}
   \centering
    \includegraphics[width=8.5cm]{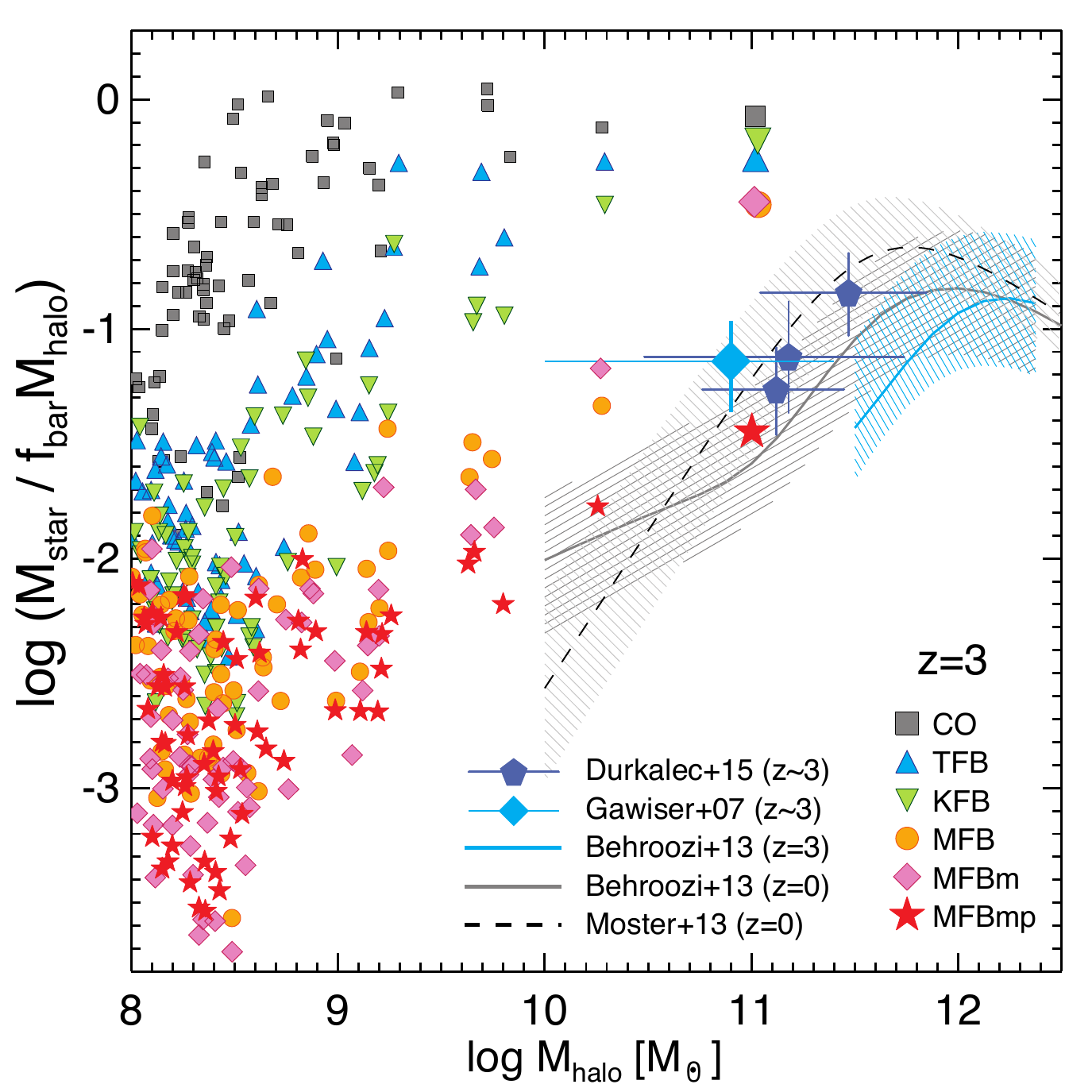} 
   \caption{Stellar mass fraction of galaxies in simulations with different feedback models 
   at $z=3$. Note that galaxies that are not accreted onto the main halo are also included. 
   Different symbols and colours correspond to different feedback schemes. 
   Empirical determinations based on the abundance matching technique are shown as 
   different lines, as indicated in the legend \citep{moster13,behroozi13}. 
    The shaded regions denote 1$\sigma$ and 2$\sigma$ uncertainties for the 
    \citet{moster13} and \citet{behroozi13} results, respectively.
   Independent measurements from the clustering analysis of 
   $z\sim3$ star-forming galaxies selected from the VIMOS ultra deep survey \citep{durkalec15} 
   and Ly$\alpha$ emitters at $z\sim3.1$ \citep{gawiser07} are shown as a blue pentagon and 
   a cyan diamond symbol, respectively. }
   It can be seen that star formation in galaxies embedded in lower-mass haloes is 
   more suppressed in general. Mechanical SN feedback is more efficient than other 
   feedback schemes at regulating star formation. This plot demonstrates that 
   momentum input from SNe alone may be able to match the empirical 
   sequence (the NutMFBmp run), provided that the ISM is highly turbulent.
   \label{fig:shmr}
\end{figure}

In Figure~\ref{fig:shmr}, we examine the stellar mass fraction of the galaxies 
($\mstar/f_{\rm bar}\mhalo$) in our simulations.  Note that galaxies outside 
the virial radius of the main halo are also included. 
For comparison, we include the empirical determinations of the stellar 
mass fraction based on the abundance matching technique \citep{moster10,guo10,behroozi13,moster13}. 
Independent measurements based on the clustering analysis of 3022 star-forming galaxies selected from 
the VIMOS Ultra Deep Survey \citep[][$z\sim3$]{durkalec15} and 162 Ly $\alpha$ emitters at $z\sim3.1$ \citep{gawiser07}
are also included as a blue pentagon and a cyan diamond with error bars, respectively.
We stress that the comparison between the empirical results and our simulations should 
be taken with caution, given that the simulated galaxy sample is quite small. 
Nevertheless, several interesting features can be gleaned from this plot.
First, a large fraction of baryons is converted into stars in massive haloes in the 
cooling run, while the fraction is small in haloes of a few times $10^8\msun$, 
as the UV background heating prevents gas from collapsing. 
We confirm that a similar trend is found when the fraction of total baryonic mass fraction
is plotted instead of the stellar mass fraction. 
Second, there is a general trend that the stellar mass fraction is lower in smaller haloes in the runs with feedback.
Third, the use of more realistic time delays does reduce stellar mass in the small haloes 
($\mhalo \lesssim 10^{9.5}\,\msun$). 
Fourth, the stellar fraction in small haloes agrees well between MFBm and MFBmp.
This means that the fraction of SN host cell mass entrained is irrelevant, 
as long as the momentum injection from SNe is high enough to unbind the star-forming cloud.
Finally, the MFBmp run produces an amount of stars consistent with the empirical sequence,
indicating that momentum input from SNe can provide enough pressure
support to regulate star formation. 
Whether this momentum can be efficiently transferred to the turbulent medium, 
as postulated in the MFBmp run, remains to be investigated, however.

\subsection{Metal enrichment}

 \begin{figure}
   \centering
    \includegraphics[width=8.5cm]{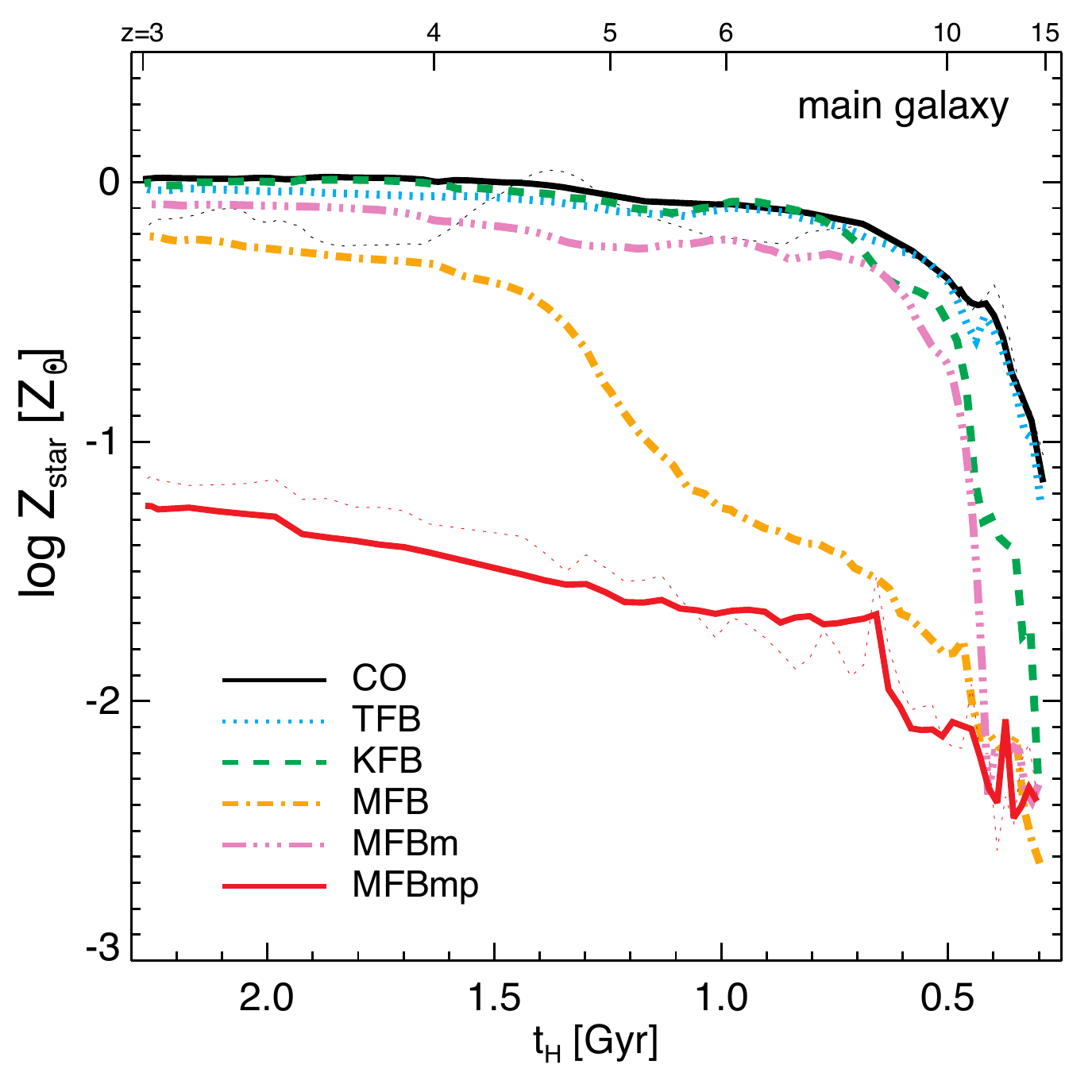} 
   \caption{Evolution of stellar metallicity in the main galaxy. Different colour-codings 
   indicate the runs with different feedback models.
   Also included as black or red dotted lines are the corresponding gas metallicities.
   Stellar metallicity is a good indicator of the overcooling problem, as demonstrated by 
   \citet{wise12b}. We find that the stellar metallicity increases rapidly as soon as a dense core 
   forms at the galaxy centre (see also Figure~\ref{fig:overcool}). Only the efficient 
   feedback model (MFBmp) forms stellar populations with a significantly sub-solar 
   metallicity. 
   }
   \label{fig:zstar_mmp}
\end{figure}

It is well established that the gas phase metallicity in galaxies increases 
with galaxy luminosity or mass \citep{garnett02,pilyugin04,tremonti04,kewley08}. 
The mass-metallicity relation is also observed in galaxies at higher redshifts 
\citep{erb06,maiolino08,mannucci09,yuan13,zahid13,wuyts14,steidel14}, 
but a systematic offset is found such that more distant galaxies are more metal-poor 
for a given mass. This suggests that metals are not instantly recycled to enrich the ISM, 
but possibly blown away through galactic winds \citep{dalcanton07}.

Stellar metallicity is an important indicator of the overcooling problem. 
\citet{wise12b} show that the metallicity of stars is very sensitive to the strength of stellar feedback.
We also confirm that the metal enrichment is directly affected by a galaxy's ability 
to regulate star formation. In the cooling run, the stellar metallicity quickly approaches 
the solar value ($Z=0.02$), as newly synthesized metals are recycled 
instantaneously within their birth clouds. Visual inspection of metallicity distributions 
within the dark matter halo shows that the intergalactic medium (IGM) is barely metal-enriched 
(Figure~\ref{fig:mmp_img}, the top second panel). This demonstrates that galaxy interactions are 
not an efficient process to re-distribute metals in this intermediate mass haloes.
The galaxies in runs with SN feedback start with a lower metallicity when the halo mass is small, 
but again most of them approach solar metallicity as soon as a massive stellar core which 
feedback cannot destroy forms in the galaxy (see Figure~\ref{fig:overcool}). The only 
exception is the NutMFBmp run in which strong outflows prevent the formation of 
such a massive core by continuously dispersing gas clumps (Figure~\ref{fig:mmp_img}).

 \begin{figure}
   \centering
    \includegraphics[width=8.5cm]{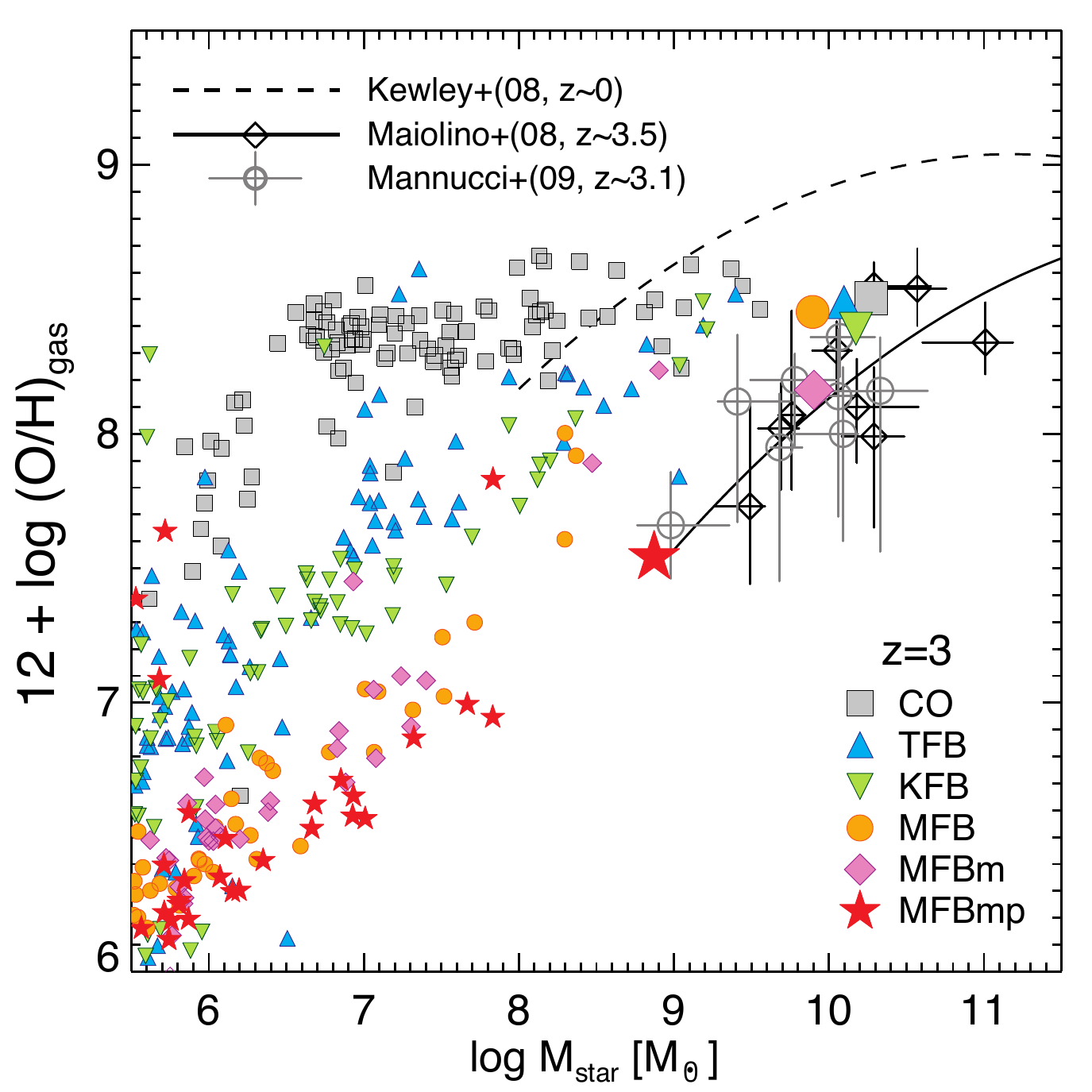} 
   \caption{The mass-metallicity relation (MZR) at $z=3$ in the Nut simulations.
   Different symbols and colours correspond to the runs with different feedback models.
   For comparison, we include the observed MZRs drawn from 9 galaxies at $z\sim3.5$ 
   \citep{maiolino08} and 8 LBGs at $z\sim3.1$ \citep{mannucci09}.  The solid line is the 
   fit to the results from \citet{maiolino08}, while the local MZR derived from the 
   local SDSS galaxies is shown as a dashed line \citep{kewley08}. 
    A fraction of simulated galaxies exhibits very enhanced metallicity compared to their 
   mean MZR sequence, because 
   SNe blow out star-forming gas temporarily and the gas metallicity 
   is governed by the enriched outflow. Note that without outflows (NutCO) the slope of 
   the MZR is shallower than the observations. The effectiveness of SN feedback determines 
   the offset from the observed MZR at $z\sim3$.}
   \label{fig:zgas}
\end{figure}

Figure~\ref{fig:zgas} compares the mass-metallicity relation (MZR) of the simulated galaxies 
from different feedback runs at $z=3$.  We compute the gas-phase metallicity by assuming 
the solar abundance (${\rm (O/H)}_{\rm \odot}=8.69$, \citealt{asplund09}).  Our minimal run (NutCO) shows a 
steep MZR in small galaxies  ($\mstar\lesssim10^7\,\msun$), as  star formation is mitigated 
by the strong UV background radiation. The MZR becomes nearly flat, however, for more 
massive galaxies, again due to efficient recycling of metals, in contradiction to observations. 
The MZR has a slope close to the observed one only when SN feedback is included. 

The normalization of the MZR in simulations is directly affected by the strength of SN feedback.
While observations at high redshifts ($z\sim3$) show that 
galaxies with $10^{9}\lesssim \mstar \lesssim 10^{9.5}\,\msun$ are metal-poor 
($\sim0.1$--$0.3\,Z_{\rm sun}$) \citep{maiolino08,mannucci09}, simulations with 
weak SN feedback (thermal and kinetic) predict metal-rich dwarf galaxy populations 
with $\sim0.3$-$1.0 \,Z_{\rm sun}$. Only the galaxies with star formation  
regulated by strong SN feedback (e.g. MFBmp) reasonably follow the observed sequence. 

It is worth noting that the simulated main galaxies with different feedback,
except the one from the MFBmp run, show similar gas-phase metallicities at $z=3$.
This is essentially because the gas metallicity saturates at around the solar value.
Although the metallicity of the ejecta adopted in this study is 2.5 $Z_{\rm sun}$,
the pristine gas is accreted continuously onto the galaxy at high redshifts, diluting the 
enriched gas. Furthermore, since stellar mass increases together with the metallicity 
\citep[see also][Figure 6]{agertz15}, the gas-phase metallicities of the main galaxies 
seem to agree with the observational data within the current measurement 
uncertainties. Given the comparison of stellar mass fraction, this means that 
the gas-phase metallicity of the massive galaxies ($\mstar\gtrsim10^{10}\,\msun$) may 
not be a good indicator for the overcooling problem, although dwarf galaxies with 
masses $10^8\lesssim \mstar\lesssim10^9\,\msun$ will still be very useful to put a 
strong constraint on galaxy formation models.

The origin of the MZR is often attributed to the fact that star formation efficiency is 
lower in smaller galaxies. Some authors argue that outflows may not be necessary to 
explain the MZR as long as the stellar fraction is a strong function of galaxy mass  
\citep{finlator08,mannucci09,calura09}. However, our simulation (NutCO) 
suggests that the fraction is always high and not a strong function of galaxy (or halo) 
mass without feedback (NutCO, see Figure~\ref{fig:shmr}). This is essentially because 
gas experiences a runaway collapse, shifting the density distribution of star-forming gas 
towards a very dense regime. Note that the typical density of star-forming clouds without 
feedback is set by the force resolution of numerical simulations, and we find that this is 
roughly $n_{\rm H}\sim10^{3.5}\,\rm{cm^{-3}}$ in our runs with 12 pc resolution (see 
also Figure~12 of \citealt{hopkins12b} for higher resolution runs where the typical density is 
larger $n_{\rm H}\gtrsim10^4\,\rm{cm^{-3}}$).
Thus, for a star formation efficiency of $\epsilon_{\rm ff}=0.02$, one would expect that 
more than 80\% of the gas cloud is converted into stars in 100 Myr. Since the Hubble time is 
much longer than this for $z=3$ galaxies, metallicity evolves similarly as a closed box 
model in the absence of feedback, except that the ``equilibrium'' metallicity is smaller 
than the metallicity of the stellar ejecta ($Z_{\rm ej}=0.05$) due to the infall of pristine gas. 
\citet{finlator08} already pointed out that in the absence of outflows the capability of 
reproducing the MZR depends on the gas consumption time scale (see their Figure 13).

\subsection{Galaxy size, circular velocity, and morphology}
\label{morph}

An important prediction from tidal torque theory is that late 
accretion carries larger specific angular momentum ($j$) than the 
gas inflow at early times  \citep{peebles69,doroshkevich70,white84,catelan96}. 
Because of this, the spin parameter of a dark matter halo 
($\lambda'=j_{\rm dm}/\sqrt{2}R_{\rm vir} V_{\rm c}$, \citealt{bullock01}) 
can be kept nearly constant across different redshifts \citep{van-den-bosch02,peirani04,bett07}. 
Numerical simulations indeed confirm that $j$ of dark matter near the 
virial radius is a factor of two to three larger than the net $j$ of the halo 
\citep[][]{kimm11b,pichon11,stewart13,danovich15}.

 \begin{figure}
   \centering
    \includegraphics[width=8.5cm]{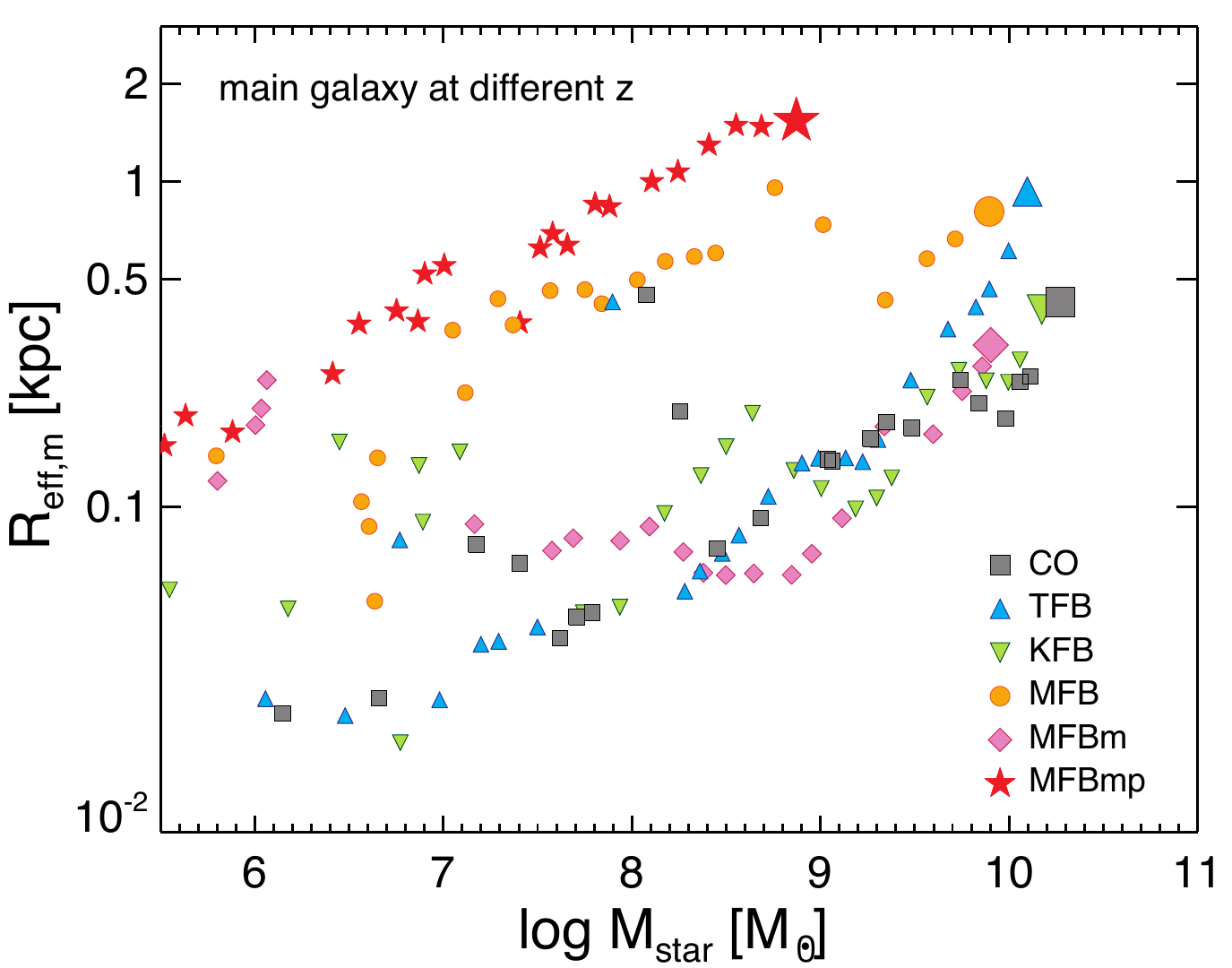} 
   \caption{Evolution of the half-mass radius ($R_{\rm eff,m}$) of the  main galaxy 
   at $3\le z \le15$ ($\Delta z \approx 0.5$). The galaxy at $z=3$ is shown as a bigger
    symbol than others. Different colour-codings and symbols are the same as in
    Figure~\ref{fig:zgas}. 
   Galaxies grow with time, as late accretion supplies gas with larger specific angular 
   momentum. Strong feedback tends to produce a more extended stellar component.
   Note that the half-mass radii can also increase temporarily due to galaxy mergers.
   The late increase in radius at late times in the run with thermal feedback (TFB) is 
   also the outcome of serendipitous mergers (see text).  
   }
   \label{fig:size}
\end{figure}

Baryons gain angular momentum by the same large-scale tidal torques as dark matter 
\citep[e.g.][]{kimm11a}, and thus galactic discs are expected to grow by late gas inflows. 
Figure~\ref{fig:size} indeed shows that the half-mass radius of the stars ($R_{\rm eff,m}$) 
in the main galaxy gets larger with time regardless of the feedback model.
One can also see that the strength of SN feedback generally leads to a larger $R_{\rm eff,m}$.
The NutMFB and NutMFBmp runs produce galaxies with $R_{\rm eff,m}=0.83\,{\rm kpc}$ and 
$1.5\,{\rm kpc}$, while the half-mass radii of the main galaxy in NutCO, NutKFB, and NutMFBm runs 
are $0.43$, $0.40$, $0.31$ kpc, respectively. 

An exception is the thermal feedback run, which produces a rather extended stellar component   
with $R_{\rm eff,m}=0.93\,{\rm kpc}$. We find that the relatively large half-mass radius 
originates from the serendipitous mergers at $5<z<6$. In this run, the merging satellite 
galaxies are more massive than those in other runs with feedback (Figure~\ref{fig:shmr}), 
and we find that the orbital angular momenta of them are coincidently aligned 
with the spin of the galactic disc. Consequently, the merger remnant becomes a more 
extended clumpy disc (Figure~\ref{fig:mmp_img}, fourth column), even though there is still a 
significant central concentration of mass at its centre.
However, it is very unlikely that such serendipitous mergers occur only in certain feedback models.
Although we have not made dwarves a focus of this study, we confirm that dwarf galaxies with 
$10^7\lesssim \mstar \lesssim 10^9\,\msun$ from the NutTFB run are in fact smaller 
($R_{\rm eff,m}\sim 100\,{\rm pc}$) than the runs with mechanical feedback 
($R_{\rm eff,m}\sim 500\,{\rm pc}$). 

It is also interesting to note that the galaxy size is closely linked with the 
star formation history and metal enrichment in stars. The most evident example 
can be seen from the NutMFB run at $4.5\lesssim z \lesssim 5.5$. 
This corresponds to the epoch during which stellar mass and metallicity increase rapidly
indicating the onset of severe runaway collapse and overcooling (Figure~\ref{fig:overcool}).
Accordingly, the galaxy gets significantly smaller during this period 
($8.5\leq \log \mstar/\msun\leq 9.5$). 

 \begin{figure}
   \centering
    \includegraphics[width=8.5cm]{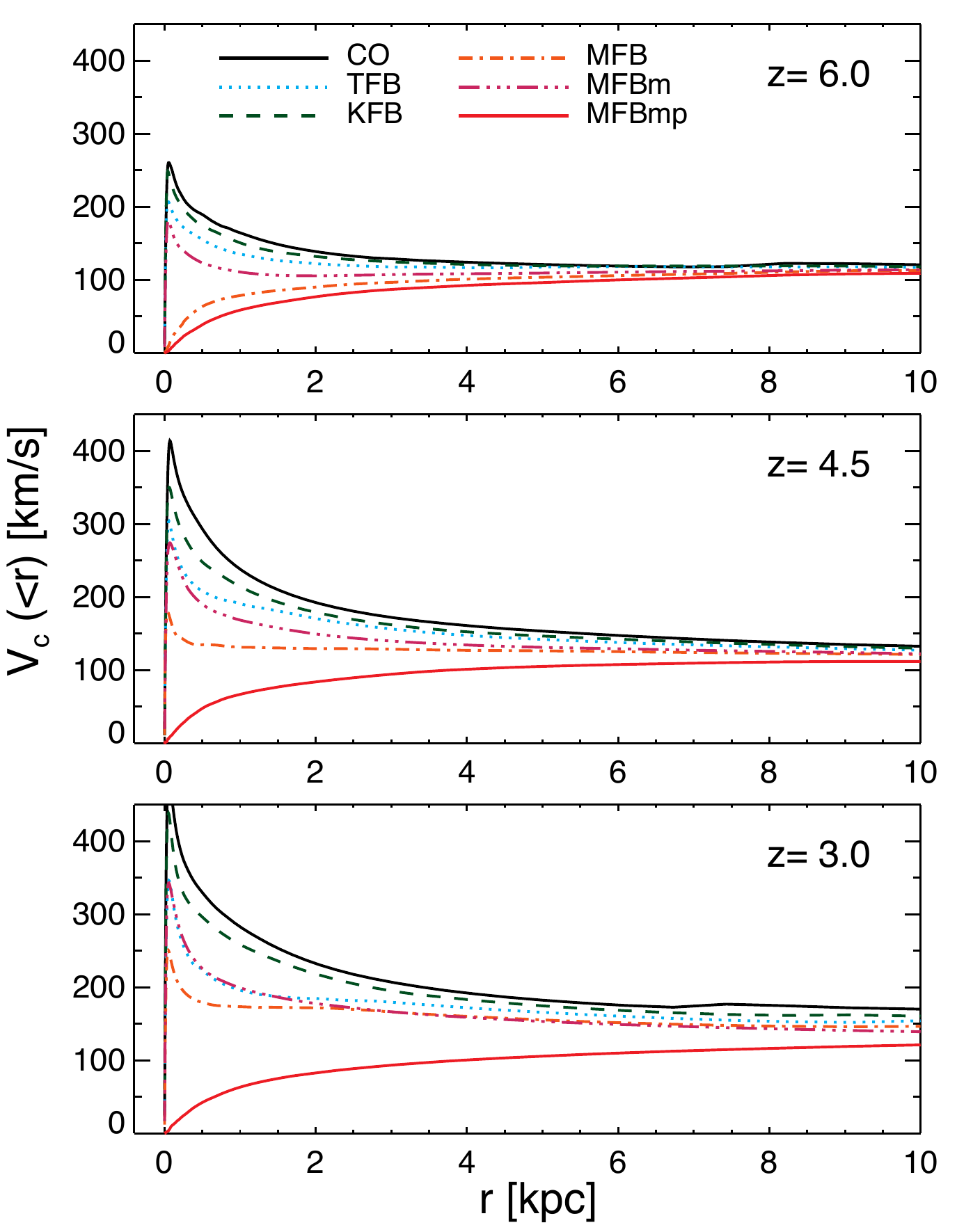} 
   \caption{Circular velocities of the main galaxy from a suite of Nut simulations. 
   The circular velocity is measured as $V_{\rm c}(<r)\equiv \sqrt{GM(<r)/r}$,
   where $M(<r)$ is the total mass (baryons+dark matter) inside radius $r$.
   Different lines correspond to runs with different feedback models, as indicated 
   in the legend. As known in the literature, the runs with weak SN feedback show a 
   peaked velocity curve towards the galaxy centre. The run with mechanical feedback 
   in a porous ISM (MFBmp) suppresses the central peak in the velocity curve.
   }
   \label{fig:vrot}
\end{figure}

Another useful galaxy quantity is the circular velocity ($V_c(<r)=\sqrt{GM(<r)/r}$).
Observed bright spiral galaxies are characterised by the smoothly varying  
baryonic rotation curves \citep[e.g.][]{kassin06}.  However, Figure~\ref{fig:vrot} shows that 
the maximum circular velocity appears at very small radii ($r\sim100\,{\rm pc}$) 
in the runs in which the gas-to-stellar conversion efficiency of the main galaxy is 
higher than the empirical estimates (CO, TFB, KFB, MFB, and MFBmp),
confirming earlier results \citep{navarro93,abadi03,governato07,scannapieco09,hummels12,agertz15}.
The peak near the centre sets in earlier and the maximum circular velocity is larger 
in simulations with less effective feedback. The velocity curve in the NutMFB run is 
smoothly rising at $z=7$, but the peak at the centre begins to emerge at $z\lesssim5$ 
again due to the runaway gas collapse (Figure~\ref{fig:overcool}).
Only the run with the mechanical SN feedback based on the assumption of a porous 
ISM (NutMFBmp) does not show the peaked circular velocity curve.
Note, however, that the {\it rotation} curve of the galaxy turns out to be very different from 
that of the circular velocity, indicating that there is no stellar disc with a well-ordered motion 
(see Figure~\ref{fig:mmp_img} for the stellar density plot in the fourth column).

 \begin{figure}
   \centering
    \includegraphics[width=8.5cm]{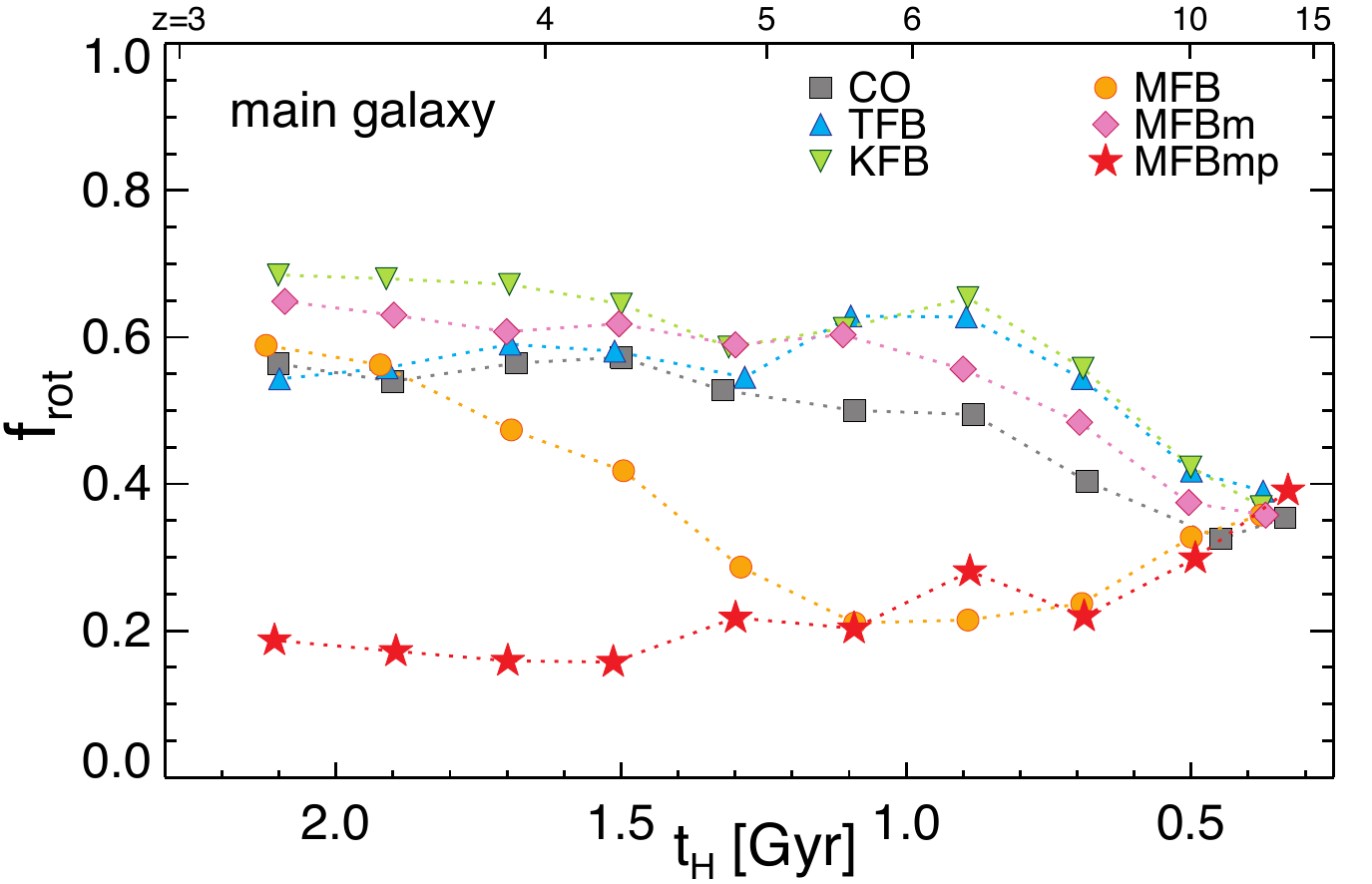} 
    \includegraphics[width=8.5cm]{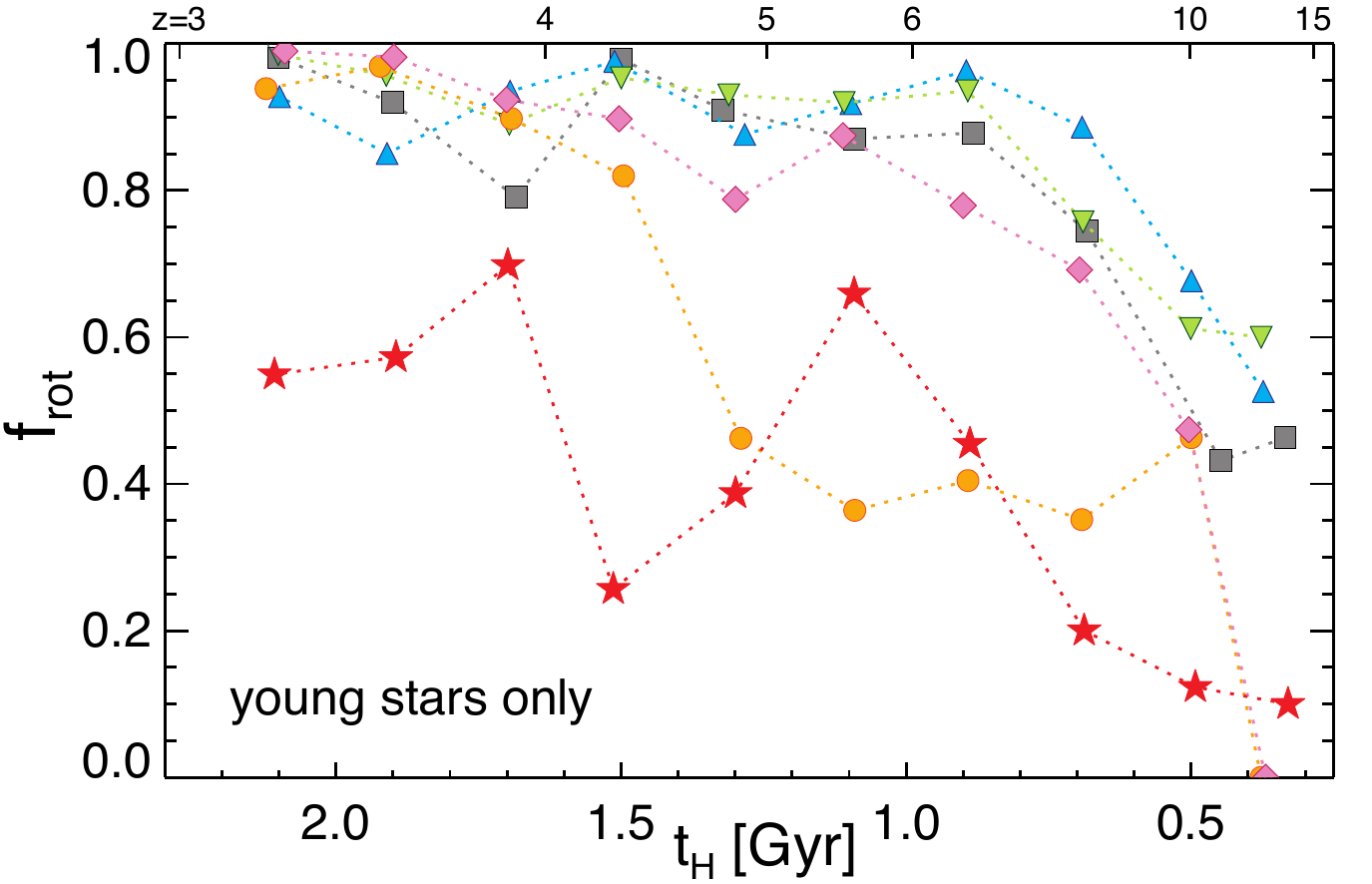} 
   \caption{Top: fraction of stars supported by rotation ($f_{\rm rot}$) in the main galaxy 
   as a function of the Hubble time. We simply assume that a star is rotationally supported 
   if more than half of its total kinetic energy is accounted for by rotational motion. 
   The different colours and symbols correspond to different feedback models, as indicated 
   in the legend. Bottom: $f_{\rm rot}$ of the stars younger than 10 Myr. The main galaxy in 
   the runs having more stars than the empirical estimates (CO, TFB, KFB, MFB, and MFBm)
   is marginally rotation-supported. Interestingly, we find that the most effective feedback 
   (MFBmp) leads to the formation of a spheroidal galaxy.
   }
   \label{fig:frot}
\end{figure}

In order to see the kinematic properties of the simulated galaxies, we quantify the fraction 
of stars supported by  rotation ($f_{\rm rot}$) in Figure~\ref{fig:frot}. We first determine 
the spin direction of the total stellar component within 0.2 virial radius and the 
rotation velocity ($v_{\rm rot}$) of each star particle along the azimuthal direction. 
Then the mass fraction of 
stars whose kinetic energy ($E_{\rm kin,\star}=mv^2/2$) is dominated by the rotational 
component ($E_{\rm rot,\star}/E_{\rm kin,\star}>0.5$) is calculated within the inner radius 
($r\le 0.2 \rvir$),
where $E_{\rm rot,\star}=mv_{\rm rot}^2/2$. $f_{\rm rot}$ may be viewed as a proxy 
for the galaxy morphology in the sense that a purely 
rotating disc would have  $f_{\rm rot}=1$, while a stellar system with only radial motions would 
have  $f_{\rm rot}=0$. Figure~\ref{fig:frot} (top panel) shows that the simulated galaxies 
with a high stellar-to-halo mass ratio are marginally supported by rotation. Approximately 
55-70 \% of the stars have orbits that are more or less aligned ($\theta<45^\circ$) with 
the spin axis of the galaxy, giving rise to geometrically thick discs (Figure~\ref{fig:mmp_img}, fourth columns)\footnote{Note that we do not reorient the 
galaxy to the edge-on configuration in Figure~\ref{fig:mmp_img} to demonstrate that 
the spin direction of the simulated galaxies may be significantly altered by the 
feedback model.}. 

On the other hand, a large fraction of stars ($\sim80\%$) in the 
main galaxy from MFBmp is supported by random motions, resulting in a spheroidal  
morphology rather than a disc (Figure~\ref{fig:mmp_img}). 
This seems to be in conflict with the previous claim that strong feedback preferentially 
blows out gas with low angular momentum \citep[e.g.][]{governato07,brook11},
but consistent with recent high-resolution simulations by \citet{roskar14} and \citet{hopkins14}.
Using the same grid-based code as ours ({\sc ramses}), \citet{roskar14} performed 
idealised simulations varying the optical depth to IR photons to investigate the 
impact of radiation pressure on galaxy properties. They find that the model with strong 
stellar feedback, which matches the empirical stellar-to-halo mass relation, leads to the 
formation of a spheroidal galaxy. \citet{hopkins14} perform cosmological simulations to 
study the formation of disc galaxies with a sophisticated stellar feedback model, including 
radiation pressure and SN feedback. They show that the morphology of the 
Milky-Way progenitor at $z=3.4$ is irregular due to strong outflows (see their Figure 1). 
Somewhat encouragingly, despite the absence of the well-ordered disc at $z\sim3$, 
they find that smooth gas accretion at later times leads to the formation of a realistic, 
extended disc. In this regard, the spheroidal morphology we obtain from the MFBmp run 
may not be problematic, but we will need to run the simulation down to $z=0$ to 
confirm this.

Although a significant fraction ($\sim30\%$) of stars has orbits misaligned with the 
overall spin of the galaxy, the stars younger than 10 Myr form out of a well organised,
gaseous disc ($f_{\rm rot}\gtrsim 0.8$) in the CO, TFB, KFB, and MFBm runs. 
This can also be seen in the projected star formation rate density distributions 
(Figure~\ref{fig:mmp_img}, fifth columns).
One may wonder why $f_{\rm rot}$ of the whole stellar population does not increase notably 
for $3\le z \lesssim 6$, given that $f_{\rm rot}$ of the young component is high ($\sim0.9$). 
This can be attributed to two factors. First, the young stars form in the core 
of the galaxy ($\sim 300$ pc) where there exists a massive bulge of 
$\sim2-4\times10^9\,\msun$. Since the motion of the young stars is mainly 
governed by the gravity of the bulge, the interaction with the spherical structure is likely 
to isotropise their motions. More importantly, the kinematically cold young stars are 
susceptible to external disturbances, such as harassment or mergers. In order to check 
this, we follow the evolution of $f_{\rm rot}$ of the young stars ($t\leq10\,{\rm Myr}$) formed at 
different redshifts ($z=4.5-4.8$, $\Delta z=0.05$) in the NutTFB galaxy during which the 
spin vector of the (young) stellar component is nearly unchanged ($\lesssim 15^{\circ}$),
while their $f_{\rm rot}$ drops to $\sim$ 0.5. We find that the evolutionary patterns of 
$f_{\rm rot}$ in the young population with different ages are well synchronised, 
following the merging histories of star clusters and/or galaxies, irrespective of their age.
This means that the young stars are continuously transformed into a bulge
 by the dynamical disturbances. 

We find that star-forming gas shows irregular morphologies when SN feedback is strong 
(see Figure~\ref{fig:mmp_img}, MFBmp) or when the galaxy is small
($z\gtrsim7$ in the MFB runs).  This is essentially because the feedback keeps 
blowing out/away a substantial amount of gas before a kinematically cold gaseous 
disc forms. 
Let us suppose that there exists a well-defined, clumpy galactic disc 
in the x-y plane. If star formation is very active, it would drive strong outflows 
along the $z$ direction destroying the star-forming clouds. Even though the gas is 
initially coplanar, the spin direction of the returning gas would no longer be 
aligned with the galactic disc, randomising the direction of the spin direction of the subsequently collapsing 
star-forming clouds. This effect is likely to be more significant if bursty star formation 
takes place in a small number of gas clouds, as in the MFBmp run 
(Figure~\ref{fig:mmp_img}). However, if a gas-rich disc becomes massive  
enough to behave like a buffer region in which the returning galactic fountain gas is first
mixed with the disc gas or other fountain gas, star formation in the well-defined disc 
is likely to continue, increasing the fraction of kinematically settled discs \citep{kassin12b}. 

 \begin{figure*}
 \begin{flushleft} 
   \includegraphics[width=5.8cm]{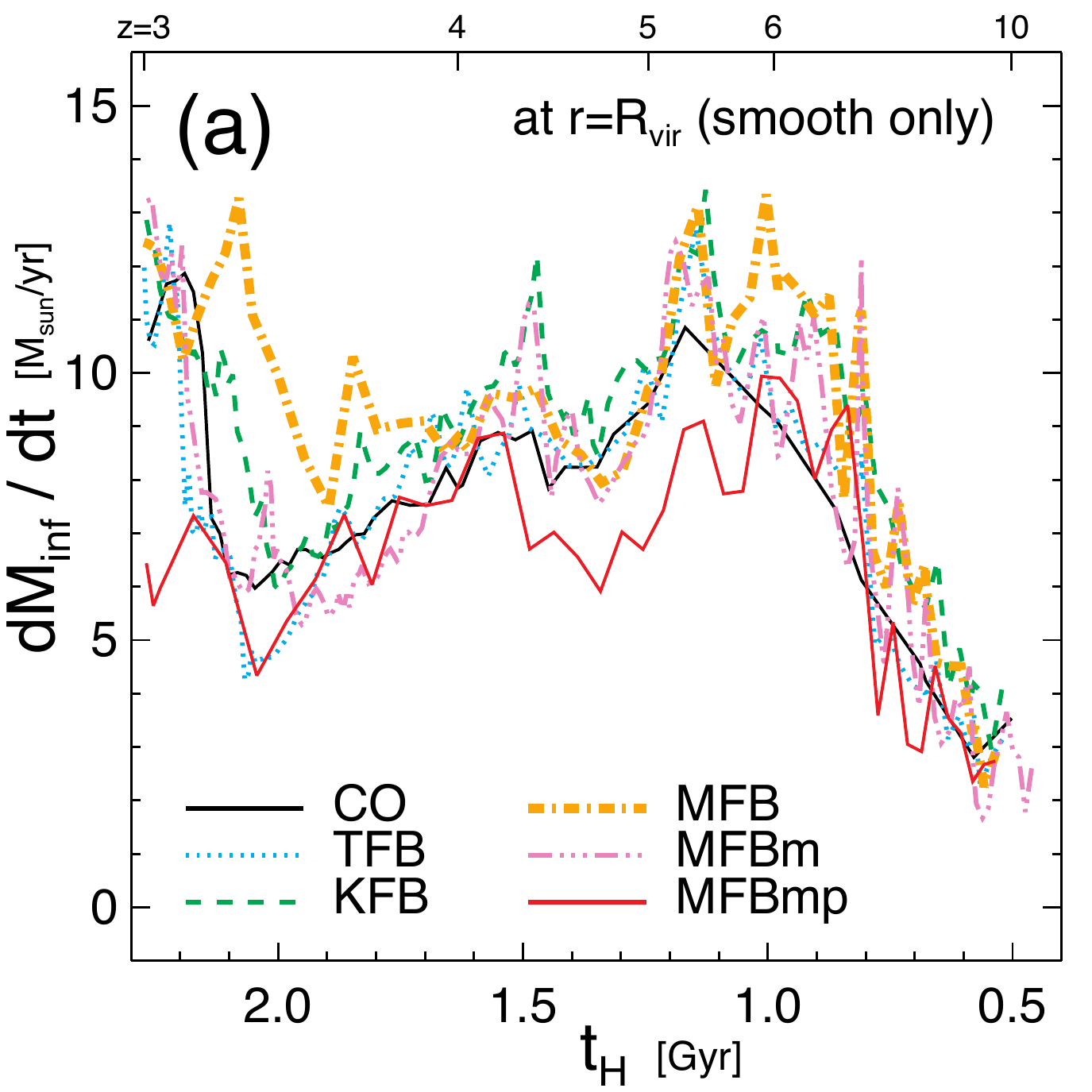} 
      \includegraphics[width=5.8cm]{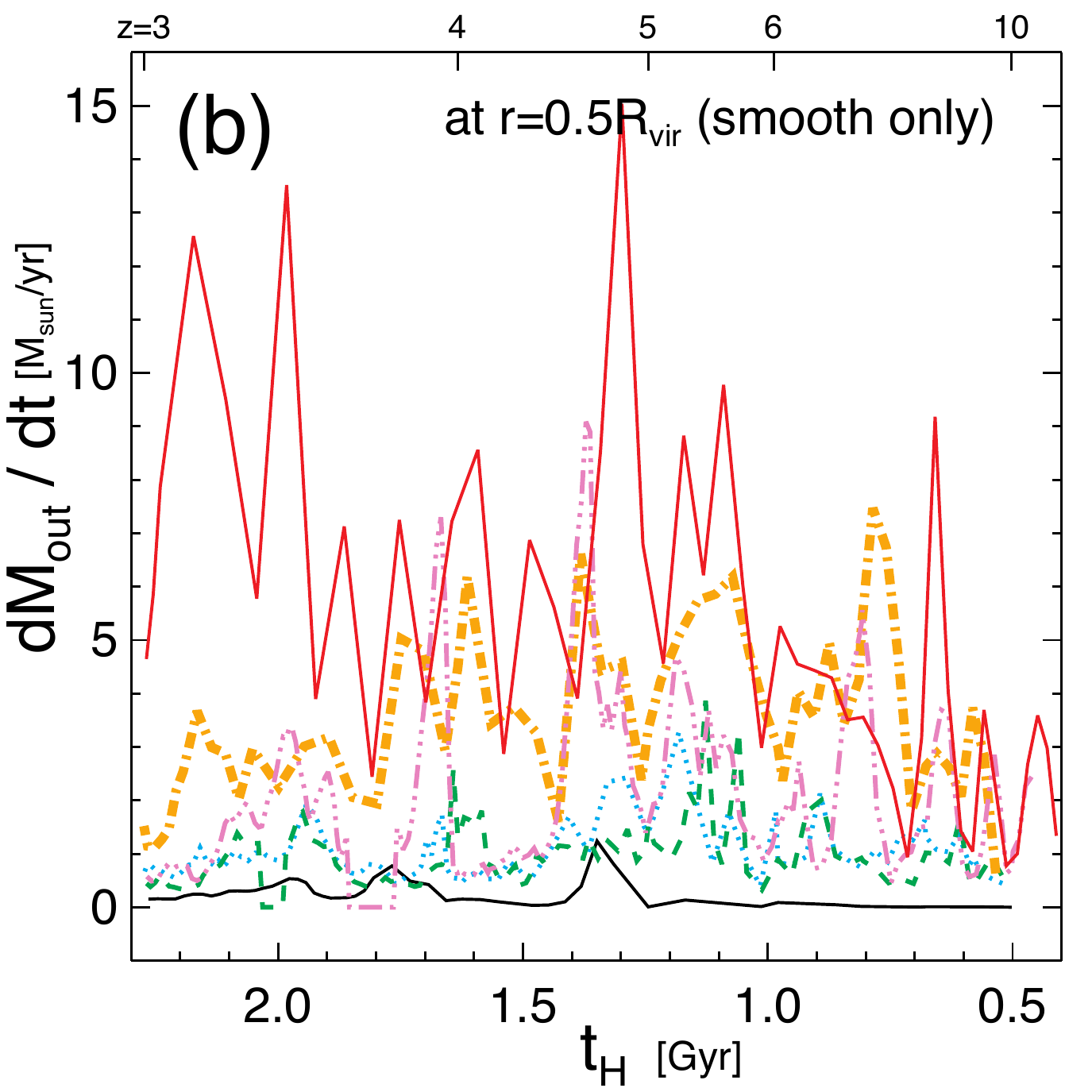} 
   \includegraphics[width=5.8cm]{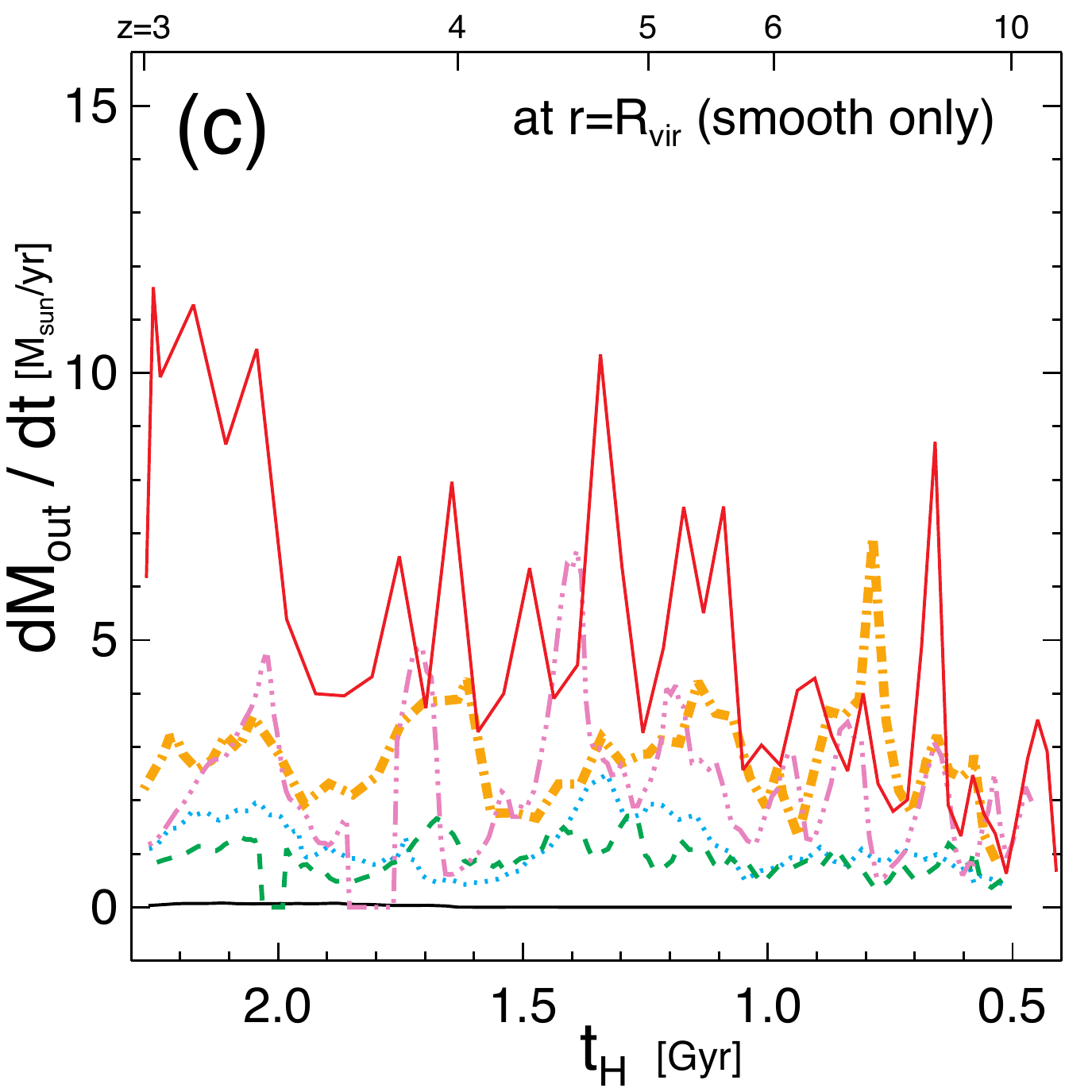} 
   \includegraphics[width=5.8cm]{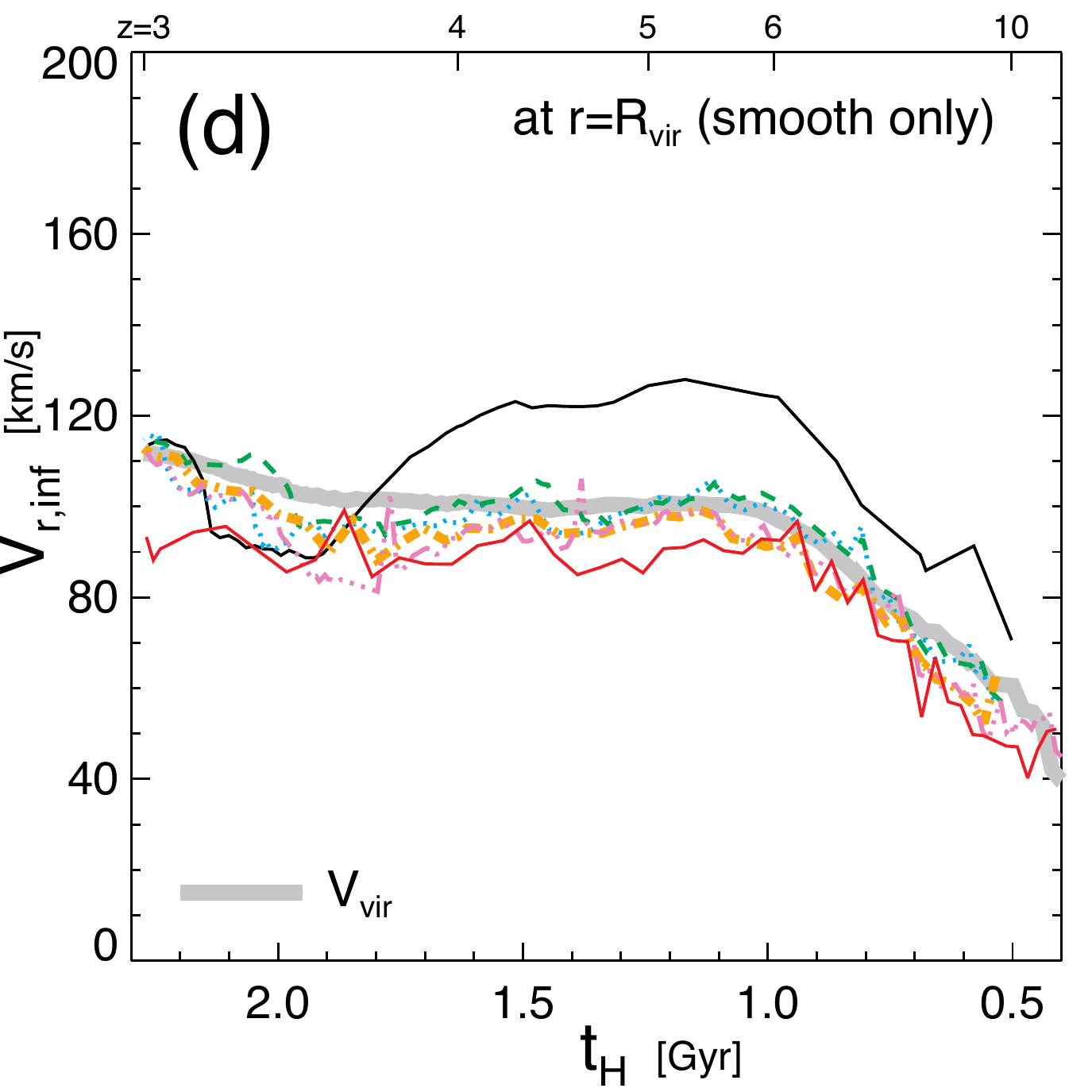} 
   \includegraphics[width=5.8cm]{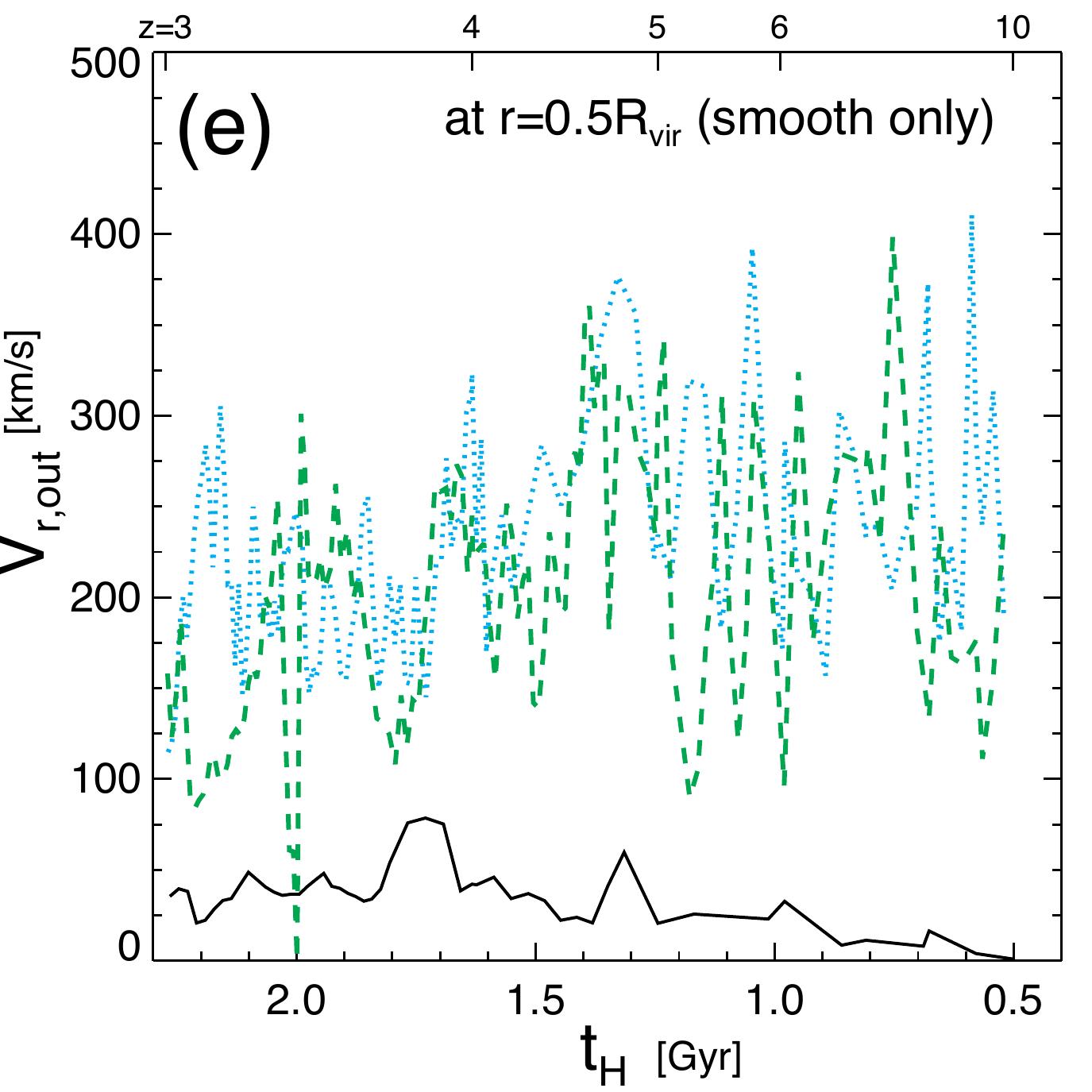} 
   \includegraphics[width=5.8cm]{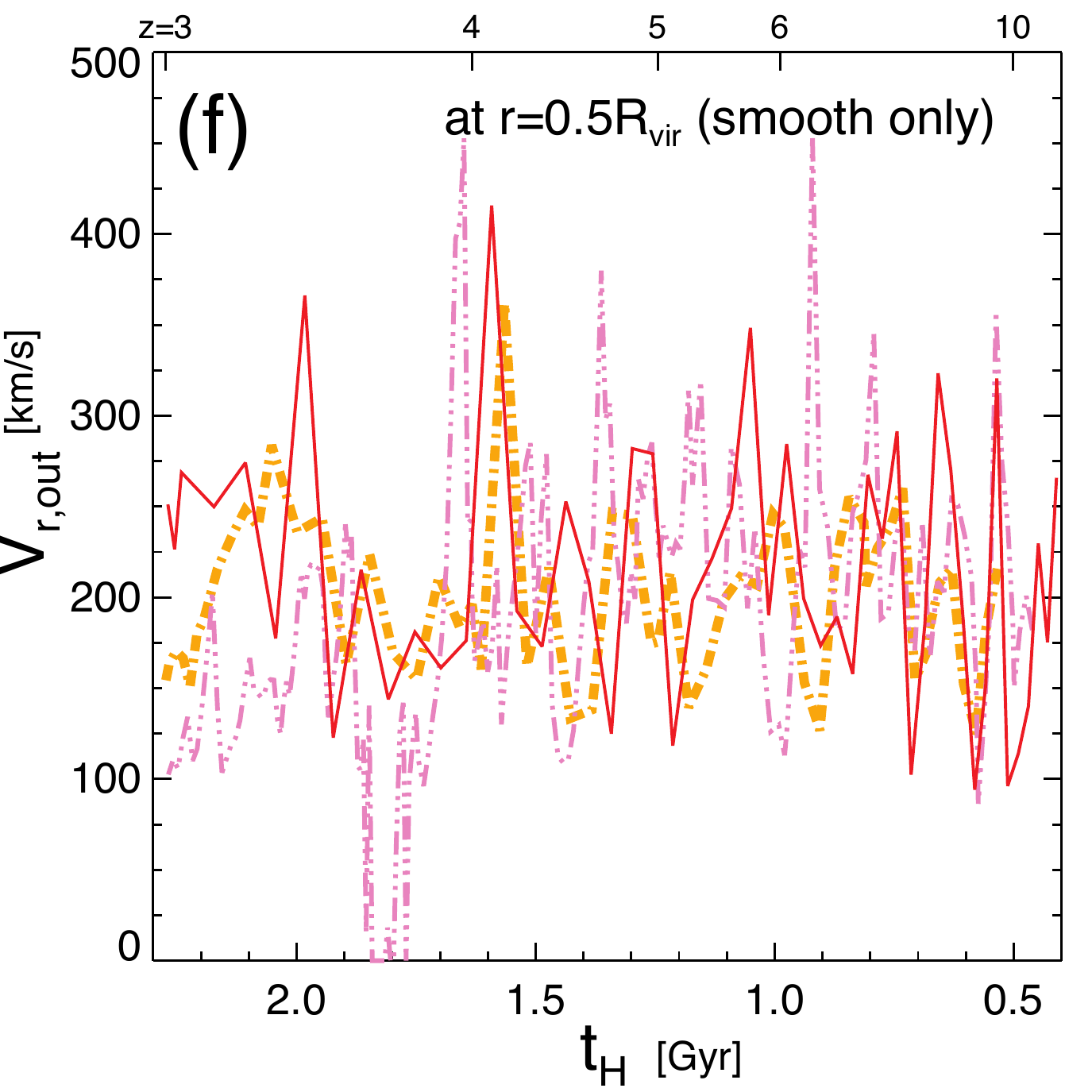} 
   \includegraphics[width=5.8cm]{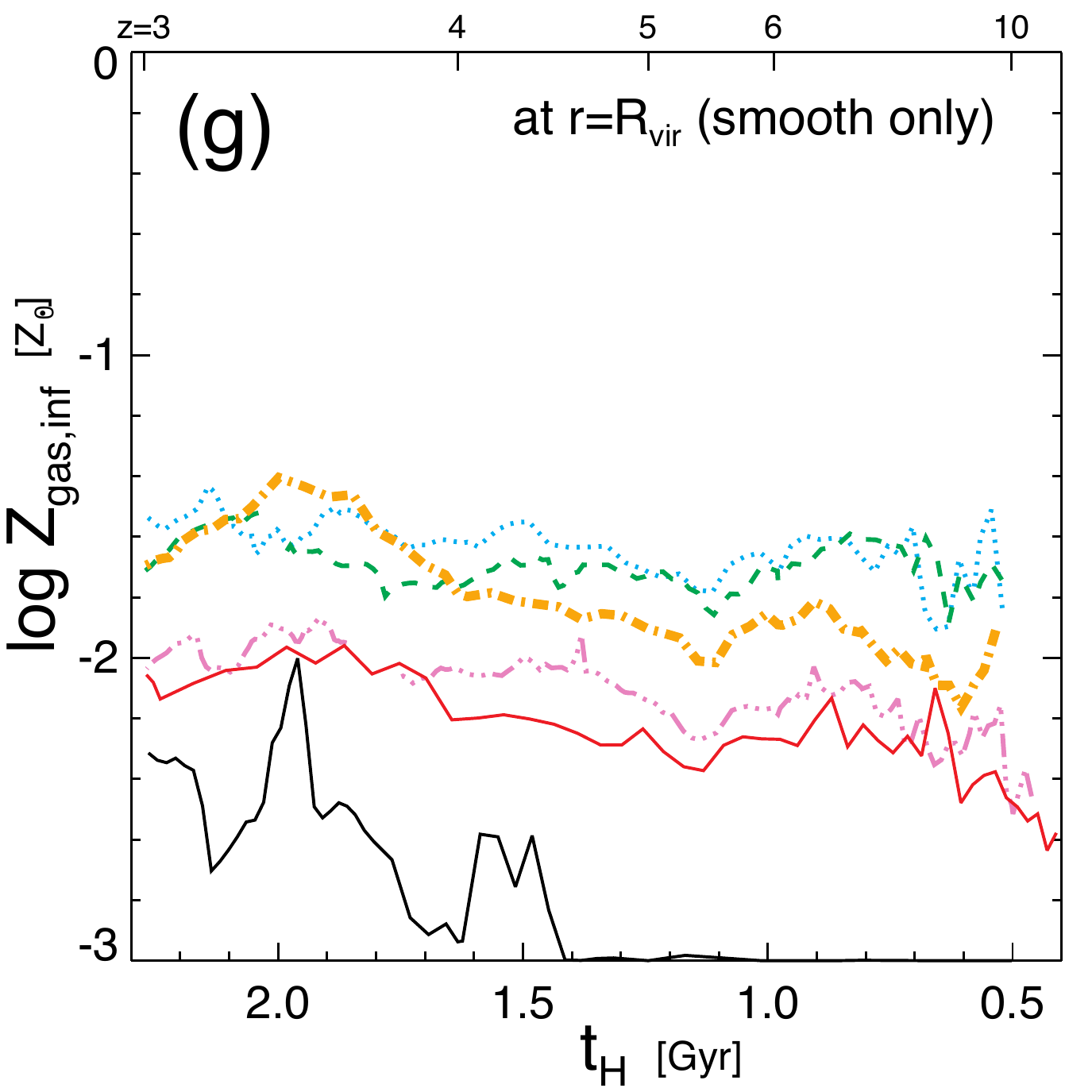} 
   \includegraphics[width=5.8cm]{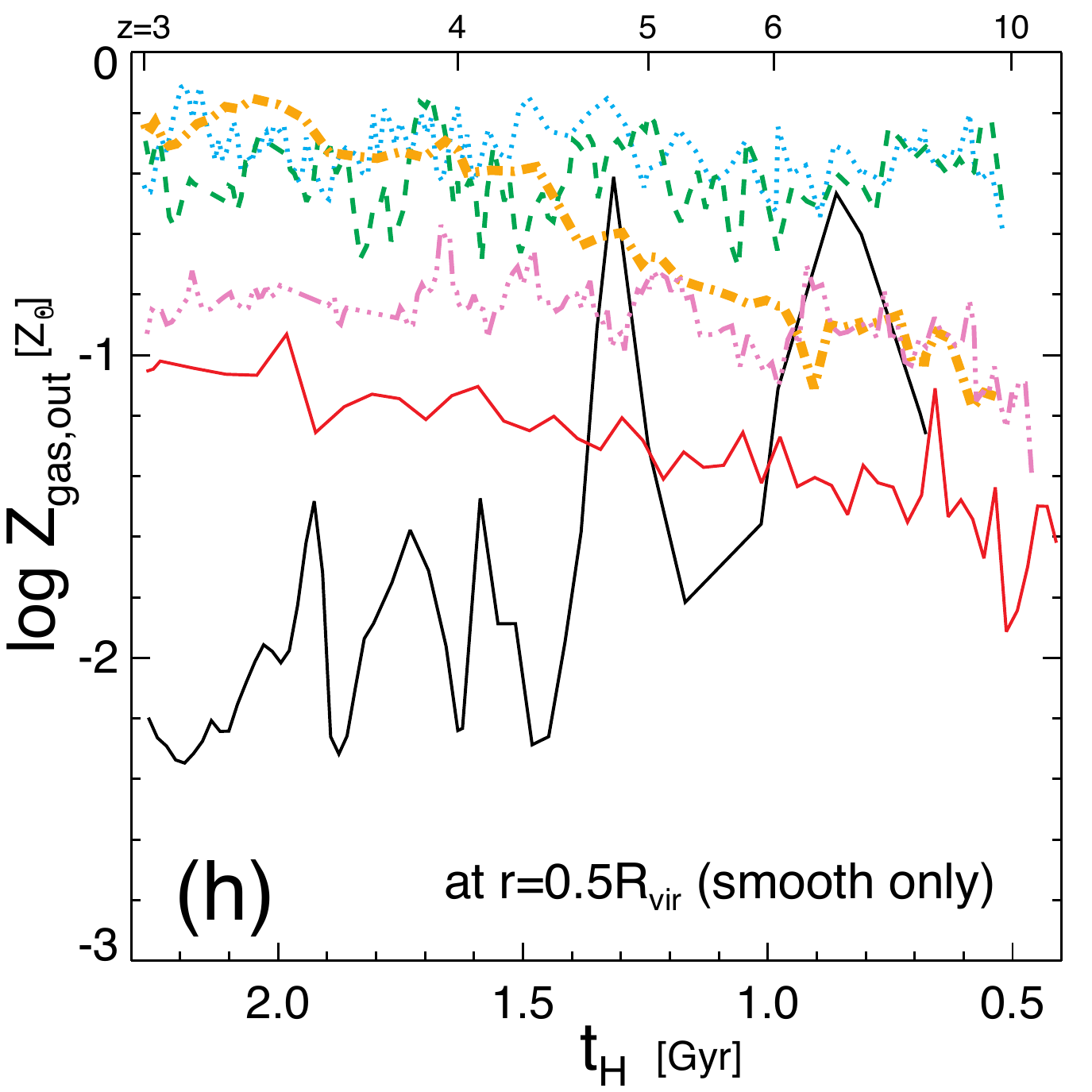} 
      \includegraphics[width=5.8cm]{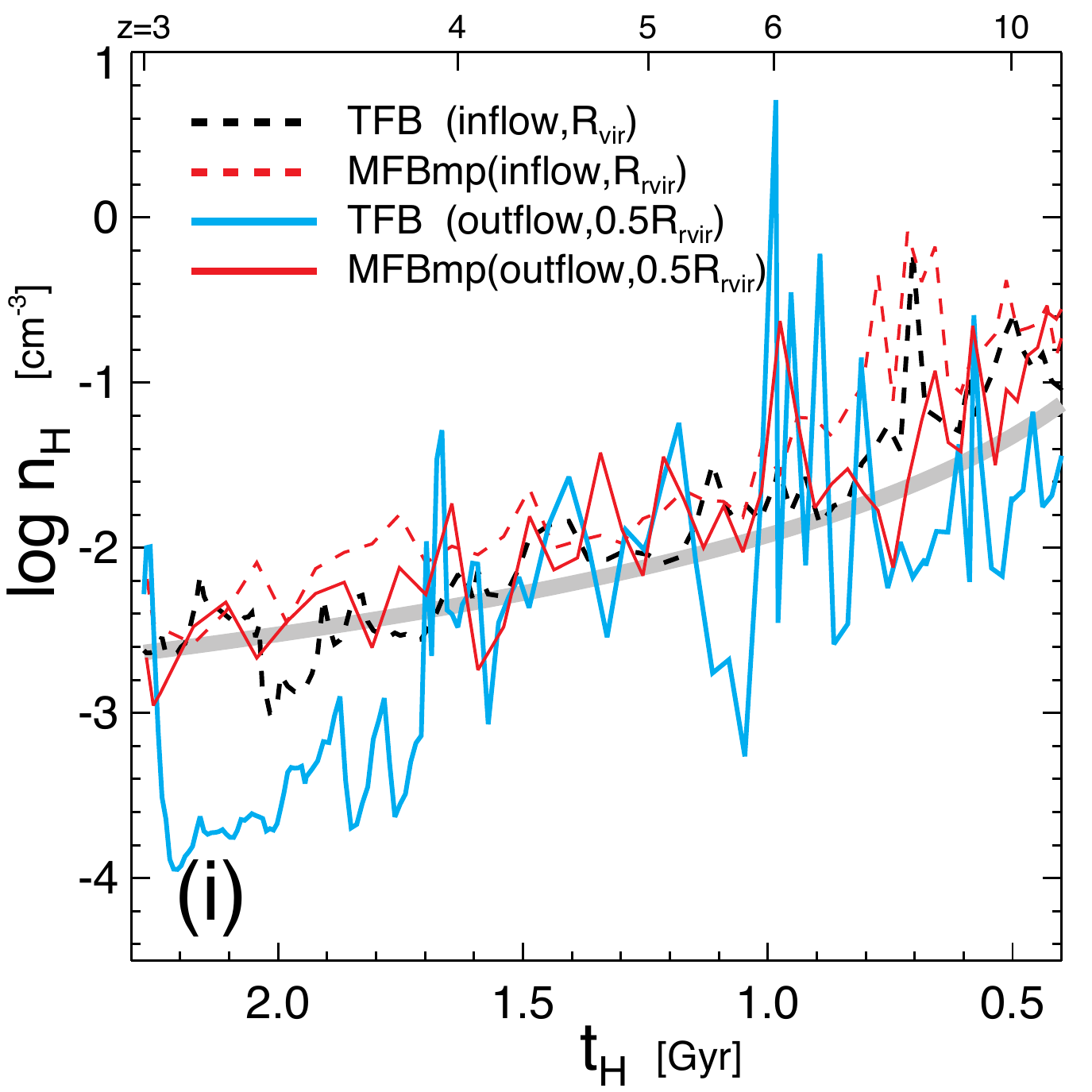} 
   \end{flushleft}
   \caption{Properties of gas inflow and outflow around the main galaxy in the simulations. 
   Different  feedback models are shown as different colour-codings and line styles.
   The top panels (a--c) display the accretion ($dM_{\rm inf}/dt$) or 
   outflow rate ($dM_{\rm out}/dt$) measured at the virial or half of the virial radius. 
   The middle (d--f) and bottom panels (g--h) show the flux-weighted radial velocity ($V_r$) 
   and metallicity ($Z_{\rm gas}$) of the inflow and outflow, respectively. Note that
   we exclude the gas within the tidal radius of satellite galaxies, hence the properties 
   represent the smooth component. Also included in the last panel (i) is the flux-weighted
   hydrogen number density of inflow(dashed lines) or outflow (solid lines) at the virial 
   radius or 0.5 \rvir, respectively (see the legend). For comparison, the mean baryonic density 
   of a virialised halo ($\approx 178 \rho_c(z)\Omega_{\rm b}/\Omega_{\rm m}$) is shown 
   as a gray line, where $\rho_c(z)$ is the critical density of the universe.
   We find that the gas accretion rate is more or less similar regardless of the feedback model,
   indicating that the cold gas filaments are impervious to SN feedback (see Figure~\ref{fig:mmp_img}, first columns). The outflow is stronger in the runs in which star formation 
   is more suppressed. However, the flux-weighted velocity and number density of the 
   outflowing component at 0.5 \rvir\ is largely indistinguishable among the different 
   feedback models. The outflow seen in the run without feedback (NutCO) is not 
   feedback-driven, but induced by mergers. Finally, we find that the inflowing gas is 
   roughly ten times    more metal-poor than outflows.}
   \label{fig:outflow}
\end{figure*}

\subsection{Galactic outflows and inflows}

We now turn our attention to the properties of gas outside the main galaxy. 
The inflow or outflow rate  is measured by computing the flux at radius $r$ as,
\begin{gather*}
\dot{M}_{\rm inf}(r)= \int d\Omega \, 4\pi r^2 \rho_{\rm gas}(r;\Omega) v_{\rm rad}(r;\Omega) \Theta(-v_{\rm rad}),\\ 
\dot{M}_{\rm out}(r)= \int d\Omega \, 4\pi r^2 \rho_{\rm gas}(r;\Omega) v_{\rm rad}(r;\Omega) \Theta(v_{\rm rad}),
\end{gather*}
where $\Omega$ is the solid angle, $v_{\rm rad}$ is the radial velocity with positive 
meaning outflows, and $\Theta$ is the Heaviside step function. Since we are mainly 
interested in the effects of SN explosions by the central galaxy, we exclude the contribution 
from the satellites by ignoring the gas within their tidal radius.

We find that the cosmic gas inflow is impervious to SN explosions 
(Figure~\ref{fig:outflow}, panel (a)). Although strong SN feedback (MFBmp) suppresses 
star formation by an order of magnitude, it cannot completely destroy the dense gas 
filaments around the virial radius (see Figure~\ref{fig:mmp_img}, first columns), 
and both the inflow rate and the net infalling velocity are only mildly reduced (panel (d)). 
Other runs with weak feedback show a comparable inflow rate as in the NutCO run. This suggests that, if a significant fraction of baryons is missing from the host dark matter halo 
of a Milky Way-like galaxy \citep[][c.f.,\citealt{werk14}]{mcgaugh10}, it is unlikely 
to be caused by the suppression of smooth gas accretion, but rather by stellar feedback
blowing gas out of the halo.

\begin{table*}
   \centering
   \caption{Simulated  properties of the most massive galaxy embedded in a 
   $1.2\times10^{11}\, \msun$ halo at $z=3$. 
   From left to right, each column corresponds to the model name, the total stellar mass of 
   a galaxy inside 0.2 \rvir\ of a dark matter halo (\mstar), star formation rate averaged 
   over the last 10 Myr ($\dot{M}_{\rm star}$), the gas-to-stellar mass ratio ($M_{\rm gas}/\mstar$) within the virial radius,
   the stellar-to-halo mass ratio ($\mstar/M_{\rm halo}$) within the virial radius, the fraction of baryons inside 
   a virial radius ($M_{\rm bar}/\mhalo$), the mean metallicity of stars ($Z_{\rm star}$), 
   the mean metallicity of gas ($Z_{\rm gas}$) within 0.2 \rvir, outflow rate measured 
   at 0.5 \rvir\ ($\dot{M}_{\rm out}$), and the half-mass radius of the main galaxy. }
   \begin{tabular}{lccccccccc}
   \hline
      Model & \mstar  & \sfr & $M_{\rm gas}^{\rm vir}$/$M_{\rm star}^{\rm vir}$ & $M_{\rm star}^{\rm vir}/M_{\rm halo}$ & $M_{\rm bar}^{\rm vir}/M_{\rm halo}$ 
      & $Z_{\rm star}$ & $Z_{\rm gas} $ & $\dot{M}_{\rm out}$ & $r_{\rm eff,m}$\\
        &  [$M_{\odot}$] & [$M_{\odot}/yr$]  &   &&& [$Z_{\odot}$]&[$Z_{\odot}$]& [$M_{\odot}/yr$] &  [kpc] \\
      \hline
      NutCO      & $1.9\times10^{10}$ & 6.2  & 0.45  & 0.147 & 0.238  &  1.03 & 0.65& 0.02 & 0.43\\
      NutTFB     & $ 1.3\times10^{10}$ & 4.3 & 0.69 & 0.096 & 0.184  & 0.95 & 0.63 & 0.50 & 0.93\\
      NutKFB     &  $1.5\times10^{10} $ & 5.7 & 0.58 & 0.115 & 0.195 & 0.98 & 0.50 & 0.03 & 0.40\\
      NutMFB    &  $7.8\times10^{9} $  &5.5 & 1.31   & 0.061 & 0.144 & 0.62 & 0.58 & 0.67 & 0.81\\
      NutMFBm &  $9.8\times 10^{9} $  &2.4 & 1.14  & 0.062 & 0.148 & 0.82 & 0.30 & 0.38 & 0.31\\
      NutMFBmp & $7.4\times10^{8}$   &0.6 & 10.2  & 0.006 & 0.071 & 0.06    & 0.07 & 1.20 & 1.5\\
      \hline
   \end{tabular}
   \label{table:props}
\end{table*}

\begin{figure}
   \centering
   \includegraphics[width=8cm]{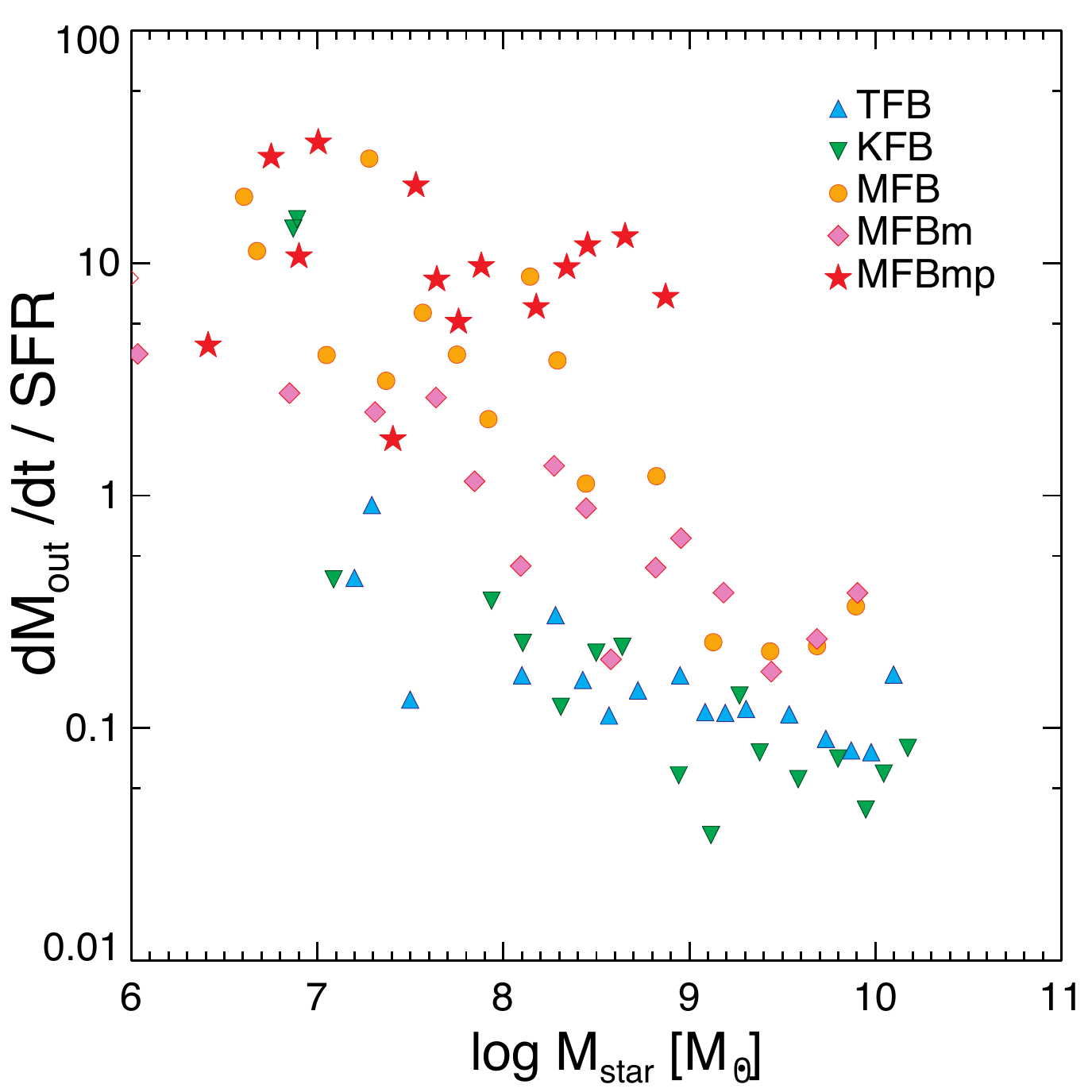} 
   \caption{The mass-loading efficiency of winds in the main galaxy at redshifts $3\le z \le 12$.
   The loading factor is computed by dividing the outflow rate measured at the virial radius  
   with the star formation rate averaged over 10 Myr. The strong SN feedback model 
   (MFBmp) leads to a high $\dot{M}_{\rm out}/{\rm SFR}$ of 5--10 when the galaxy stellar
   mass is greater than $10^8\,\msun$. Such a high mass loading is normally  
   required to reproduce the stellar mass functions in large-scale 
   hydrodynamic simulations 
   \citep[e.g.][]{oppenheimer10}. }
   \label{fig:loading}
\end{figure}

The more effective a feedback model is at reducing star formation,
the larger the outflow rate it produces (panel (b)). While the runs with thermal and kinetic 
feedback (TFB and KFB) blow gas out of the virial 
sphere at a rate of 1--2 $\msunyr$, the mechanical feedback runs  show 
more significant outflows of 1-5 $\msunyr$ (MFB and MFBm) or 5-10 $\msunyr$ (MFBmp). 
The outflow rates measured at smaller radii (0.5 \rvir) tend to be larger than those  
at the virial radii (panel (c)), indicating that the outflow is launched more or less 
ballistically.
Although we do not include the figure in this paper, we find that a large 
fraction ($\sim60\%$ ) of baryons escapes from the halo in the MFBmp run,
resulting in $(\mstar+\mgas)/\mvir=0.071$ at $z=3$ (see Table~\ref{table:props}).
Note that this is a factor of three lower than that in the cooling run 
($(\mstar+\mgas)/\mvir=0.238$)\footnote{
\citet{faucher-giguere11} also found a similar feature in 
their SPH simulation without feedback, and attributed this to the delayed gas infall due to 
photoionization by the UV background. However, we find that the main halo in the NutCO 
run shows the high fraction even before the UV background is instantaneously turned on at z=8.5.
We argue that the high baryon fraction is due to the fact that baryons and dark 
matter particles settle into a dark matter 
halo in a different way \citep{kimm11a}. Whereas gas agglomerates at the centre of a 
halo once accreted, collision-less dark matter particles pass through the central region 
first and populate the outskirts or regions beyond the virial radius of the halo 
depending on their angular momentum. 
Thus, it may not be surprising that the baryon fraction is higher than $f_b$  
if one measures the fraction in the most baryon-rich region (i.e. halo centre). }.

Due to the enhanced outflow and reduced star formation in the MFBmp run, 
a high mass-loading factor ($\eta=\dot{M}_{\rm out}/\dot{M}_{\rm star}$) of 5--10 is 
observed for galaxy stellar masses of $10^8\lesssim M_{\rm star}\lesssim10^9\,\msun$ 
(Figure~\ref{fig:loading}). This is roughly two orders of magnitude larger than 
those of weaker feedback runs ($\eta\sim0.1$, TFB or KFB). Here we use 
the outflow rate at the virial radius, and thus the mass-loading factor could be even 
higher if it is measured at smaller radii. We note that such a high wind efficiency 
in dwarf-sized galaxies is required in large-scale cosmological simulations to 
match the stellar mass functions \citep{oppenheimer10}. 
\citet{hopkins12a} also show using idealised simulations with parsec-scale resolutions that 
the mass loading is $\eta\sim10$ for their Sbc galaxy where the maximum circular velocity 
of the dark matter halo ($86\,\kms$) is comparable to our main galaxy at $3\lesssim z \lesssim6$.
The comparison is by no means sufficient to demonstrate the success of the models, however, 
and future works will have to directly compare to observable signatures of the outflows, 
such as low-ionization absorption lines \citep{rupke05,martin05}.

\begin{figure*}
   \centering
   \includegraphics[width=7cm]{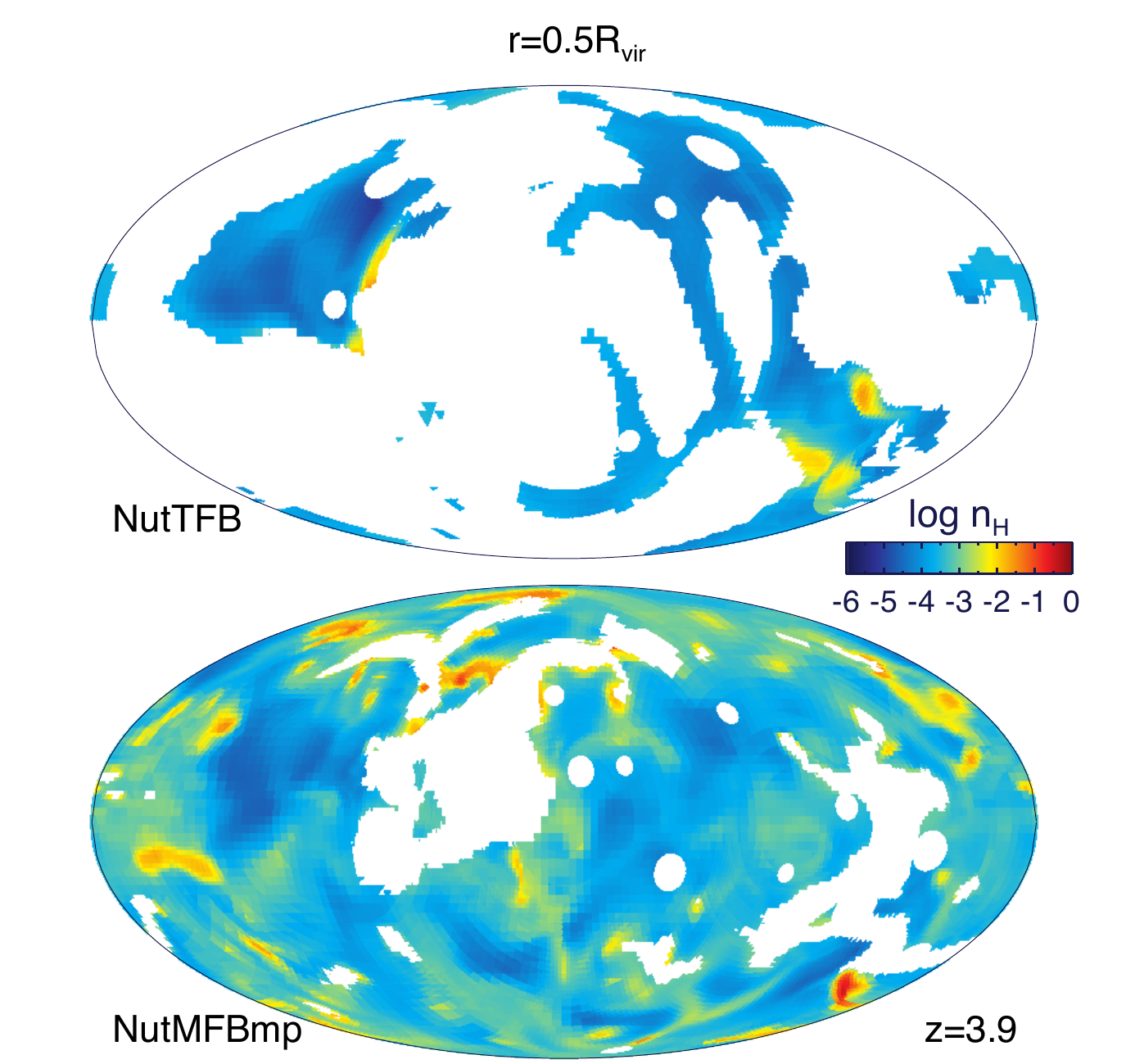} 
      \includegraphics[width=7cm]{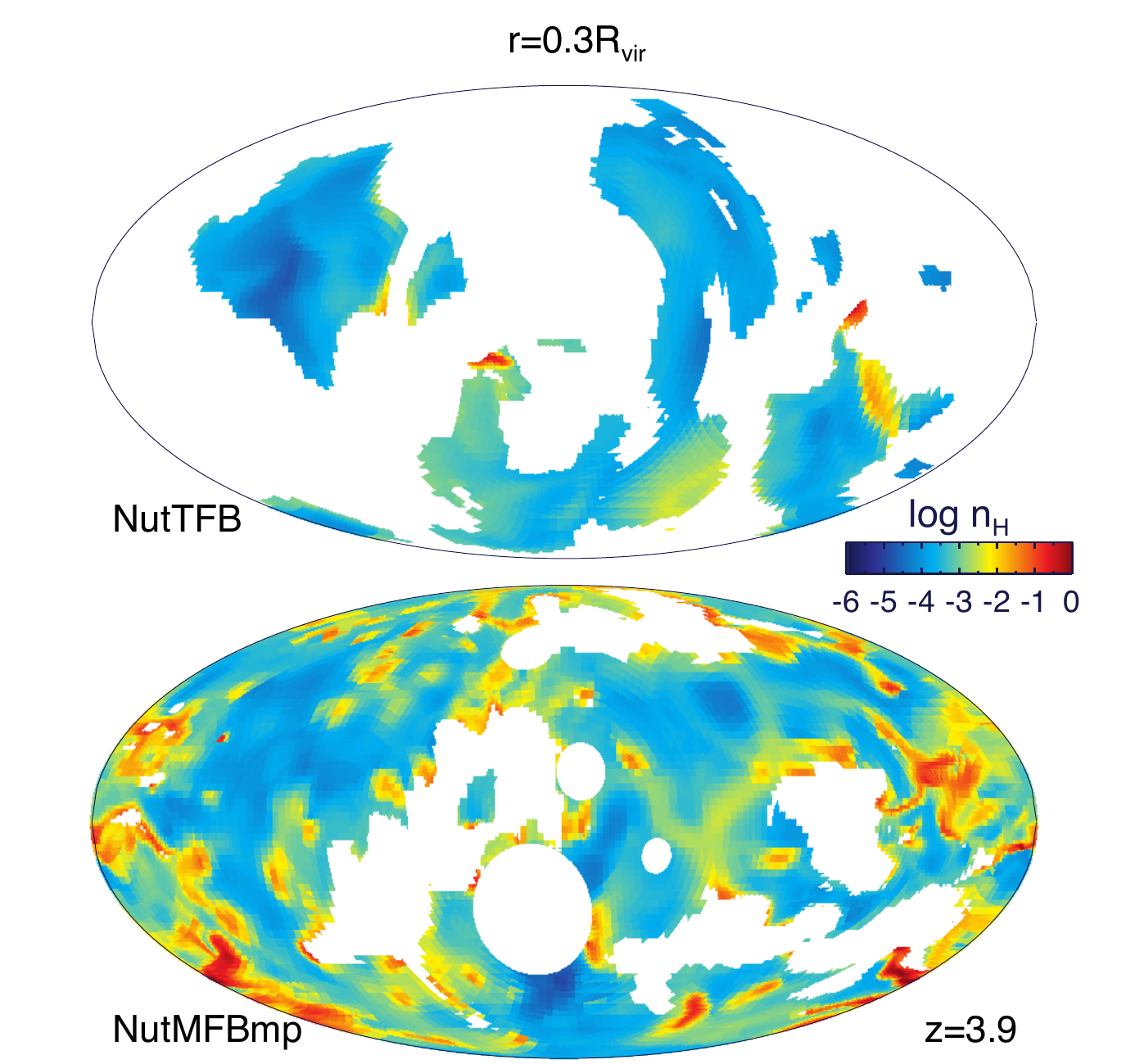} 
   \caption{Differences in the opening angle of the outflow 
   between the thermal SN feedback (top) and mechanical feedback run 
   with the assumption of a highly porous ISM (bottom).
   The panels display the distribution of hydrogen number density on a sphere of 
   0.5 \rvir\ (=13.3 kpc, left) at $z=3.9$ when the 
   flux-weighted outflowing velocity and density are very similar 
   in the two runs. The right panels correspond to the density maps at smaller 
   radii ($0.3 \rvir = 8.0$ kpc). The gas belonging to satellites galaxies or infalling gas 
   is removed for clarity. The plot demonstrates that the outflow in the MFBmp run 
   is stronger than the TFB run, because it subtends a larger solid angle than the former. 
   At smaller radii, the NutMFBmp run drives
   winds composed of many gas clumps with $n_{\rm H}\gtrsim 0.1\,{\rm cm^{-3}}$. 
   }
   \label{fig:opening}
\end{figure*}

We also find that the physical properties of the outflowing gas are  
generally similar in all feedback runs. Figure~\ref{fig:outflow} (panels e,f, and i) shows
the flux-weighted velocity and hydrogen number density of the outflow in different runs.
The galactic winds (panels e,f) are faster (150--300 \kms) than the inflow (panel d),
and their density at 0.5 \rvir\ is comparable to the mean baryonic density of a virialised halo
($\approx 178 \rho_c(z)\Omega_{\rm b}/\Omega_{\rm m}$),  regardless of the feedback model used (panel i), where $\rho_c(z)$ is the critical density of the universe.
This suggests that it is the opening angle of the outflow that determines the overall outflow rate.
To substantiate this, we compare the distribution of the outflowing gas between NutTFB
and NutMFBmp in Figure~\ref{fig:opening}.  The plot shows the {\sc Healpix} map 
\citep{gorski05} of hydrogen number density on a sphere of 0.5 \rvir\ at $z=3.9$ 
when the outflow velocity of the two runs is very similar.
For clarity, both the infalling gas and the gas belonging to satellites are removed. 
It is evident from the figure that the outflow rate in the MFBmp run is larger,
as it subtends a larger solid angle than the one in the thermal feedback.

Note that the typical density shown in panel (i) (Figure~\ref{fig:outflow}) 
is sensitive to the radius at which it is measured. For example, it can increase up to 0.1--1 ${\rm cm^{-3}}$ 
in the MFBmp run at $0.3\rvir$, as small gas clumps that are  blown away from 
the galaxy can contribute to the outflows (Figure~\ref{fig:opening}, the bottom right panel).
These gas clumps with $n_{\rm H}>0.01\,{\rm cm^{-3}}$ 
move outwards more slowly ($\sim 40 \,\kms$) than the mean outflow velocity ($\sim90\,\kms$) 
at 0.3 \rvir, and thus we expect that many of these clumps do not escape from 
the halo \citep[see also][]{cooper08}.  On the other hand, the runs with weak 
feedback (NutTFB or NutKFB)  cannot form many outflowing clumps, 
as SN explosions cannot transfer the correct amount of momentum to the ISM.

Figure~\ref{fig:outflow} (panel h) shows that the outflowing gas is metal-rich, 
but generally sub-solar at $z\ge3$ in the main halo. This means that SN ejecta entrains 
a considerable amount of gas, which is at least 1.5 times the ejecta mass, even in the 
weak feedback runs (TFB and KFB), given that the assumed metallicity of the SN ejecta 
is 2.5 solar. Thus, a small mass-loading factor of $\eta\sim0.1$ seen in Figure~\ref{fig:loading} 
arises simply because not all SN explosions drive large-scale winds. 
Because of the significant entrainment, the metallicities of the outflow are comparable 
to the gas-phase metallicity of the main galaxy (Figure~\ref{fig:zgas}). For example, 
the main galaxy in the MFBmp run with a mass loading factor of $\eta\sim10$ reveals 
a significantly smaller metallicity of $\sim 0.1 Z_{\odot}$ than the ejecta metallicity.
The cooling run displays somewhat complex behaviour, as the outflow in this case is 
driven by mergers and interactions with satellite galaxies.

As  accretion is dominated by the cold filamentary gas in our main halo at $z\ge3$ 
\citep{kimm11a}, the metallicity of the inflow is found to be an order of magnitude smaller 
($0.005-0.03\,Z_{\odot}$) than that of the outflow. The different metallicity between the inflow
and outflow demonstrates that the outflow is not efficiently mixed with the cold flows 
(see Figure~\ref{fig:mmp_img}, second columns).
Indeed, the metallicities turn out to be different even for the MFBmp run where the 
outflow rate is comparable to the infall rate.  
Recently, \citet{crighton13} analyse the absorption features of a $z=2.44$ 
system with the background quasar at $z=2.66$, and find a cool ($T<20,000\,{\rm K}$) 
component with 0.007--0.015 solar metallicity. Their neutral column density ($N_{\rm H}=10^{19.50\pm0.16}\,{\rm cm^{-2}}$) as well as the low metallicity are surprisingly similar 
with the properties of the inflow in our simulations, supporting 
their claim that it is a direct detection of the cold flow around a LBG.

\section{Discussion}
\label{sec:discussion}

\subsection{Overcooling problem and a viable solution}

\begin{figure}
   \centering
   \includegraphics[width=8cm]{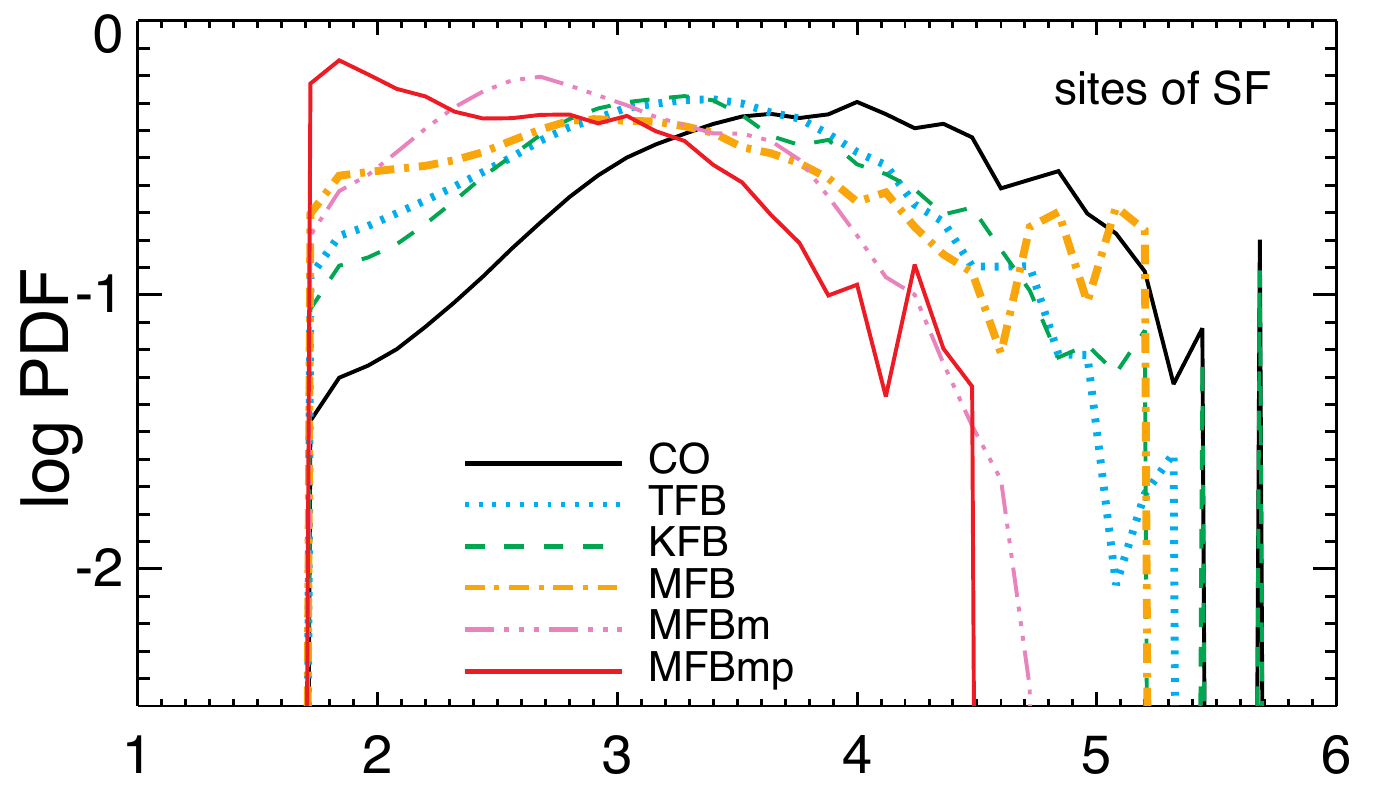} 
      \includegraphics[width=8cm]{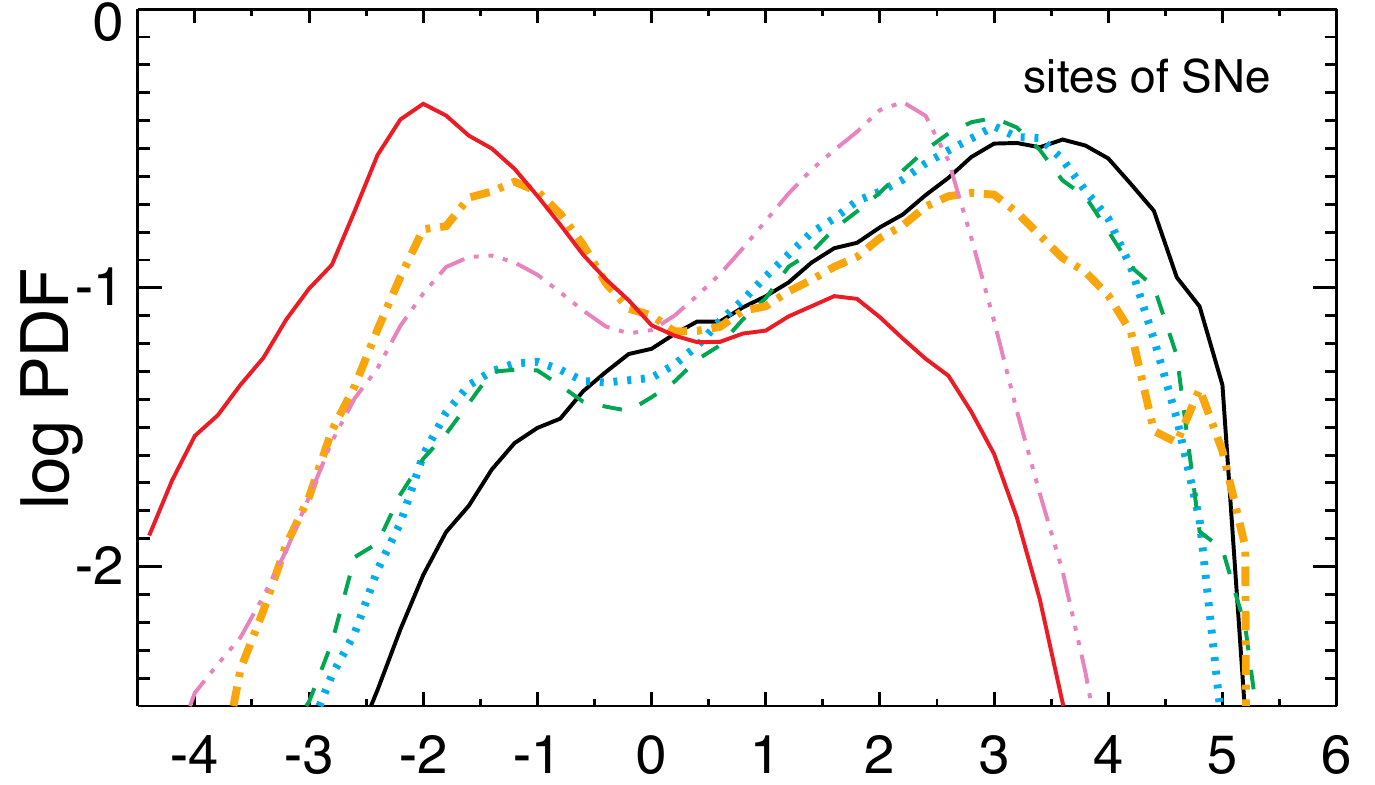} 
           \includegraphics[width=8cm]{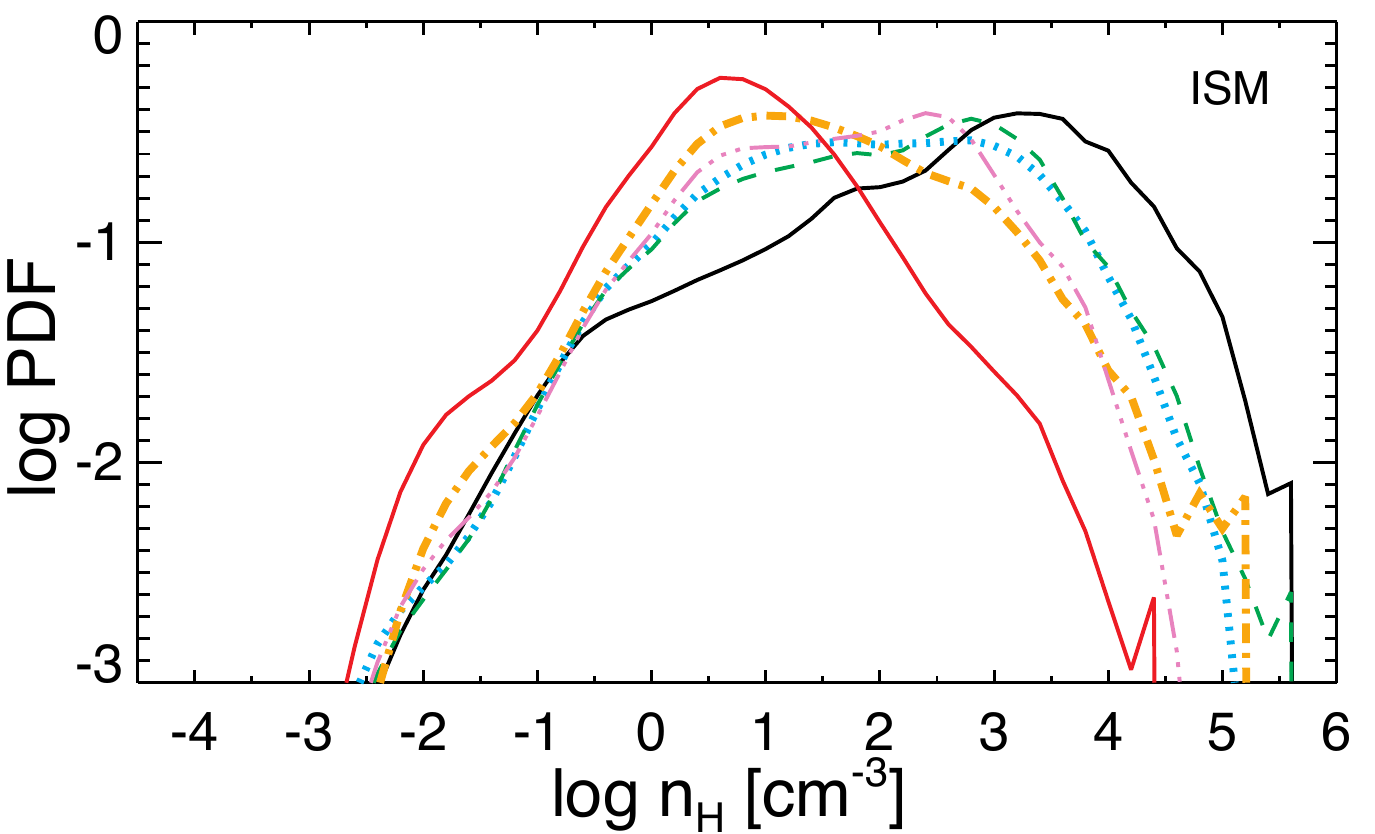} 
   \caption{Local environments at which star formation occurs (top) or SNe 
   explode (middle). The PDFs of the density at which stars form ($n_{\rm H,form}$) are 
   obtained by computing $M_{\rm gas}/t_{\rm ff}$ of the ISM in the main galaxy
    at $3\le z \le 10$. Similarly, we compute the PDFs of the density at which SNe 
    explode ($n_{\rm H,SN}$) using potential SN particles in the main galaxy (see text).
    Note that although there is no energy input, the NutCO run also has SNe 
    that release metals at 10 Myr after the birth of a star particle.
   Also included in the bottom panel is the mass-weighted PDF of the gas within a small
   radius [$r=\min\left(0.2\rvir,{\rm2\,kpc}\right)$].
   The realistic time delay of SNe commencing at 3 Myr suppresses the formation of 
   very dense gas ($n_{\rm H}\gtrsim 10^{4}\,{\rm cm^{-3}}$, MFBm and MFBmp),
   allowing SNe to explode at lower density ($\nH\lesssim 300\,\cmq$) than the weak 
   feedback runs (TFB or KFB).    More SNe emerge in lower density environments 
   when SN feedback becomes stronger.
   }
   \label{fig:sites}
\end{figure}

In the previous section, we have shown that a large amount of gas is converted into stars 
at the centre of a galaxy if the feedback cannot regulate star formation effectively.
The overcooling problem is a well known issue and reported in many earlier 
works based on thermal SN feedback \citep{katz92} or a kinetic feedback model \citep{dubois14}. 
In this work, we demonstrated that the use of the momentum available 
at the end of the adiabatic phase (NutMFB and NutMFBm) cannot eliminate 
the overcooling problem,
even though it better suppresses star formation in small galaxies.
More importantly, we find that star formation histories of galaxies predicted from 
hydrodynamic simulations are very sensitive to the details of how momentum from SNe
is distributed \citep[see also][]{Hennebelle14}. 

In order to understand why some feedback models are more effective at blowing 
gas away than others, we compute the probability distribution functions (PDFs) of 
the density at which stars primarily form ($n_{\rm H,form}$, top panel) and 
the density at which SNe explode ($n_{\rm H,SN}$, middle panel) in Figure~\ref{fig:sites}.
The PDFs of $n_{\rm H,form}$ are obtained by computing the actual probability 
($\propto \mgas/t_{\rm ff}$) of star formation in the ISM of the main galaxy 
from $\sim$50 snapshots at $3\le z \le10$. 
The sites of SN explosions are chosen as regions where young 
stars with the age $9\le t\le 10\, {\rm Myr}$ are located for the TFB, MFB, and CO runs, 
or by taking into account the realistic SN rates based on the metallicity and age of 
each star in the case of the MFBm and MFBmp runs at $3\le z\le10$ \citep{leitherer99}.
Also included in the bottom panel is the mass-weighted PDF of the ISM, 
which we simply calculate for gas within a radius $r_{\rm ISM}$ 
[$\equiv\min\left(0.2\rvir, {\rm 2\,kpc}\right)$].
The figure shows that in the absence of SN feedback, the majority of stars are born 
in a very dense medium ($n_{\rm H,form}>10^3\,{\rm cm^{-3}}$). Because the stars do
not undergo violent dynamical events (except for mergers), most of the metals are simply
distributed around their birth cloud (Figure~\ref{fig:mmp_img}). 

The inclusion of thermal feedback lowers the density peak of the PDF for 
star-forming clouds to $n_{\rm H,form}\sim10^3\,{\rm cm^{-3}}$,
and allows $\approx 10\%$ of SNe to emerge at densities $n_{\rm H,SN} \le 1\, {\rm cm^{-3}}$. 
However, most of the SNe explode in the dense ISM ($n_{\rm H,SN}\sim10^3\,{\rm cm^{-3}}$).
Note that the shell formation radius at this density is only $\approx$ 2 pc 
for metal-poor gas \citep[$Z=0.01\, Z_{\odot}$,][]{thornton98}.
Given that the radius should be resolved by at least three computational 
cells to capture the final momentum from SN \citep{kim15}, it is not surprising 
that thermal feedback fails to regulate star formation in the simulation with 12 pc resolution.
Only SNe that explode in an ISM with $n_{\rm H} \le 2\,\cmq$ may 
approximate the Sedov-Taylor solution with 12 pc resolution, but this requires 
a very stringent refinement that a cell should not contain more than 110 \msun\ of gas.

The kinetic feedback scheme by \citet{dubois08} is originally designed to 
alleviate the overcooling problem by distributing a boosted momentum of 
$p_{\rm SN}=\sqrt{(1+\eta_w)2 E_{\rm SN} M_{\rm ej}}$ to the neighbouring cells of a SN, 
where $\eta_w = {\rm min}(\eta_{w,{\rm max}}, M_{\rm host}/M_{\rm ej})$ and $M_{\rm host}$ is the gas mass
of the host cell of a SN. 
However, if the adiabatic phase of SN expansion is not resolved,
then radial momentum cannot build up and so the impact of SNe
is essentially limited by the initial momentum injected to the ISM.
Furthermore, the assumption used in the model that the mass loading 
only comes from the gas in the host cell of a SN reduced the estimation 
of the loading factor ($\eta_w$) in a low density medium or in regions where 
there are multiple SNe, compared to the MFB run where neighboring gas is 
also taken into account to calculate the swept-up mass.
As a result, the sites of star formation and supernova explosion in the KFB run 
turn out to be quite similar to those of the TFB run. 

The most notable difference in the mechanical feedback runs (MFB, MFBm, and MFBmp) 
from others is that a significant fraction of SNe explodes in low densities 
of $n_{\rm H}\le 1\,{\rm cm^{-3}}$ (40\%, 26\%, and 78\%, respectively).
The PDFs of $n_{\rm H,SN}$ are bimodal in these runs, which reflect the fact 
that the low-density gas ($n_{\rm H}\sim10^{-2}\,{\rm cm^{-2}}$) becomes more 
volume-filling in the main galaxy. This low-density gas has a wide distribution of 
temperature ranging from $10^4$ to a few times $10^7$ K, indicating that the 
multiphase ISM is naturally generated with SN explosions \citep{cox74,mckee77}. 
\citet{kim13} simulate SN-regulated star formation by injecting the final momentum 
and unlike us, they find that the ISM is composed of cold and warm 
($T\lesssim10^4\,{\rm K}$) gas. We attribute this difference to the 
fact that SNe in their simulations are assumed to emerge instantaneously 
as soon as a dense gas cloud of $n_{\rm H}=100\,{\rm cm^{-3}}$ forms, 
whereas the time delay of 3--40 Myr allows SNe to explode not only in the dense environment, but also in a low-density 
medium in our simulations. 

Figure~\ref{fig:sites} demonstrates that a realistic time delay for SNe   
is crucial for preventing gas from collapsing to very high densities 
($n_{\rm H}>10^{4}\,{\rm cm^{-3}}$) (the MFBm and MFBmp runs).
This is essentially because more frequent explosions keep stirring dense gas.
The bottom panel indicates that, in the case of MFBm, the mass fraction for 
gas with densities in the range $100\lesssim \nH \lesssim 10^4\,\cmq$ increases,
lowering to  $\nH\lesssim300\,\cmq$ the typical density at which a SN explodes.
However, the gas mass contained in a 12 pc cell with a density of 100 \cmq 
($\approx 5500 \,\msun)$ already exceeds the mass swept up by the end of the 
adiabatic phase ($\approx1300\,\msun$, for $0.1\,Z_{\odot}$). Thus,
the velocity gained from an individual SN would be only $\sim 2\,\kms$ 
if the final momentum of  $2.4\times10^5\,\msun\, \kms$ (Equation~\ref{psn_one}) is 
equally distributed to 48 neighboring cells of the same density.
Given that the local escape velocity ($\sim 8\,\kms$) is greater than this, 
it is very unlikely that a single SN would disrupt the entire cloud.
One way to unbind the gas is for collective winds to continually lift up the gas.
This may be plausible because the gas experiencing a runaway collapse will always 
produce young stars. On the other hand, the ram pressure due to gas accretion 
onto a collapsing gas cloud is likely to counterbalance the expanding motion.
The MFBm run suggests that such collective events do not occur frequently 
enough to unbind the gas.
We note, however, that this may simply be due to the fact 
that the structure of the ISM in our simulation is intrinsically smooth at our 
resolution limits (12 -- 48 pc), and that fast outflows along low-density channels 
are not resolved.

Small-scale turbulence simulations show that the structure of the ISM is 
complex \citep[e.g.][]{padoan97,ostriker01,molina12,federrath12}, and that the 
volume-filling gas density is smaller than the mean density by 
$\exp{\left[-0.5\ln\left(1+\beta b^2\mathcal{M}^2/(1+\beta)\right)\right]}$, 
where $\beta$ is the ratio of thermal to magnetic pressure, $b$ is the turbulence 
forcing parameter ($0.3\le b\le1$), and $\mathcal{M}$ is the Mach number.
As aforementioned, it is likely that our simulations underestimate the extent to which SNe channel 
through a lower density medium on scales of 12--48 pc \citep{iffrig15,kim15}, 
i.e. our resolution scale, hence, too much gas is likely to be entrained along with 
the SN ejecta from the host cell.
Reducing the mass loading from the host cell, as in the MFBmp run, 
increases the outflow velocity in low-density neighbours, 
and the input momentum during the snowplough phase increases 
slightly ($\propto n_{\rm H}^{-2/17}$).

\begin{figure}
   \centering
   \includegraphics[width=8.6cm]{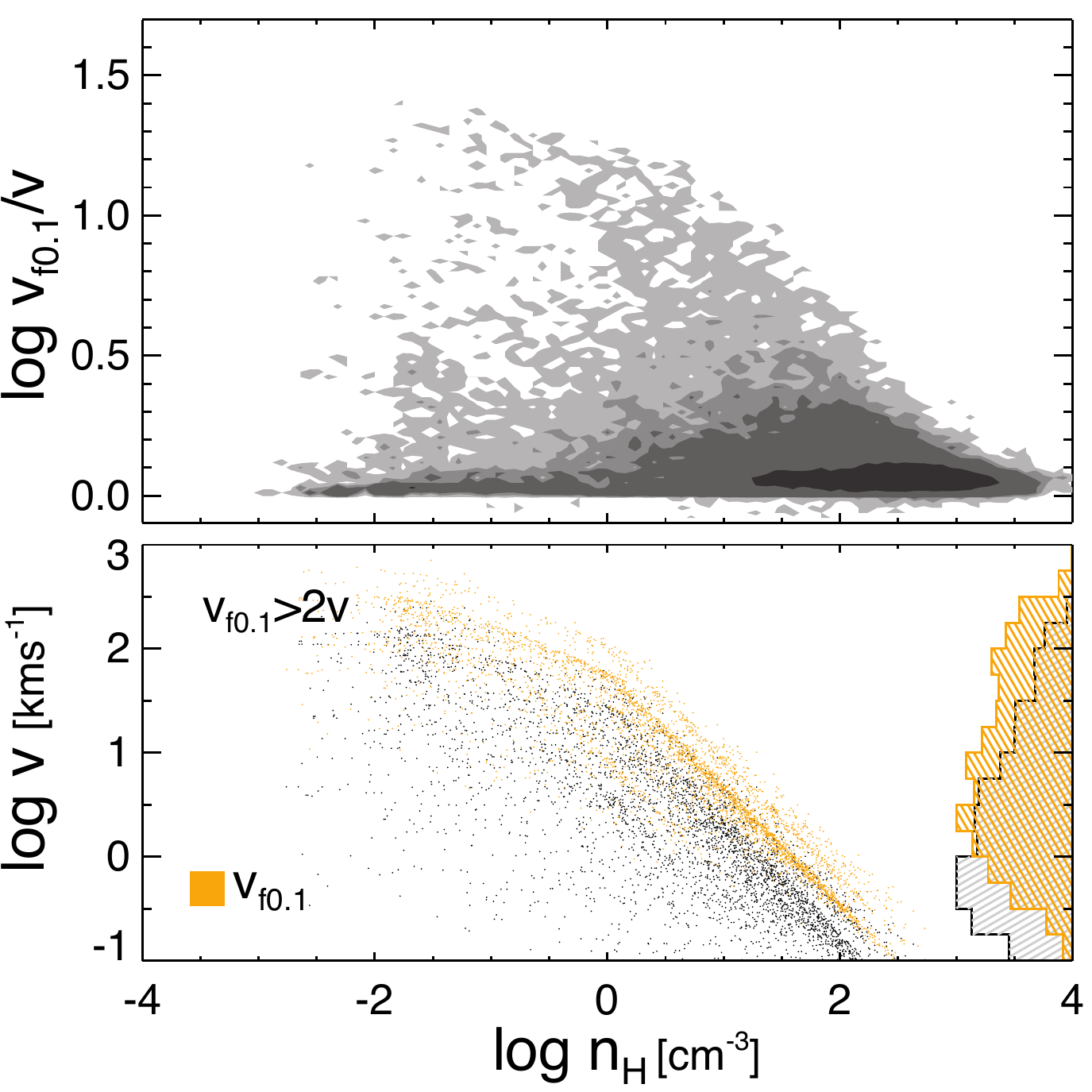} 
   \caption{Expected increase in the outflow velocity of gas when the ISM is assumed 
   to be highly porous. The top panel shows the ratio between the outflow velocity 
   estimated assuming that all the gas in the host cell of the SN is entrained ($v$) 
   and the velocity  in the case where the volume-filling density is ten times lower 
   than the mean density ($v_{\rm f0.1}$). Approximately 10\% of the cells that are 
   directly affected by the SN explosions would have a twice faster outflow along 
   low-density channels than the fiducial case. The bottom panel represents the 
   velocities of the cells with $v_{\rm f0.1} > 2 v$. Black and orange points denote 
   $v$ and $v_{\rm f0.1}$ in physical units, respectively. The knee present around 
   $\nH\sim1\,\cmq$ corresponds to the transition between the adiabatic and 
   snowplough phase. 
   }
   \label{fig:vboost}
\end{figure}

Figure~\ref{fig:vboost} illustrates an example of how much we expect the outflow 
velocity of each cell to be boosted when we assume that $f_{\rm w,host}=0.1$ 
 (i.e. only 10\% of the gas from the host cell is entrained) instead of $f_{\rm w,host}=1$. 
 We use the ISM structure of the main galaxy at $z=3$ from the cooling run 
 to compare the velocity for the case where $f_{\rm w,host}=0.1$ ($v_{\rm f0.1}$) 
 to that where $f_{\rm w,host}=1$ ($v$).  
We assume that young stars with $3\le t \le 40$ Myr are SNe. The top panel shows that 
$v_{\rm f0.1}$ can be a factor of $\sim10$ larger than $v$. 
Note that such a significant increase is rare because it can only occur around a very 
sharp edge of a dense clump where there is a rapid spatial variation of the density 
($\rho_{\rm nbor} /\rho_{\rm host} \sim 0.01$).
We still find, however, that $\approx 10\%$ of the gas cells that are 
directly affected by SN explosions would have a velocity more than double their value 
in the case in which the ISM structure is assumed to be smooth on the resolution scale.
This occurs in gas cells where $\rho_{\rm nbor} /\rho_{\rm host} \sim 0.1$.
Of the 10\%, the fraction of cells accelerated more than the typical escape velocity 
of the local GMCs (i.e. $10\,\kms$) increases from 25\% to 65\% for 
$0.1\lesssim \nH \lesssim 10 \,\cmq$. Thus, there is a higher probability that the gas is 
blown away through the lower density channels.
On the other hand, in dense regions with $\nH\gtrsim100\,\cmq$, the velocity change 
is not significant in general, as the local ISM structure is more or less smooth 
in the NutCO run and the mass contained in the neighbour is already significant, 
compared to that entrained from the host cell of SN. Even though only a small fraction 
of gas gains a substantial velocity, 
the comparison between MFBm and MFBmp indicates that 
this plays an important role in dispersing the star-forming gas (Figure~\ref{fig:sites}).
We do not find any sign of overcooling in the MFBmp run. 
This leads us to conclude that {\em the propagation of SN momentum through lower-density 
channels may be crucial in modelling star formation in galaxies, providing a viable 
solution to the overcooling problem}.

\subsection{Comparison to previous work}

To single out the impact from SNe, our simulations do not include any form 
of radiation pressure. However, other feedback processes come into play 
before the first SN ($t<3$ Myr), and the role of SNe in this work may be 
somewhat exaggerated. Photoionisation heating can reduce the 
density by an order of magnitude around young stars by over-pressurising the 
medium \citep{krumholz07,sales14}. Although the determination of the IR optical 
depth is not straightforward \citep{krumholz12,davis14,rosdahl15a}, 
radiation pressure can also drive additional turbulence in optically thick regions. 
Including the additional feedback mechanisms may further reduce 
star formation than the model with SN feedback alone \citep[e.g.][]{agertz13} or 
the importance of SNe may simply be mitigated, as the early processes substitute their role.
Bearing this in mind, we compare our results with previous studies in this section.

Using an AMR code, {\sc Enzo} \citep{bryan14}, \citet{hummels12} performed 
16 cosmological simulations of Milky Way-like galaxies with a few hundreds of pc resolution. 
They included SN feedback by heating the host cell of SN.
Because the thermal energy is quickly radiated away, more than 50\% of the baryons ends up 
turning into stars, leading to a too peaked rotation curve \citep[see also][]{joung09}.
The resulting star formation history is found to be smooth, in agreement with 
the results from the NutTFB run. Although there are several differences in numerical methods 
(e.g. hydrodynamic solver) between the two studies, this suggests that adopting a higher 
resolution does not solve the angular momentum catastrophe  \citep{navarro00,governato04} 
so long as the spatial resolution is larger than the shell formation radius. 

To compensate for the finite resolution ($\sim$50--300 parsecs) of cosmological simulations,
several studies adopted the cooling suppression model \citep{stinson06,governato07,brooks11,governato10,avila-reese11,guedes11,teyssier13,shen14}. 
As gas is prevented from collapsing onto the centre of dwarf to Milky way-sized galaxies by 
the over-pressure from SNe, galaxies in these studies showed a smoothly rising rotation curve
in general, as in observations.
Because they only presented the results at low redshifts ($z\le1$), it is difficult to make a 
direct comparison with our work, but star formation histories in the Milky Way-sized galaxies 
appear rather smooth at high redshifts ($z\gtrsim2$)
\citep[][c.f. \citealt{shen14} for more episodic SF in dwarf galaxies]{guedes11},
compared to our MFBmp run or \citet{hopkins14}.
\citet{hummels12} also examined the model in which the radiative cooling is turned off for 50 
Myr in a $10^{12}\,\msun$ halo, and found that the central peak in the rotation curve is reduced, 
but still too many stars formed ($\sim60\%$), compared to the local \mstar--\mhalo\ 
relation \citep[e.g.][]{moster10}.

Recently, \citet{hopkins11,hopkins12a} investigated the effects of momentum transfer from 
the UV--IR photons on the ISM in four different types  of isolated galaxies 
(SMC-like dwarf, Milky Way-like spiral, gas-rich star-forming galaxies hosted by a
$\sim10^{11}\,\msun$ halo, and star-bursting disc in a $10^{12}\,\msun$ halo) with an 
entropy-conserving SPH code \citep{springel05}. 
They carried out a series of simulations, varying the feedback model, with parsec scale 
resolution, and concluded that the radiation pressure is mainly responsible for the 
generation of galactic winds in the actively star-forming, gas-rich galaxies, 
whereas SN explosions play a major role in relatively quiescent galaxies 
(SMC and MW-like galaxies). However, their SN feedback model is based on the initial 
energy of $10^{51}\,{\rm erg}$,  which is likely to have underestimated its impact. 
Our Nut galaxy is similar to their gas-rich, star-forming galaxy (Sbc), but we find that the 
SN explosions are capable of driving a strong outflow if the ISM is assumed to be 
porous (MFBmp). 
As mentioned, the wind efficiency ($\dot{M}_{\rm out}/\dot{M}_{\rm star}$) from the 
MFBmp run (Figure~\ref{fig:loading}) is comparable or slightly larger than the star-forming 
galaxy (Sbc) in \citet{hopkins12a}. In a subsequent paper, \citet{hopkins14} incorporate the 
final momentum from the SN blast wave in a dense medium, in a very similar spirit to 
our mechanical feedback scheme, and concluded that both SNe and radiation pressure 
are essential for reproducing the observed relations, such as \mstar--\mhalo\ relation or 
Kennicutt-Schmidt law \citep{kennicutt98,bigiel08}.

By including SN, stellar winds, and momentum transfer from radiation
 in the \ramses\ code \citep{teyssier02}, \citet{agertz13} systematically examined 
 their relative importance in an isolated cloud of $10^{6}\,\msun$ and a disc galaxy 
 embedded in a halo of mass $10^{12}\,\msun$. Because of their finite resolution, 
 they used the the local size and mass relation of cluster/clumps to estimate the 
 surface density, which is distinct from \citet{hopkins11}, who directly measured it 
 from their simulations. \citet{agertz13} pointed out that the inclusion of early 
 feedback is important for disrupting the dense gas in star-forming regions. 
 Applying the radiation pressure model to cosmological simulations, 
\citet{agertz15} argue that reproducing a realistic disc galaxy requires two conditions: 
i) radiation pressure should be included on top of the energy variable that dissipates away 
slowly on a 10 Myr timescale, ii) star formation should be modelled, such that a large 
fraction ($\epsilon=10\%$) of gas is converted into stars on a local dynamical timescale.
They find that both the \mstar-\mhalo\ relation and the mass--metallicity relation are 
well accounted for if these conditions are met simultaneously. It is interesting to note that 
when a low star formation efficiency  ($\epsilon=1\%$) is used, their radiation pressure 
model could not slow down star formation in the Milky Way-like galaxy and converted
$50\%$ of the baryons into stars.
The need for a high $\epsilon$ appears to suggest that star formation should be 
bursty for radiation pressure to control the growth of galaxies.
Given that our MFBmp run achieved such bursty SFH with 
a relatively low star formation efficiency ($\epsilon=2\%$), the requirement for a high 
$\epsilon$ may be model-dependent.

Several authors also studied the role of radiation pressure, and come to a different 
conclusion. Using the {\sc ART} code \citep{kravtsov97}, \citet{ceverino14} argue that 
radiation pressure from the ionising photons alone provide enough energy to unbind 
star-forming gas clumps, making the role of multiply scattered IR photons to be less 
significant. However, the results are not straightforward to compare with other studies, 
because they model the radiation pressure as a non-thermal term that does not dissipate 
away until the density becomes lower than the star formation threshold ($\nH=1\,\cmq$). 
\citet{agertz13} tested a similar model by using a ``feedback energy variable'', and 
showed that the resulting star formation histories are quite sensitive to the dissipation 
time scale as well as the fraction of the energy that is deposited into the non-thermal 
term. By contrast, \citet{aumer13} claim that a high IR optical depth of $\sim20$ is 
requisite at $z>2$ to form a realistic disc galaxy in SPH simulations. They could 
reasonably match not only the \mstar--\mhalo\ relation but also the size and 
metallicity evolution in 16 dwarf to intermediate-size galaxies resimulated 
from the Aquarius project \citep{springel08} and some simulations presented 
in \citet{oser10}. The different conclusions on the role of radiation pressure by 
UV to IR photons call for further tests based on hydrodynamics calculations fully 
coupled with radiative transfer \citep{krumholz12,davis14,rosdahl15a}.

\section{Conclusions}
We have investigated the impact of SN feedback on the evolution of the progenitor 
of a Milky Way-like galaxy, embedded in a $10^{11}\,\msun$ halo 
at high redshifts ($z\ge3$). For this purpose, we have carried 
out a suite of zoom-in cosmological hydrodynamic simulation with high spatial (12 pc) 
and mass ($m_{\rm dm}=5.5\times10^4\,\msun$, $m_{\rm star}=610\,\msun$) resolution 
using the adaptive mesh refinement code, \ramses, varying the SN feedback model.  
Our main results can be summarised as follows. 

\begin{enumerate}

\item Our simulations confirm that too many stars would form when the feedback from SN 
cannot drive strong outflows. The cooling run without SN feedback (NutCO) turns 85\% of 
baryons into stars inside 0.2 \rvir\  of the main halo ($\mhalo\approx10^{11}\,\msun$), 
resulting in a much smaller gas-to-stellar mass ratio (Figure~\ref{fig:sf}) than 
observations at high redshifts. The run with thermal feedback (NutTFB) also 
shows the overcooling problem, as the cooling radius of the SN blast wave is still 
under-resolved even in our zoom-in simulations. Inclusion of the SN explosion in a kinetic 
form with a small mass loading of $\eta_w\le10$ (NutKFB) does not greatly help either, 
as shock-heated gas is again subject to artificial radiative losses and 
the subsequent generation of radial momentum 
during the adiabatic phase is not guaranteed. Nevertheless, the increase in the input 
momentum ($\sqrt{1+\eta_w}$) used in the kinetic feedback scheme 
helps suppress star formation in low-mass haloes 
($\mhalo\lesssim10^{10}\,\msun$) by an order of magnitude, compared to the cooling 
run (Figure~\ref{fig:shmr}).

\item The mechanical feedback models (NutMFB, MFBm, and MFBmp) that transfers
the correct amount of radial momentum from the adiabatic to snowplough phase is able 
to substantially suppress star formation in small haloes ($\mhalo\lesssim10^{10}\,\msun$).
The galaxy stellar mass of the haloes is reduced by roughly two orders of magnitude 
from that found in the cooling run (Figure~\ref{fig:shmr}).

\item The prediction of star formation histories in the main galaxy depends on 
the details of how momentum from SN is distributed. The model in which SNe 
emerge relatively late (10 Myr, MFB) could not prevent a runaway collapse of a 
massive cloud at the galaxy centre during frequent mergers of clumps (at $z\lesssim 5$, 
Figure~\ref{fig:overcool}). This led to a rapid increase in galaxy stellar mass 
from $z\sim5$. The final stellar mass at $z=3$ was over-predicted, compared with the 
typical observational estimates from the clustering analysis of LAEs or the abundance 
matching technique.

\item Inclusion of a realistic time delay for SNe, commencing at 3 Myr, efficiently 
suppresses the build-up of very dense gas ($\nH\gtrsim10^{4}\cmq$) in the main galaxy 
(Figure~\ref{fig:sites}, MFBm). But because gas outflow velocities become smaller,
distributing SN explosions in time makes a galaxy more susceptible to the overcooling 
problem than the case where at least10 SNe explode simultaneously 
(MFB). As a result, star formation is not strongly regulated from high redshifts 
($z\gtrsim5$), and follows that of the weak feedback models 
(TFB or KFB, Figure~\ref{fig:sf}).

\item The observed relations, such as the mass-metallicity relation and stellar mass to 
halo mass ratio, are reproduced by the SN feedback model with the realistic time delay,
provided that the ISM is highly porous (MFBmp). The latter assumption is motivated by 
the observation that galaxies at high redshifts are more turbulent than their local 
counterpart \citep{forster-schreiber09}. If the ISM is highly turbulent, the volume-filling 
density of the medium would be systemically smaller than the mean. This means that 
taking the uniform gas density at the computational grid scale (as in MFBm) is 
likely to underestimate the wind velocities along low-density channels 
by entraining too much gas from the host cell of SN. We find that reducing the 
volume-filling density to 10\% (equivalent to the supersonically turbulent ISM with the 
Mach number of $\sim$ 10) suffices to prevent the overcooling issue, yielding a mass 
loading that is ten times larger than the star formation rate ($\eta\sim10$, 
Figure~\ref{fig:loading}), as postulated in the momentum-driven winds \citep{oppenheimer10}. 
Although the degree of turbulence is arbitrarily chosen, this demonstrates 
that the momentum input from SNe {\it alone} may be able to regulate star formation, 
as observed. However, future works based on resolved SN feedback will be required to draw 
a firm conclusion on the relative role of different stellar processes
on the evolution of galaxies though.

\item The MFBmp model shows stronger outflow than the weak feedback 
models (e.g. TFB, Figure~\ref{fig:outflow} (b) and (c)). The difference originates from 
the opening angle of the outflow on the sky (Figure~\ref{fig:opening}), not the density 
or velocity, which are found to be insensitive to the feedback prescriptions 
(Figure~\ref{fig:outflow} (e), (f), and (i)). 
On the other hand, metallicity of the outflow is lower in the model with stronger feedback 
and generally follows that of the galactic gas. 
The fact that outflow metallicity ($0.07 \lesssim Z/Z_{\odot} \lesssim 0.7$) is much lower 
than the metallicity of the ejecta ($2.5 Z_{\odot}$) indicates 
that the outflow is efficiently mixed with the ISM before it is blown away.
The strong outflow in this MFBmp model leads to the formation of a spheroid with 
$\mstar=7.8\times10^8\,\msun$ rather than a well-ordered stellar disc at $z=3$ (Figure~\ref{fig:frot}).

\item The cold filamentary inflow around the Nut halo ($\mhalo\approx10^{11}\,\msun$) is impervious to 
SN feedback. The accretion rate at \rvir\ is found to be more or less similar, irrespective of the strength of the feedback (Figure~\ref{fig:outflow} (a)).
The metallicity of the inflow ($Z=0.005-0.02\,Z_{\odot}$) is an order of magnitude 
smaller than that of the outflow, and not very sensitive to the feedback model.

\end{enumerate}

\section*{Acknowledgements}
We are grateful to Sam Geen, Chang-Goo Kim, and Eve Ostriker for fruitful discussions,
and Romain Teyssier for making his code {\sc ramses} publicly available. 
We also thank the anonymous referee for suggestions, which improved the clarity of the paper. 
Computing resources were provided by the NASA High-End Computing (HEC) 
Program through the NASA Advanced Supercomputing (NAS) Division at Ames 
Research Center, the DiRAC facility jointly funded by STFC and 
the Large Facilities Capital Fund of BIS, and the Horizon-UK program through DiRAC-2 facilities. 
The research is supported in part by NSF grant AST-1108700 and NASA grant NNX12AF91G 
and in part by grant  Spin(e)  ANR-13-BS05-0005 of the french ANR.
JD and AS's research is supported by funding from Adrian Beecroft, the Oxford Martin 
School and the STFC.

\small
\bibliographystyle{mn2e}
\bibliography{refs}

\end{document}